\documentclass[oldversion]{aa}
\usepackage{graphicx}
\usepackage{txfonts}
\usepackage{natbib}
\usepackage{fancyheadings}
\usepackage{longtable}
\usepackage{graphicx}
\usepackage{txfonts}
\usepackage{subfigure}
\usepackage[toc,page]{appendix}
\bibpunct{(}{)}{;}{a}{}{,} % to follow the A&A style 

\newcommand{\chisq}{\mbox{$\chi^{2}$}}
\newcommand{\sqdeg}{\mbox{deg$^{2}$}}

\newcommand{\vi}{\mbox{$V\!-\!I$}}

\newcommand{\jks}{\mbox{$J\!-\!K_{\rm s}$}}
\newcommand{\yks}{\mbox{$Y\!-\!K_{\rm s}$}}
\newcommand{\ks}{\mbox{$K_{\rm s}$}}

\newcommand{\dmo}{\mbox{$(m\!-\!M)_{0}$}}
\newcommand{\av}{\mbox{$A_{V}$}}

\newcommand{\evi}{\mbox{$E_{V\!-\!I}$}}

\newcommand{\feh}{\mbox{\rm [{\rm Fe}/{\rm H}]}}

\newcommand{\Msun}{\mbox{$M_{\odot}$}}

\newcommand{\comment}[1]{}
\newcommand{\beq}{\begin{equation}}
\newcommand{\eeq}{\end{equation}}
\newcommand{\beqa}{\begin{eqnarray}}
\newcommand{\eeqa}{\end{eqnarray}}

\newcommand{\avhs}{\mbox{$A_{V}^{\rm HS}$}}
\newcommand{\avcs}{\mbox{$A_{V}^{\rm CS}$}}
\newcommand{\avtw}{\mbox{$A_{V}^{\rm TW}$}}
\newcommand{\dmotw}{\mbox{$(m\!-\!M)_{0}^{\rm TW}$}}
\newcommand{\dmobftw}{\mbox{$(m\!-\!M)_{0}^{\rm BFTW}$}}
\newcommand{\dmoc}{\mbox{$(m\!-\!M)_{0}^{\rm centre}$}}
\newcommand{\RAc}{${\alpha_{\rm c}}$}
\newcommand{\DECc}{${\delta_{\rm c}}$}
\newcommand{\SFRt}{\mbox{${\rm SFR}(t)$}}
\newcommand{\logtyr}{\mbox{$\log(t/{\rm yr})$}}
\begin{document}
\title{The VMC Survey}
\subtitle{IV. The LMC star formation history and disk geometry from
  four VMC tiles \thanks{Based on observations made with VISTA at ESO under program ID 179.B-2003.}}

%\author{Stefano Rubele\inst{1,2}
%\and Leandro Kerber\inst{3}
%\and L\'eo Girardi\inst{1}
%\and Maria-Rosa Cioni\inst{4,5,\star}
%\and Paola Marigo\inst{2}
%\and Simone Zaggia\inst{1}
%\and \\ Kenji Bekki\inst{6}
%\and Richard de Grijs\inst{7,8}
%\and Jim Emerson\inst{9}
%\and Martin A.T. Groenewegen\inst{10}
%\and Marco Gullieuszik\inst{10}
%\and \\ Valentin Ivanov\inst{11}
%\and Brent Miszalski\inst{12,13}
%\and Joana M. Oliveira\inst{14}
%\and Ben Tatton\inst{14}
%\and Jacco Th. van Loon\inst{14}
%}

\author{S. Rubele\inst{1,2}
\and L. Kerber\inst{3}
\and L. Girardi\inst{1}
\and M.-R. Cioni\inst{4,5,\star}
\and P. Marigo\inst{2}
\and S. Zaggia\inst{1}
\and \\ K. Bekki\inst{6}
\and R. de Grijs\inst{7,8}
\and J. Emerson\inst{9}
\and M.A.T. Groenewegen\inst{10}
\and M. Gullieuszik\inst{10}
\and \\ V. Ivanov\inst{11}
\and B. Miszalski\inst{12,13}
\and J.M. Oliveira\inst{14}
\and B. Tatton\inst{14}
\and J.Th. van Loon\inst{14}
}

\institute
{Osservatorio Astronomico di Padova -- INAF, Vicolo dell'Osservatorio
5, I-35122 Padova, Italy 
\and Dipartimento di Astronomia, Universit\`a di Padova, Vicolo 
dell'Osservatorio 2, I-35122 Padova, Italy 
\and Universidade Estadual de Santa Cruz, Rodovia Ilh\'eus-Itabuna, km. 16 -- 45662-000 Ilh\'eus, Bahia, Brazil
\and University of Hertfordshire, Physics Astronomy and Mathematics,
Hatfield AL10 9AB, UK
\and University Observatory Munich, Scheinerstrasse 1, 81679 M\"unchen, Germany
\\ $^{\star}$Research Fellow of the Alexander von Humboldt Foundation
\and ICRAR M468, The University of Western Australia, 35 Stirling Highway, Crawley, WA 6009, Australia
\and Kavli Institute for Astronomy and Astrophysics, Peking University, Yi
He Yuan Lu 5, Hai Dian District, Beijing 100871, China
\and Department of Astronomy and Space Science, Kyung Hee University,
Yongin-shi, Kyungki-do 449-701, Republic of Korea
\and Astronomy Unit, Queen Mary University of London, Mile End Road, London E1 4NS, UK 
\and Royal Observatory of Belgium, Ringlaan 3, B-1180 Brussels, Belgium 
\and European Southern Observatory, Av. Alonso de C\'ordoba 3107, Casilla 19, Santiago, Chile
\and South African Astronomical Observatory, PO Box 9, Observatory, 7935, South Africa
\and Southern African Large Telescope Foundation, PO Box 9, Observatory, 7935, South Africa
\and Lennard-Jones Laboratories, Keele University, ST5 5BG, UK
}

\date{Received August 2011 / Accepted}

%%%%%%%%%%%%%%%%%%%%%%%%%%%%%%%%%%%%%%%%%%%%%%%%%%%%%%%%%

\abstract {We derive the star formation history (SFH) for several
  regions of the Large Magellanic Cloud (LMC), using deep
  near-infrared data from the VISTA near-infrared $YJ\ks$ survey of
  the Magellanic system (VMC). The regions include three
  almost-complete 1.4~\sqdeg\ tiles located $\sim\!3.5^\circ$ away
  from the LMC centre in distinct directions. They are split into
  $21.0\arcmin\times21.5\arcmin$ (0.12~\sqdeg) subregions, and each of
  these is analysed independently. To this dataset, we add two
  $11.3\arcmin\times11.3\arcmin$ (0.036~\sqdeg) subregions selected
  based on their small and uniform extinction inside the 30~Doradus
  tile. The SFH is derived from the simultaneous reconstruction of two
  different colour--magnitude diagrams (CMDs), using the minimization
  code StarFISH together with a database of ``partial models''
  representing the CMDs of LMC populations of various ages and
  metallicities, plus a partial model for the CMD of the Milky Way
  foreground.  The distance modulus \dmo\ and extinction \av\ is
  varied within intervals $\sim\!0.2$ and $\sim\!0.5$~mag wide,
  respectively, within which we identify the best-fitting star
  formation rate \SFRt\ as a function of lookback time $t$,
  age--metallicity relation (AMR), \dmo\ and \av.
  Our results demonstrate that VMC data, due to the combination of
  depth and little sensitivity to differential reddening, allow the
  derivation of the space-resolved SFH of the LMC with unprecedented
  quality compared to previous wide-area surveys. In particular, the
  data clearly reveal the presence of peaks in the \SFRt\ at ages
  $\logtyr\simeq9.3$ and $9.7$, which appear in most of the
  subregions.  The most recent \SFRt\ is found to vary greatly from
  subregion to subregion, with the general trend of being more intense
  in the innermost LMC, except for the tile next to the N11 complex.
  In the bar region, the \SFRt\ seems remarkably constant over the
  time interval from $\logtyr\simeq8.4$ to $9.7$.  The AMRs, instead,
  turn out to be remarkably similar across the LMC.  Thanks to the
  accuracy in determining the distance modulus for every subregion --
  with typical errors of just $\sim\!0.03$~mag -- we make a first
  attempt to derive a spatial model of the LMC disk. The fields
  studied so far are fit extremely well by a single disk of
  inclination $i=26.2\pm2.0^\circ$, position angle of the line of
  nodes $\theta_0=129.1\pm13.0^\circ$, and distance modulus of
  $\dmo=18.470\pm0.006$~mag (random errors only) up to the LMC centre.
  We show that once the \dmo\ values or each subregion are assumed to
  be identical to those derived from this best-fitting plane,
  systematic errors in the \SFRt\ and AMR are reduced by a factor of
  about two.  }

  \keywords {Magellanic Clouds -- Galaxies: evolution -- Surveys --
    Infrared: stars - Hertzsprung-Russell (HR) and colour--magnitude
    diagrams }

\authorrunning{Rubele et al.}
\titlerunning{The SFH across the LMC from VMC data}

\maketitle

%%%%%%%%%%%%%%%%%%%%%%%%%%%%%%%%%%%%%%%%%%%%%%%%%%%%%%%%%%%
\section{Introduction}
\label{sec_intro}

The VISTA near-infrared $YJ\ks$ survey of the Magellanic system
\citep[VMC; ][]{Cioni11} is performing deep near infrared imaging in
the filters $Y$, $J$ and \ks\ for a wide area across the Magellanic
system, using the VIRCAM camera \citep{Dalton_etal06} on VISTA
\citep{Emerson_etal06}. One of VMC's main goals is the derivation of
the complete spatially resolved star formation history (SFH) across
the system. To this aim, the survey has been designed so that the
photometry reaches magnitudes as deep as the oldest main sequence
turn-off (MSTO), for both LMC and SMC, with signal-to-noise ratios of
$\sim\!10$. Extensive pre-survey simulations of VMC images and its
SFH-recovery by \citet{Kerber09} indicated that this goal could be
reached even in the most crowded areas of the LMC bar.\footnote{This
  is true except for the few very highly extincted regions.}

\begin{figure}
\resizebox{\hsize}{!}{\includegraphics{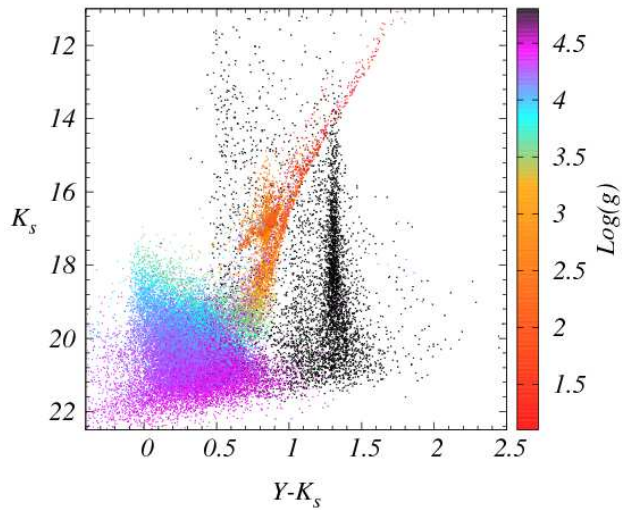}}
\caption{Simulated CMD for a 0.037~\sqdeg\ region of moderate stellar
  density in the LMC.  The black points show the distribution of the
  Milky Way foreground.  The coloured dots show the LMC stars
  according to their surface gravity (in c.g.s. units). }
\label{spop}
\end{figure}

Although the derivation of the SFH from photometry down to the old
MSTO is now routinely performed for most galaxies in the Local Group
\citep[e.g.,][and references therein]{Orban08, Noel09, Williams09,
  Hidalgo11, Weisz11}, this is effectively the first time that such a
method has been tried using data from a near-infrared wide-area
survey. The use of near-infrared filters has the obvious advantage
that the CMDs are less affected by extinction and reddening (and their
differential effects) than optical ones. On the other hand, they are
much more affected by the presence of foreground Milky Way stars.
Fig.~\ref{spop} shows an example of a simulated colour--magnitude
diagram (CMD) for the LMC field, derived as in \cite{Kerber09} using
the specifications of VMC, in which the different colours code the
stellar gravities of the observed stars. The presence of two
almost-vertical strips composed of Milky Way dwarfs is striking: there
is a prominent one at $\yks\simeq1.3$~mag, and a less defined one just
redward of $\yks\simeq0.5$~mag. These features partially overlap the
red giant branch and helium burning sequences of LMC stars.
Fortunately, they exhibit also a smooth and well understood behaviour
as a function of Galactic coordinates, and do not affect in any way
the interpretation of the stars on the LMC main sequence and turn-off.

First results from the VMC survey are described in \citet{Cioni11},
\citet{Miszalski_etal11, Miszalski_etal11a}, and
\citet{Gullieuszik_etal11}. In this paper, we present the recovery of
the SFH for part of the LMC using the first season of VMC data. The
data and their preparation for this work are briefly described in
Sect.~\ref{sec_data}.  Sect.~\ref{sec_sfh} presents the basic results
regarding the star formation rate as a function of lookback time,
\SFRt, and the age--metallicity relation, AMR. The minimization method
employed allows us to estimate also the distance and extinction for
each examined subregion. The geometry inferred for the LMC disk is
discussed in Sect.~\ref{sec_dist_av}. In Sect.~\ref{sec_discussion} we
adopt this revised geometry and revise the results for the \SFRt\ and
AMR, obtaining a significant reduction of the error bars. Finally, we
draw some general conclusions, and compare our results with previous
works.
 
%%%%%%%%%%%%%%%%%%%%%%%%%%%%%%%%%%%%%%%%%%%%%%%%%%%%%%%%%%%
\section{The VMC data}
\label{sec_data}

\begin{figure}
%\begin{minipage}{0.85\textwidth}
%\rotatebox{-90}{
%\includegraphics[trim=0cm 0cm 0cm 0cm, clip=true, totalheight=0.4\textwidth, angle=270]{tileSFH.ps}
\resizebox{\hsize}{!}{\includegraphics[angle=270,origin=c]{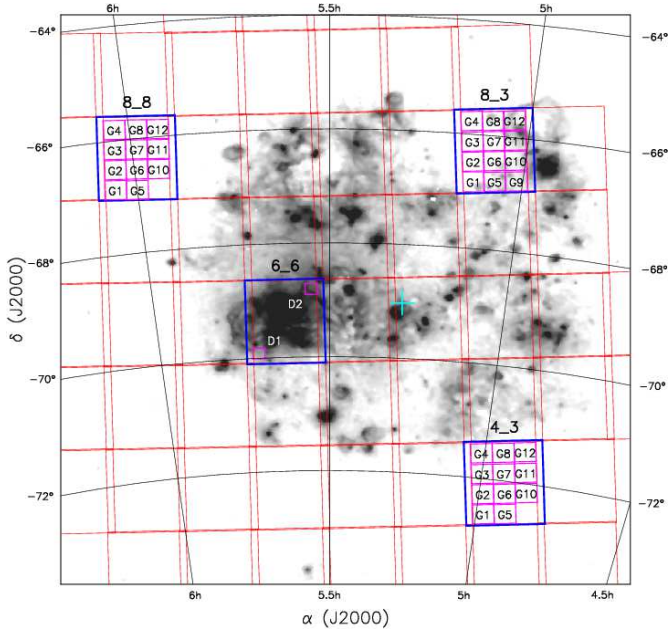}}
%}
%\resizebox{0.45\hsize}{!}{\includegraphics{8_3.RBG_KJY_sub.eps}}
%%%%%\resizebox{0.4\hsize}{!}{\includegraphics{8_3.RBG.ps}}
%\end{minipage}
%\hfill
%\begin{minipage}{0.49\textwidth}
%\resizebox{\hsize}{!}{\includegraphics{completeness_all_colour_3.ps}}
%\end{minipage}
%\hfill
\caption{An H$\alpha$ image of the central LMC from
  the Southern H-Alpha Sky Survey Atlas \citep[in gray;][]{shassa},
  with VMC tiles marked by red rectangles. The blue and magenta
  rectangles show the positions of the tiles and subregions,
  respectively, considered in this work.  The cyan cross marks the
  centre of the LMC as derived by \citet{Nikolaev_etal04}. 
}
\label{field}
\end{figure}

The VMC survey and its initial data are thoroughly described in
\citet{Cioni11}, to which we refer for all details.  We have recovered
the SFH from VISTA data for three VMC tiles located around the main
body of the LMC and for which the VISTA imaging is most complete.
Details about these tiles, and their sub areas, are presented in
Table~\ref{tab_tiles}.  Moreover we have used 2 small subregions in
the more central 6\_6 tile, which comprises 30~Doradus.  The
subregions were selected on the basis of the extinction maps derived
by Tatton et al. (in prep.), as having a small and almost-constant
extinction. Fig.~\ref{field} shows the location of all LMC tiles of
the VMC survey (red rectangles). The black rectangles mark those
used in this work.
   
\begin{table*}
  \caption{VMC tiles used in this work.}
\label{tab_tiles}
\begin{tabular}{ccccccccc}
\hline
\hline
Tile name & $\alpha$ (J2000) & $\delta$ (J2000) & completion in \ks-band & comments\\
\hline
4\_3 & 04:55:19.5 & $-$72:01:53 & 63\% & \\
6\_6 & 05:37:40.0 & $-$69:22:18 & 100\% & 30~Dor field\\
8\_3 & 05:04:53.9 & $-$66:15:29 & 75\% & \\
8\_8 & 05:59:23.1 & $-$66:20:28 & 90\% & South Ecliptic Pole region\\
%GAIA calibration field\\
\hline
\end{tabular}
\end{table*}

We used the v1.0 VMC release pawprints. The pawprints were processed
by the VISTA Data Flow System \citep[VDFS,][]{Emerson_etal04} in its
pipeline \citep{Irwin_etal04} and retrieved from the VISTA Science
Archive
\citep[VSA,][]{Hambly_etal04}\footnote{http://horus.roe.ac.uk/vsa/}.
We combined the calibrated pawprints into deep tiles with the SWARP
tool \citep{Bertin02}. The 4\_3, 8\_3, and 8\_8 tiles, covering areas
of $\sim 1.4$~\sqdeg\ each, were subdivided into twelve subregions of
$21.0\arcmin\times21.5\arcmin$ ($\sim 0.12$~\sqdeg), as illustrated in
Fig.~\ref{field}. In the 6\_6 tile, the selected subregions are
smaller ($11.3\arcmin\times11.3\arcmin$ each, or $\sim 0.036$~\sqdeg).

The pawprints contributing to corner subregion G9 (see
Fig.~\ref{field}) include a contribution from the ``top'' half of
VIRCAM detector number 16 which is known to show a significantly worse
signal-to-noise ratio than the other detectors because its
pixel-to-pixel quantum efficiency seems to vary on short timescales
making accurate flatfielding impossible. The effect is negligible in
\ks, small in $H$ and becomes more noticeable in the bluer bands, i.e.
$J$, $Y$ and $Z$.  Therefore subregions G9 of tiles 4\_3 and 8\_8 are
not further considered. 8\_3 is used as its signal-to-noise in G9, in
the $Y$ band, was just $\sim16$~\% smaller than in neighbouring
subregions.

\subsection{Photometry and Artificial Star Tests}

We performed point spread function (PSF) photometry using the IRAF
DAOPHOT packages \citep{daophot}, generating photometric catalogues
and CMDs. We used the PSF package to produce the PSF model (variable
across the subregion), and the ALLSTAR package to perform the
photometry using a radius of three pixels (2.5 pixels in the case of
the 6\_6 tile). We checked that our PSF photometry produced results
consistent with those provided by VSA for the bulk of the observed
stars. For the SFH-recovery work, PSF photometry was preferred to
aperture photometry because it produced deeper catalogues, especially
in the case of the highly crowded regions in the 6\_6 tile.

\begin{figure}
\centering
%\begin{minipage}{\textwidth}
%\end{minipage}
%\begin{minipage}{\textwidth}
\resizebox{0.48\hsize}{!}{\includegraphics{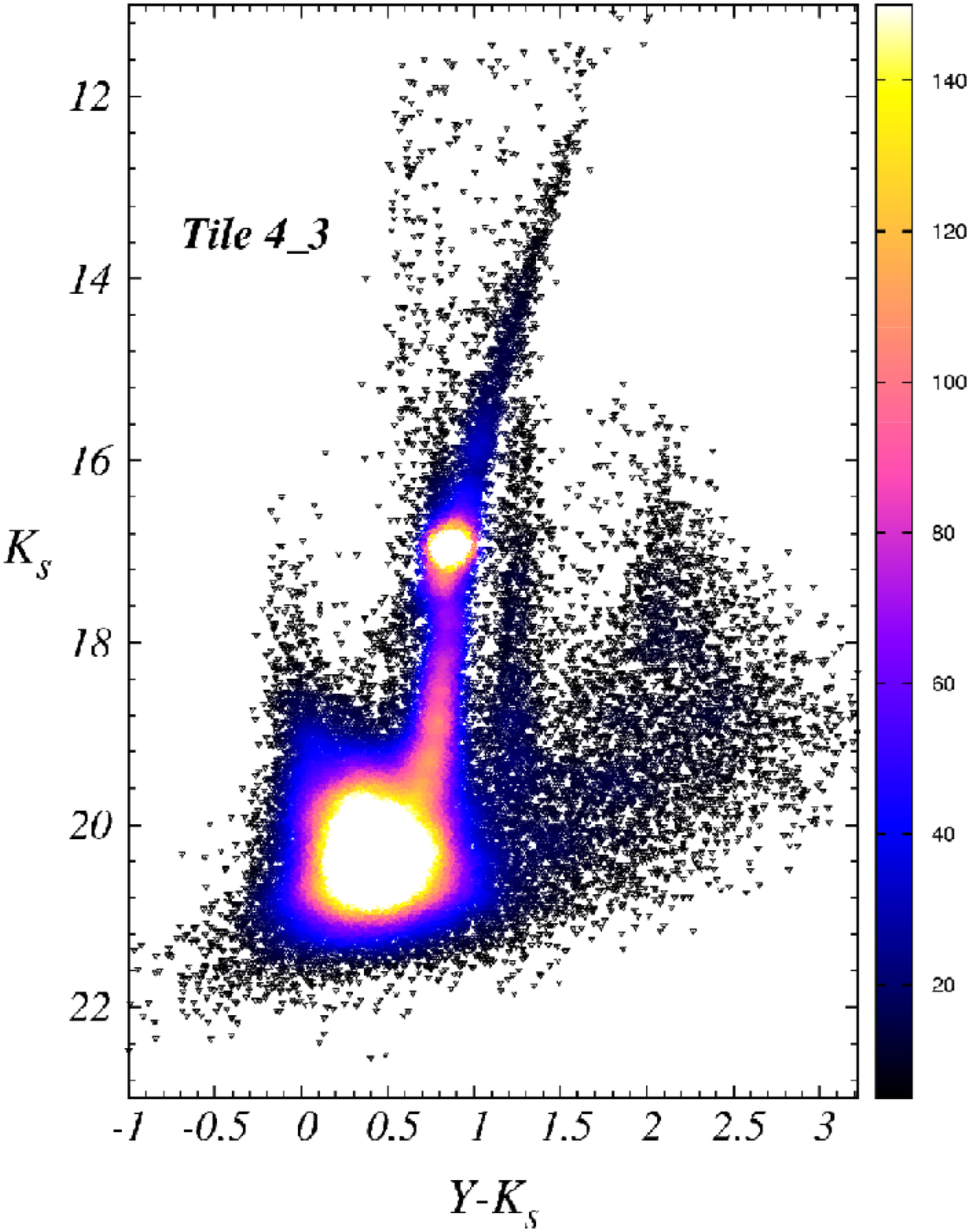}}
\resizebox{0.48\hsize}{!}{\includegraphics{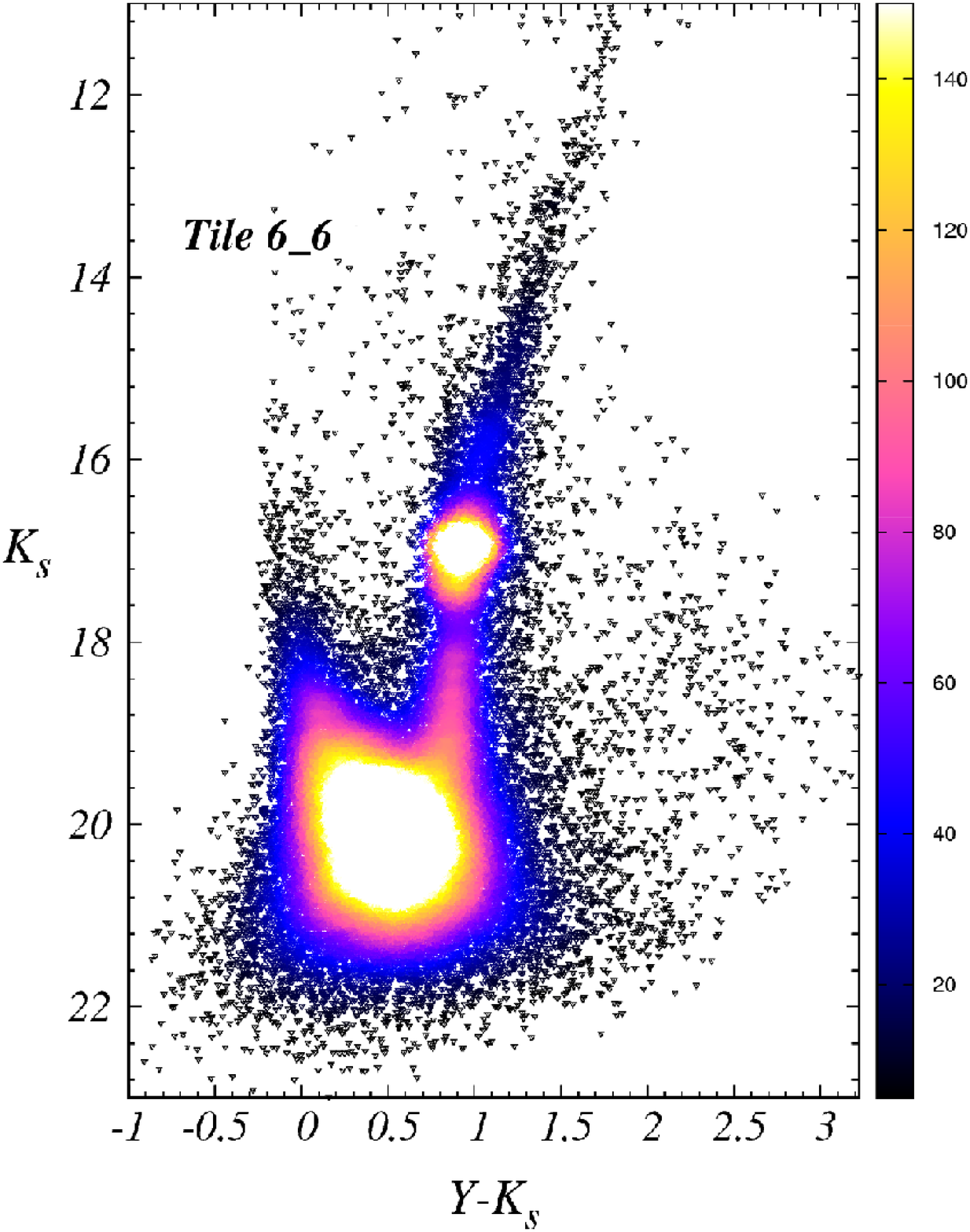}}\\
\resizebox{0.48\hsize}{!}{\includegraphics{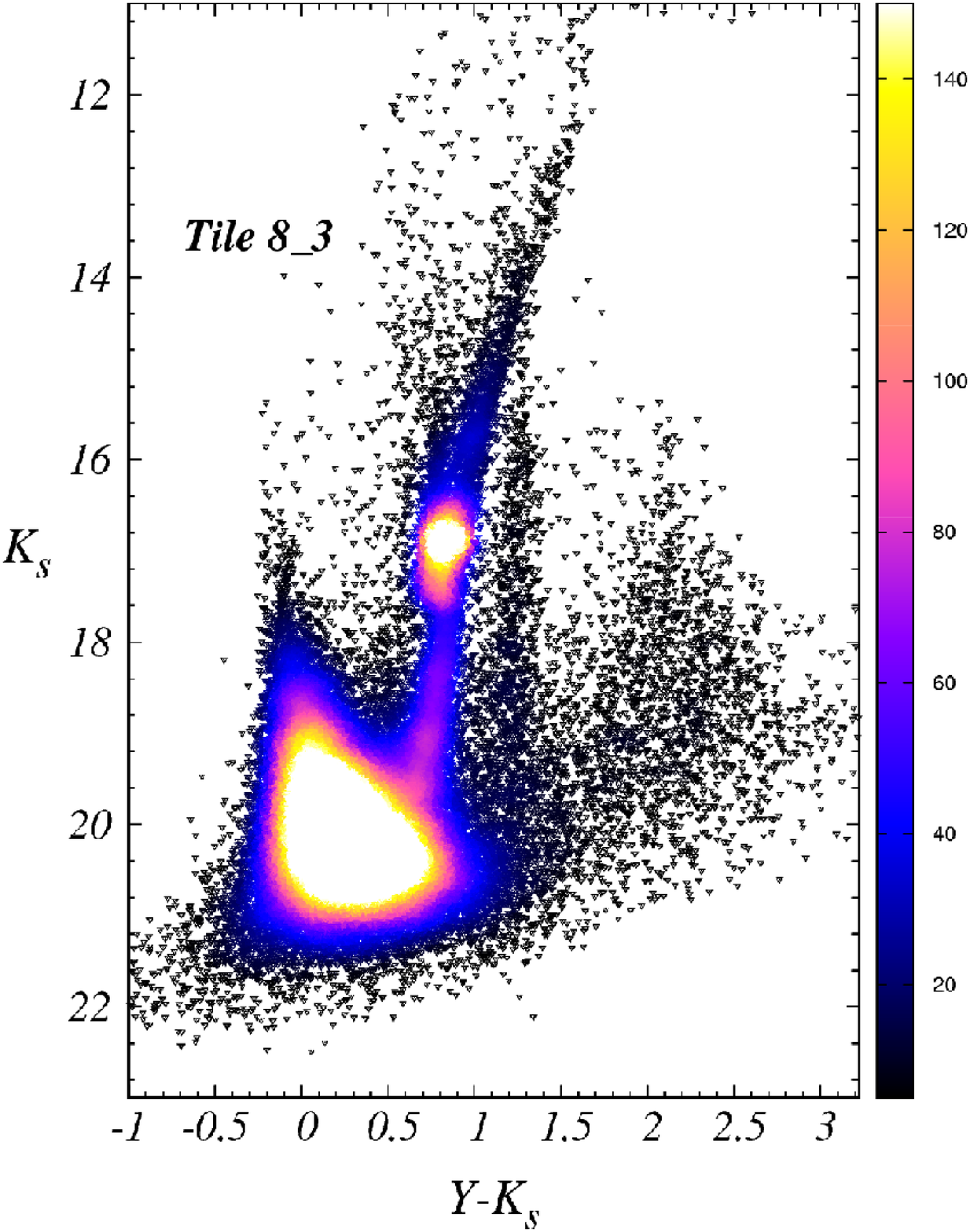}}
\resizebox{0.48\hsize}{!}{\includegraphics{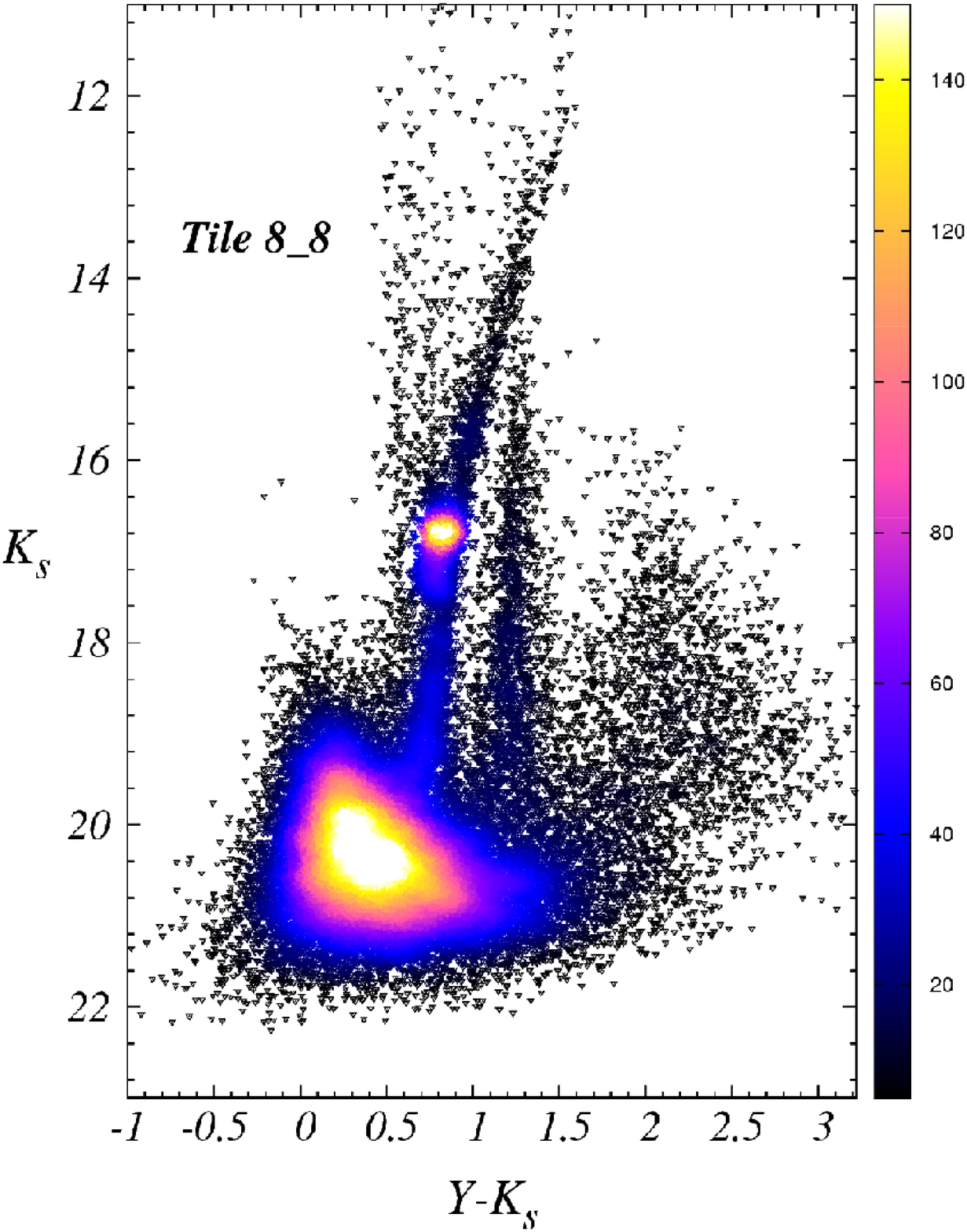}}\\
%\end{minipage}
\centering
\hfill
\caption{Examples of $\ks$ vs.\ $\yks$ CMDs in subregions of the 4\_3
  (top left), 6\_6 (top right), 8\_3 (bottom left) and 8\_8 (bottom
  right) tiles. An arbitrary color scale is used to highlight the CMD
  regions with a higher density of stars.}
\label{cmd}
\end{figure}

Fig.~\ref{cmd} shows some examples of $\ks$ vs.\ $\yks$ CMDs, for the
subregion G3 of tiles 8\_8, 8\_3, and 4\_3, and for the subregion D2
in the 6\_6 tile.

We have recovered the SFH using two CMDs simultaneously, namely $\ks$
vs.\ $\yks$ and $\ks$ vs.\ $\jks$. We recall that the contamination by
compact galaxies can be mostly prevented by simply eliminating objects
with $\jks>0.88$~mag and $\yks>1.56$~mag \citep[see][]{Kerber09} from
our data.  This is done later in our analysis (see
Sect.~\ref{sec_sfh}).

We have run large numbers of artificial star tests (ASTs) to estimate
the incompleteness and error distribution of our data for each
subregion and in every part of the CMD. For each subregion we ran
$\sim\!2.8\times10^6$ ASTs as described in \cite{Rubele11}, using a
spatial grid with 30 pixels width and with a magnitude distribution
proportional to the square of the magnitude. This latter choice allows
us to better map completeness and errors in the less complete regions
of the CMD.

\begin{figure}
%\begin{minipage}{\textwidth}
\resizebox{\hsize}{!}{\includegraphics{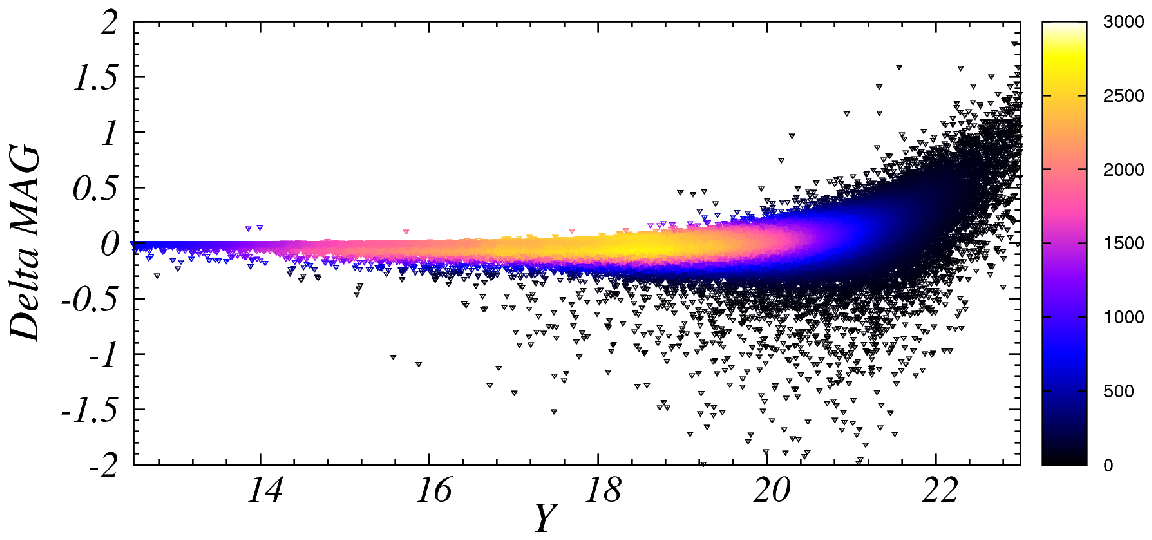}}
% \end{minipage}
\hfill
%\begin{minipage}{\textwidth}
\resizebox{\hsize}{!}{\includegraphics{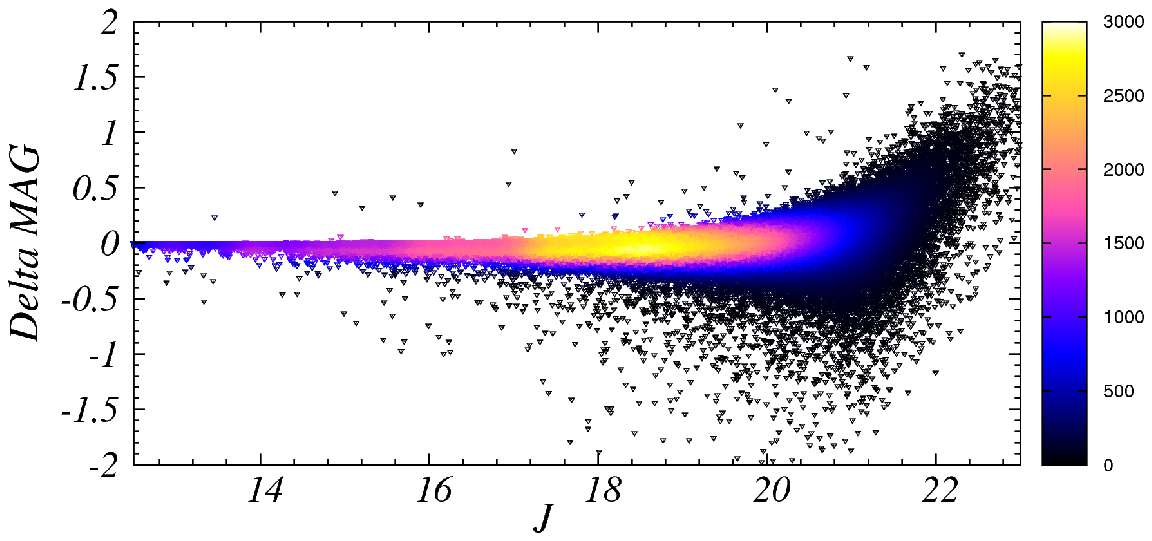}}
%\end{minipage}
\hfill
%\begin{minipage}{\textwidth}
\resizebox{\hsize}{!}{\includegraphics{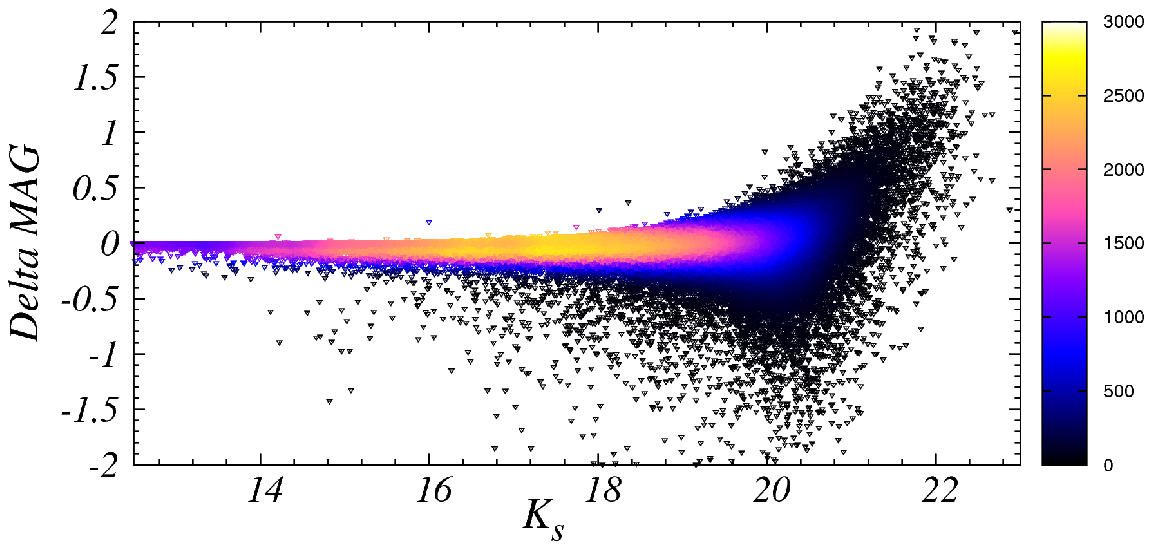}}
%\end{minipage}
\hfill
\caption{Examples of error distributions derived from the ASTs, in
  $Y$, $J$, $\ks$ filters for a subregion of the 8\_3 tile.  An
  arbitrary color scale is used to highlight the regions with a
  higher density of stars.}
\label{ast}
\end{figure}

\begin{figure}
\centering
\resizebox{\hsize}{!}{\includegraphics{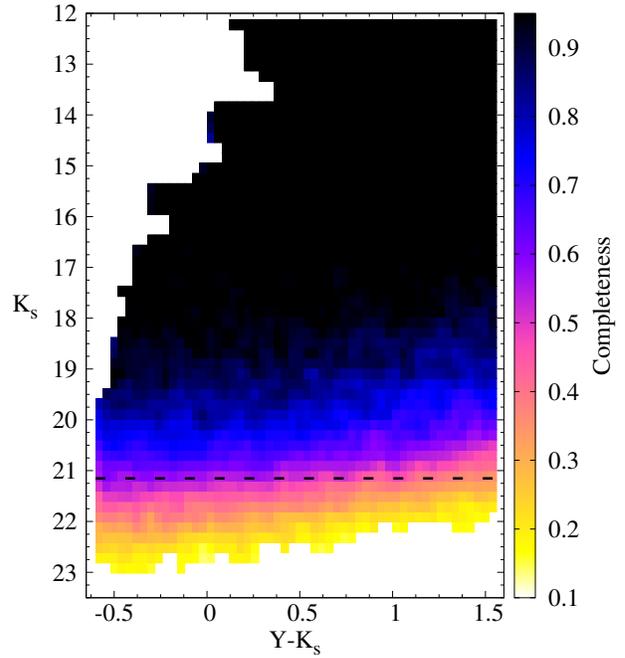}}
\caption{Example of completeness map derived from the ASTs on a $\ks$
  vs. $\yks$ CMD. The colours code the completeness level.  The black
  dashed line shows the $5\sigma$ depth (or ${\rm S/N}=5$) in the
  $\ks$ band.}
\label{completeness}
\end{figure}

Figure~\ref{ast} shows an example, based on an 8\_3 tile subregion, of
the error distribution derived from ASTs in $Y$, $J$, and $\ks$ versus
the difference between the output and input magnitude.
Fig.~\ref{completeness} shows an example, for a tile 8\_8 subregion,
of the completeness across the \ks\ vs.\ \jks\ CMD.

\subsection{Converting models and data to the same zeropoints}
\label{vegamags}

For the present work, we have converted large sets of theoretical
stellar evolutionary models \citep[see][and references
therein]{Marigo_etal08} to the VISTA Vega magnitude system (Vegamag)
which is itself derived from the 2MASS system, and where Vega has a
magnitude equal to 0 in all filters. The procedure is thoroughly
discussed in \citet{Girardi_etal02, Girardi_etal08}. The filter
transmission curves employed are the official ones\footnote{
  http://www.eso.org/sci/facilities/paranal/instruments/vista/inst}.
The model isochrones, extinction coefficients, and other miscellaneous
data, are retrievable via the web interface at 
http://stev.apd.inaf.it/cmd.

\begin{figure*}
%\begin{minipage}{\textwidth}
\resizebox{0.31\hsize}{!}{\includegraphics{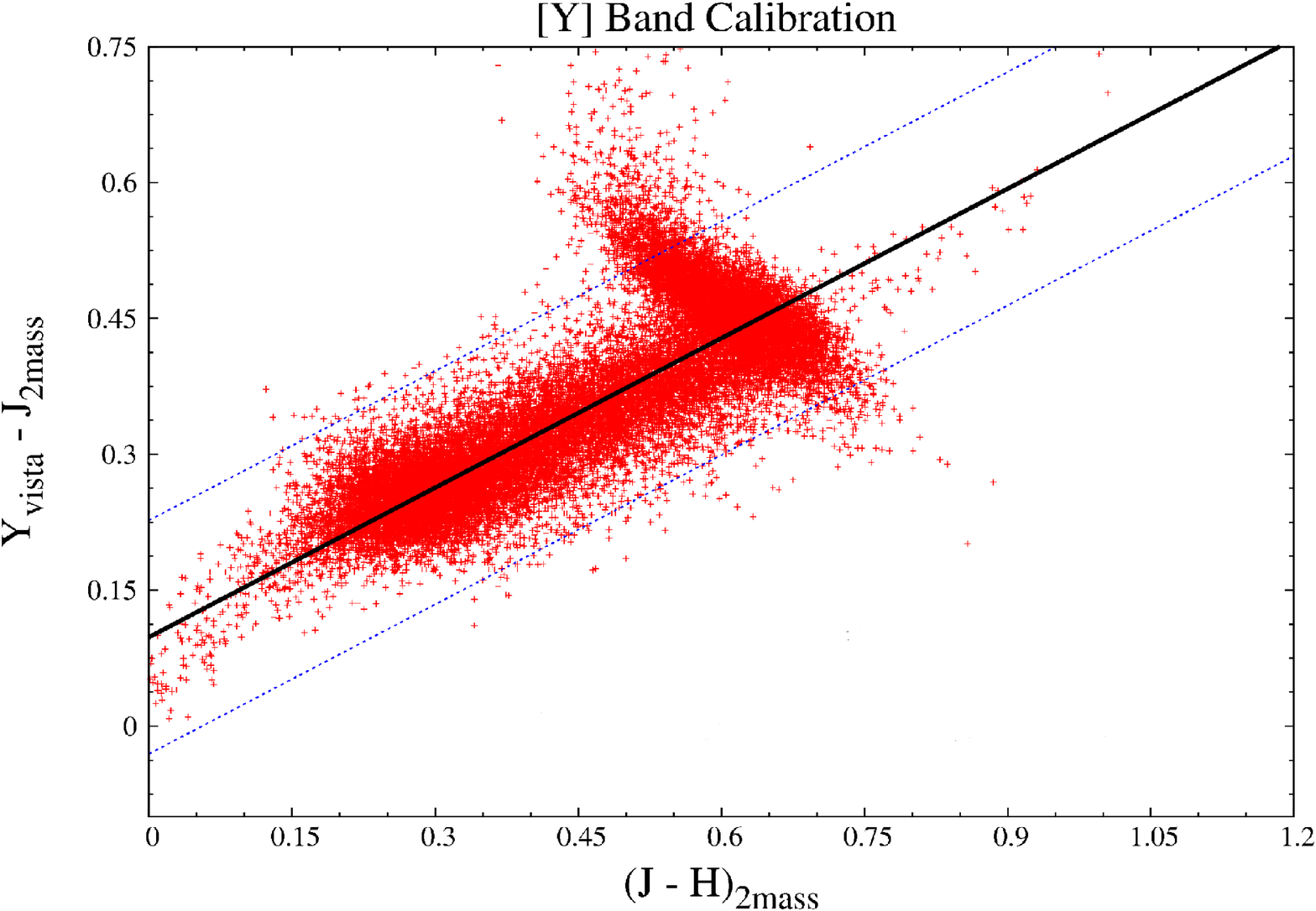}}
\hfill
\resizebox{0.31\hsize}{!}{\includegraphics{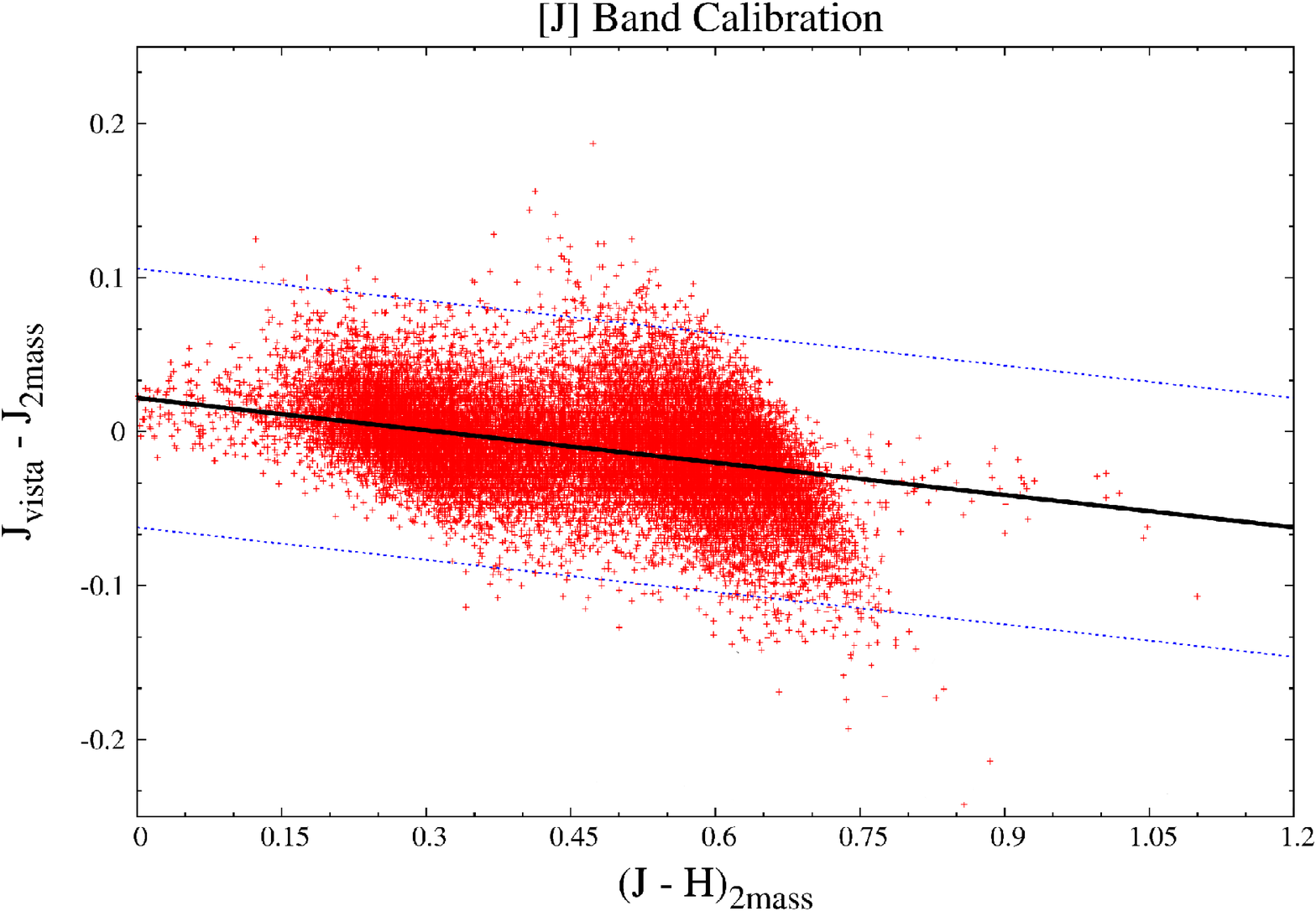}}%\\
\hfill
\resizebox{0.31\hsize}{!}{\includegraphics{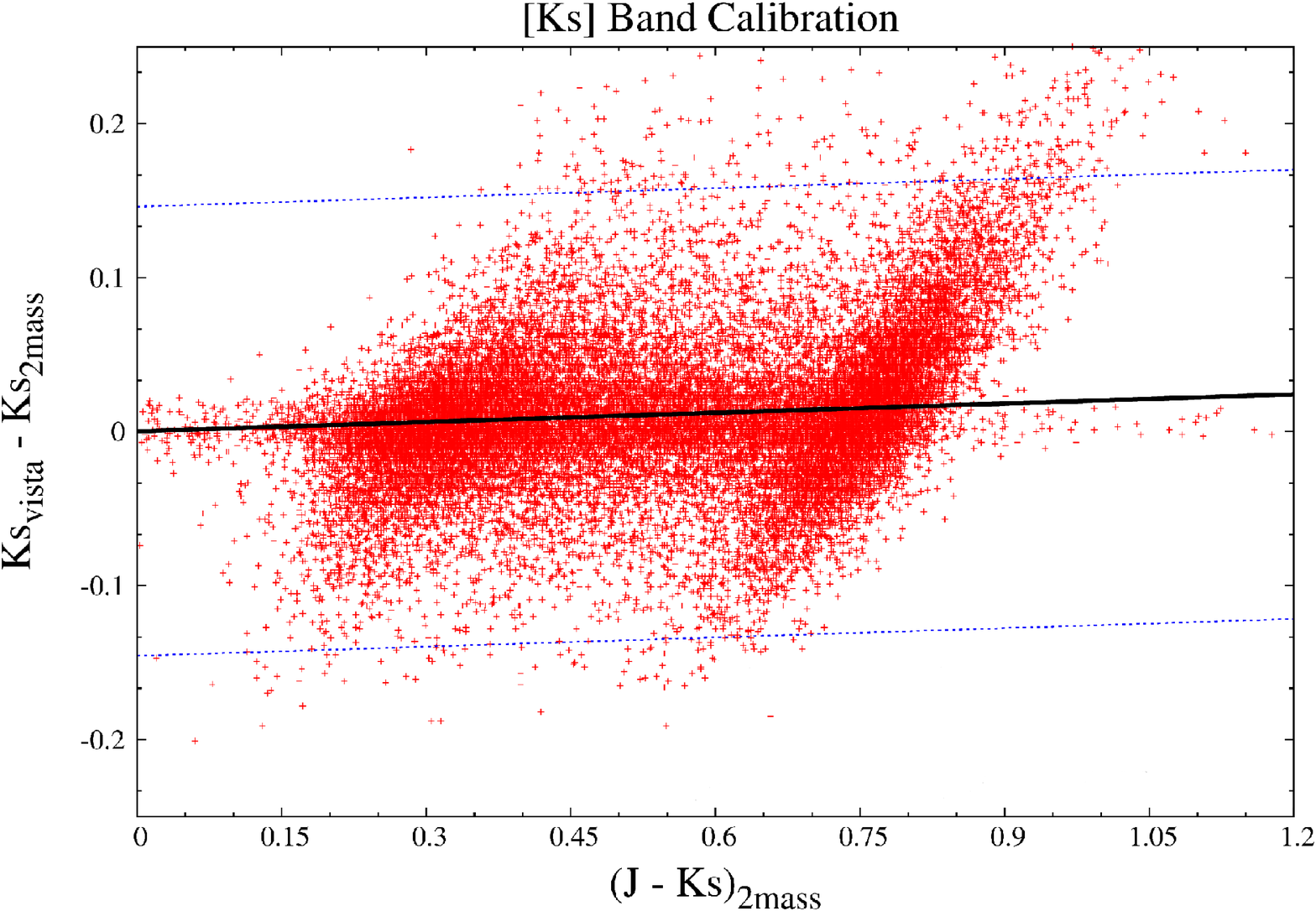}}
%\end{minipage}
\caption{ Calibration to the Vegamag system. The points show Milky Way
  model stars generated with the TRILEGAL model for an area of
  $\sim\!1.4$~\sqdeg, and convolved with the typical photometric
  errors of 2MASS. The dashed lines show the limits for the $3\sigma$
  clipping. The continuous lines are the best-fitting linear relations
  with a fixed slope as in Eq.~(\ref{eq_calib}).}
\label{photo_all}
\end{figure*}

The photometric calibration of v1.0 VISTA pipeline processed pawprints
by the pipeline processing group uses a similar approach to that for
WFCAM data \citep{Hodgkin_etal10} although the details of the colour
equations are different for VISTA, and VISTA, like 2MASS, has a \ks\
filter whereas WFCAM has $K$.

The present zero point (ZP)
calibration\footnote{http://casu.ast.cam.ac.uk/surveys-projects/vista/technical/vistasensitivity}
is based on using the 2MASS stars with $0 < (\jks)_{\rm 2MASS} < 1$
which fall on each detector in each pawprint to calculate their
magnitudes on the VISTA system using the following linear fits for
stars distributed all across the sky:
\begin{eqnarray}
 Y_{\rm VISTA}-J_{\rm 2MASS} &=& 
 0.550\,(J\!-H\!)_{\rm 2MASS}\nonumber\\
 J_{\rm VISTA}-J_{\rm 2MASS} &=&
 -0.070\,(J\!-\!H)_{\rm 2MASS}\label{eq_calib} \\
 %H_{\rm VISTA}-H_{\rm 2MASS} &=&0.060(J-H)_{\rm 2MASS}\nonumber\\
 \ks_{\rm VISTA}-\ks_{\rm 2MASS} &=& 
 0.020\,(J\!-\!K_{\rm s})_{\rm 2MASS}\nonumber
\end{eqnarray}
with a small correction for mean extinction in the general area. The
derived magnitudes of the 2MASS stars in the VISTA system are then
used to calibrate each detector at each pointing by deriving the
median ZP of each detector at that time.  The $0 < (\jks)_{\rm 2MASS}
< 1$ colour selection ensures that very blue and very red stars are
not used in calibrating each tile.

The resulting VISTA magnitudes should then be on a Vegamag system
assuming the colour relations used are accurate. However deviations
from the true Vegamag zero points would result if these relations are
not strictly linear, but have second-order terms, or inaccurate
coefficients. Such second-order terms have been well characterized in
WFCAM data by \citet{Hodgkin_etal10}, but are not yet available for
VIRCAM.

The current photometric calibration may thus not be precisely on the
Vegamag system, especially in $Y$ where the greatest extrapolation
from 2MASS $J$ is required. Therefore, before performing any detailed
data--model comparison, we choose to convert the models to the same
ZPs as the v1.0 data. This is done internally, without modifying the
isochrones being distributed at http://stev.apd.inaf.it/cmd. Indeed,
this procedure is the more convenient since future VISTA data releases
may have different definitions of their ZPs.

To correct the observational ZPs we proceed as follows.  We take the
present stellar models and build the theoretical counterpart of the
calibration data, using the TRILEGAL code \citep{Girardi_etal05} to
simulate field stars in the VISTA Vegamag system, and in 2MASS (using
\citealt{MaizApellaniz07} zeropoints).  Photometric errors are added
to the 2MASS data\footnote{Photometric errors are dominated by 2MASS
  since it represents by far the shallowest data in this case.},
following the error distributions inferred from \cite{bon04} in
low-density regions of the sky. The derived simulations are shown in
Fig.~\ref{photo_all}; they show colour ranges and distributions in the
colour-colour plots which are very similar to those found in the real
WFCAM data \citep[see e.g.][]{Hodgkin_etal10}.  Then, to the simulated
data we fit a set of lines with the same slopes as in
Eqs.~(\ref{eq_calib}). The outcome of the fitting is a set of
constants, which are then interpreted as the ZP offsets between the
VDFS pipeline v1.0 calibration and the Vegamag system of the stellar
models. The offsets we find are equal to 0.099~mag in $Y$, 0.021~mag
in $J$ and 0.001~mag in \ks, with respect to the v1.0 calibration.
These offsets are subtracted from our models before starting the
SFH-recovery work.

\begin{figure}
  \resizebox{\hsize}{!}{\includegraphics{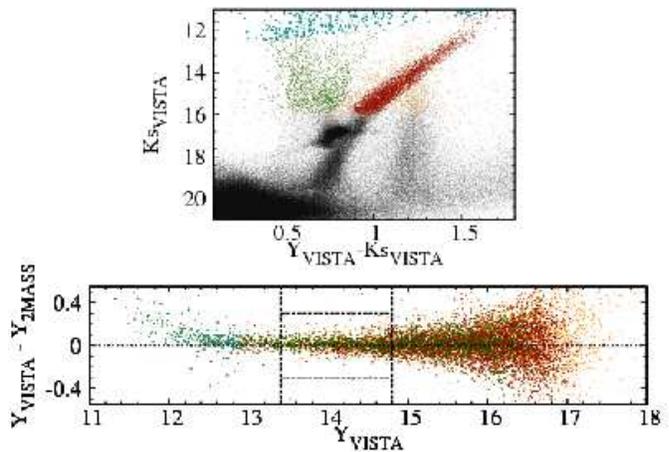}}
  \caption{Verification of Eq.~\ref{eq_calib} when applied to real LMC
    data. The {\bf top panel} shows stars in the 8\_8 tile, in the
    \ks\ vs.\ \yks\ diagram of VMC.  Different colours are used for
    likely LMC giants (brown), likely Milky Way dwarfs (green), stars
    bright enough to be partially saturated (cyan), and stars for
    which 2MASS photometry is not available (black). The {\bf bottom
      panel} shows the difference between the $Y$ magnitude as
    measured from VMC data, and as inferred from 2MASS photometry via
    the first Eq.~(\ref{eq_calib}), as a function of $Y$, for stars
    with both VISTA and 2MASS photometry.  The vertical dashed lines
    denote the magnitude interval for which this comparison is the
    most significant, as discussed in the text.}
\label{fig_offset}
\end{figure}

The bulk of the stars shown in Fig.~\ref{photo_all} are dwarfs in the
solar vicinity. We check if applying this calibration method to the
LMC result in any bias/offset in the calibration because of the nature
of the stellar objects in the LMC tiles. To this aim we use the stars
in the 8\_8 tile for which we have both 2MASS and VISTA magnitudes, as
ilustrated in Fig.~\ref{fig_offset}. The top panel shows the \ks\ vs.
\yks\ diagram from VMC, which is used for a rough classification into
likely Milky Way dwarfs and likely LMC giants.  Then, we limit the
sample to stars of intermediate brightness, so as to avoid including
long-period variables and partly saturated objects in VMC at
$\ks<13.4$~mag, and stars with large photometric errors in 2MASS at
$\ks>14.8$~mag.  The bottom panel plots the difference between the $Y$
magnitudes derived from VMC data and those derived from application of
Eq.~\ref{eq_calib} to 2MASS photometry, as a function of $Y$.  The
$1\sigma$ dispersion at $13.4<\ks<14.8$ is equal to $\simeq0.065$~mag
for both dwarfs and giants, and is likely dominated by the errors in
2MASS photometry.  The offset between the two estimates of $Y$-band
magnitude is just 0.005~mag for the likely LMC giants, and 0.019~mag
for the likely Milky Way dwarfs. In both cases the offset is much
smaller than the dispersion in the data, and less than the random
errors we claim in the determination of the LMC distance moduli (see
Sect.~\ref{sec_dist_av}).  Moreover, what is particularly reassuring
in this comparison is that differences in the mean $Y$-band offsets
between dwarfs and giants are just $\sim0.01$~mag, despite the
different color ranges comprised by both samples. We conclude that
application of Eq.~\ref{eq_calib} in VMC field 8\_8 does not appear to
introduce any significant systematic error in the calibration.

%%%%%%%%%%%%%%%%%%%%%%%%%%%%%%%%%%%%%%%%%%%%%%%%%%%%%%%%%%%
\section{The SFH recovery}
\label{sec_sfh}
Our SFH-recovery work is largely based on \citet{Kerber09}, to which
we refer for more details and basic tests of the algorithms. Here, we
briefly mention the particular assumptions adopted in the present
work. To recover the SFH we used a set of ``stellar partial models''
(SPMs), which are model representations of stellar populations
covering small intervals of ages and metallicities, and observed at
the same conditions of completeness and crowding as the real data.
They are distributed over 14 age intervals that cover from
$\log(t/{\rm yr})=6.6$ to $10.15$. They also follow five different
AMRs located parallel on a $\feh$~vs.~age (or lookback time) plot.
These AMRs cover those observed for stellar clusters
\citep{Olszewski_etal91, MG03, Grocholski_etal06, Kerber_etal07} and
field stars \citep{HZ04, HZ09, Cole_etal05, Carrera_etal08}.  Hence,
LMC populations are described as linear combinations of 70 distinct
SPMs. Table~\ref{tab_amr} shows their central values of $\log(t/{\rm
  yr})$ and \feh; they are also indicated as green starred points in
Figs.~\ref{sol43}--\ref{sol88}. The width of each SPM is 0.15 dex in
\feh\ and 0.2~dex in $\Delta \log(t)$, except for the 3 youngest age
bins where we used $\Delta \log(t)$ widths of 0.6, 0.4 and 0.4~dex,
respectively, and for the oldest age bin where we used 0.15~dex.
Inside the age and \feh\ intervals of each SPM, the star formation
rate is assumed to be constant.

The LMC populations are simulated using the \citet{Chabrier01}
log-normal initial mass function, plus a 30~\% binary fraction. The
simulated binaries are non-interacting systems with primary/secondary
mass ratios evenly distributed in the interval from 0.7 to 1.

\begin{table}
\caption{Grid of stellar partial models used in the SFH recovery.}
\label{tab_amr}
\centering
\begin{tabular}{c|ccccc}
\hline
\hline
$\log(t/{\rm yr})$ & $\feh_1$ & $\feh_2$ & $\feh_3$ & $\feh_4$ & $\feh_5$ \\
\hline
6.9 & $-$0.10 & $-$0.25 & $-$0.40 & $-$0.55 & $-$0.70 \\
7.4 & $-$0.10 & $-$0.25 & $-$0.40 & $-$0.55 & $-$0.70 \\
7.8 & $-$0.10 & $-$0.25 & $-$0.40 & $-$0.55 & $-$0.70 \\
8.1 & $-$0.10 & $-$0.25 & $-$0.40 & $-$0.55 & $-$0.70 \\
8.3 & $-$0.10 & $-$0.25 & $-$0.40 & $-$0.55 & $-$0.70 \\
8.5 & $-$0.10 & $-$0.25 & $-$0.40 & $-$0.55 & $-$0.70 \\
8.7 & $-$0.10 & $-$0.25 & $-$0.40 & $-$0.55 & $-$0.70 \\
8.9 & $-$0.10 & $-$0.25 & $-$0.40 & $-$0.55 & $-$0.70 \\
9.1 & $-$0.25 & $-$0.40 & $-$0.55 & $-$0.70 & $-$0.85 \\
9.3 & $-$0.25 & $-$0.40 & $-$0.55 & $-$0.70 & $-$0.85 \\
9.5 & $-$0.25 & $-$0.40 & $-$0.55 & $-$0.70 & $-$0.85 \\
9.7 & $-$0.40 & $-$0.55 & $-$0.70 & $-$0.85 & $-$1.00 \\
9.9 & $-$0.70 & $-$0.85 & $-$1.00 & $-$1.15 & $-$1.30 \\
10.075 & $-$1.00 & $-$1.15 & $-$1.30 & $-$1.45 & $-$1.60 \\
\hline
\end{tabular}
\end{table}

To the 70 SPMs corresponding to LMC populations, we add a partial
model that describes the Milky Way foreground. The latter is simulated
with the TRILEGAL code \citep{Girardi_etal05} using the standard
calibration for the Milky Way components, and for the same central
coordinates and area as the VMC observations.

All SPMs including the LMC ones are built with the aid of the TRILEGAL
code, in the form of well populated and deep photometric catalogues.
They are then displaced by a true distance modulus, \dmo, and the
extinction plus reddening implied by the $V$-band extinction,
\av.\footnote{Note that \av\ is a mean value that includes both
  extinction internal to the LMC, and from the Milky Way foreground.}
Subsequently, the SPMs are degraded by applying the distributions of
completeness and photometric errors derived from the ASTs.

\begin{figure*}
\resizebox{0.24\hsize}{!}{\includegraphics{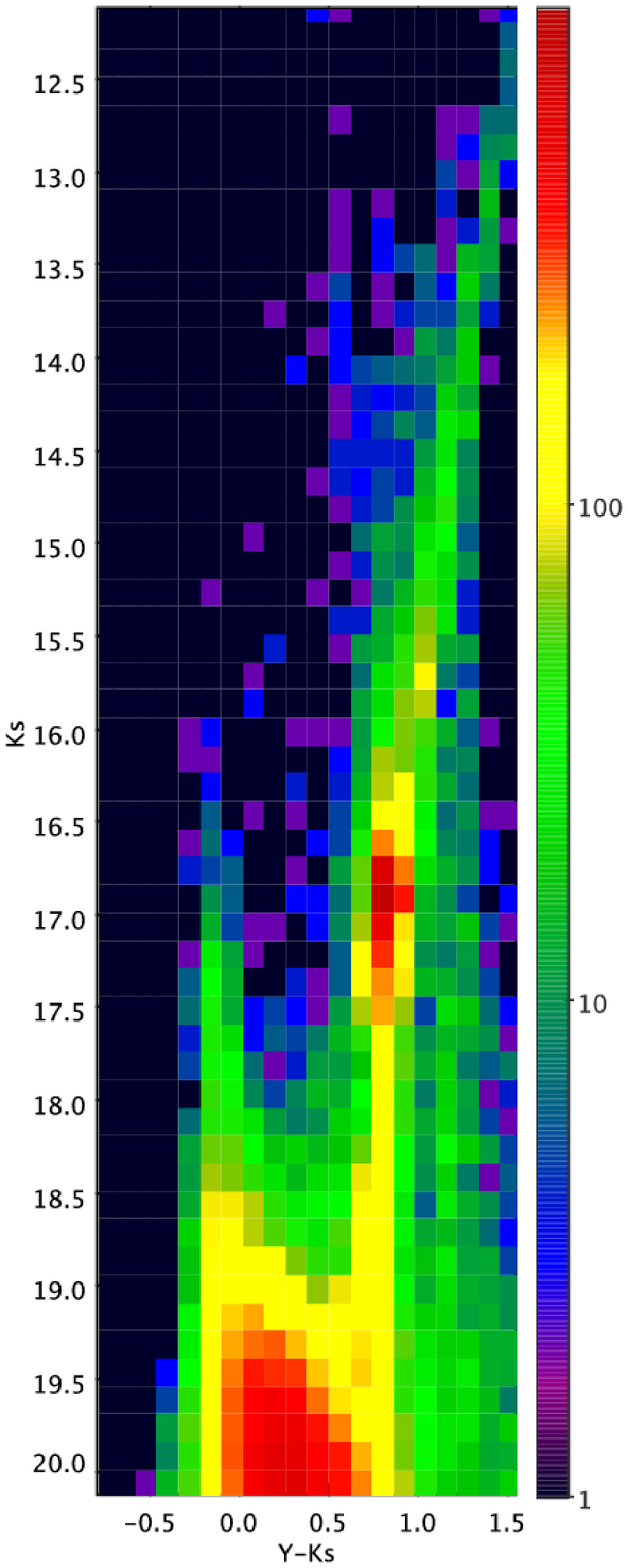}}
\resizebox{0.24\hsize}{!}{\includegraphics{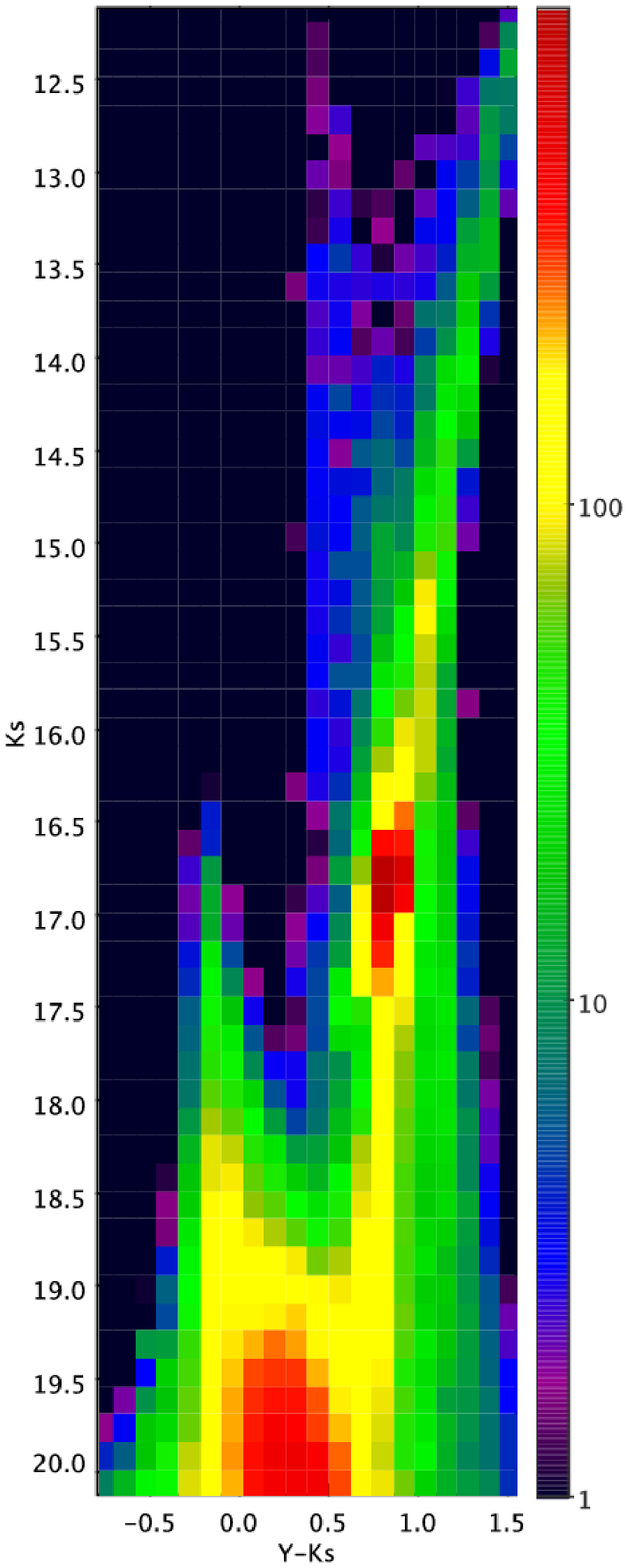}}
\resizebox{0.24\hsize}{!}{\includegraphics{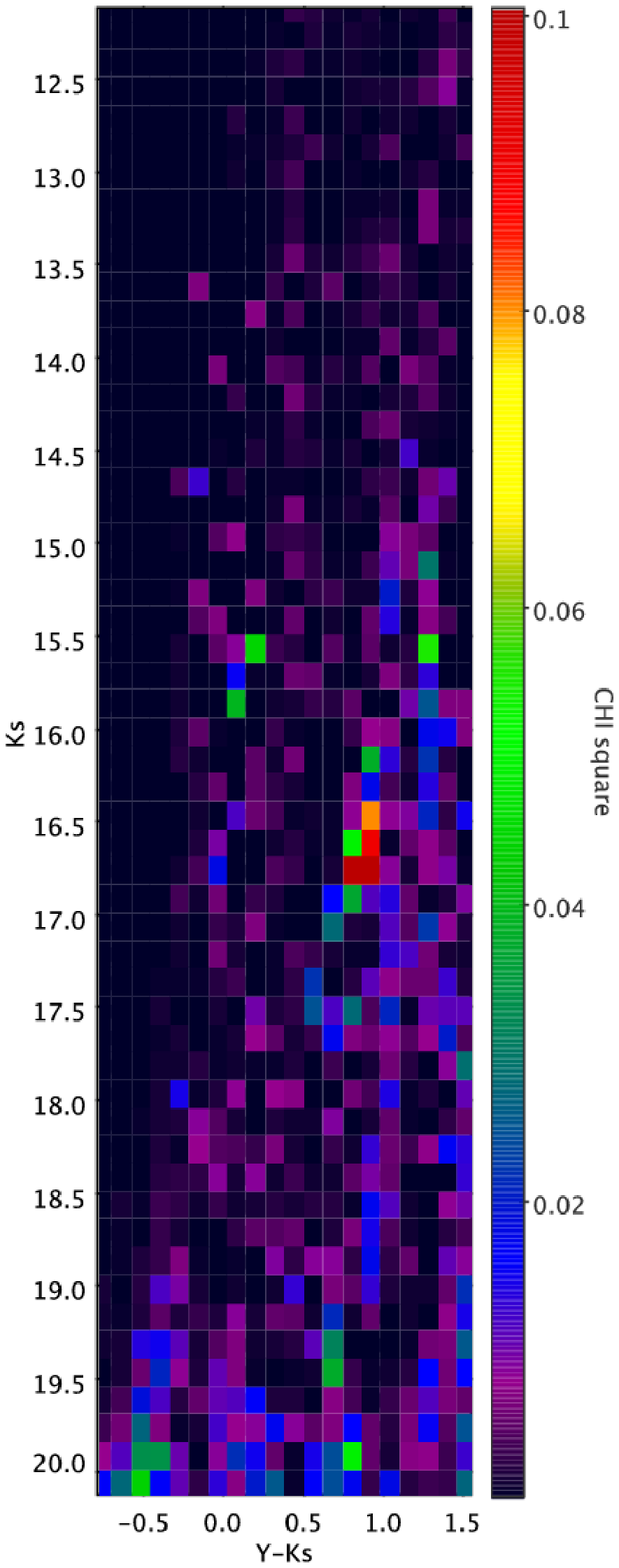}}
\resizebox{0.24\hsize}{!}{\includegraphics{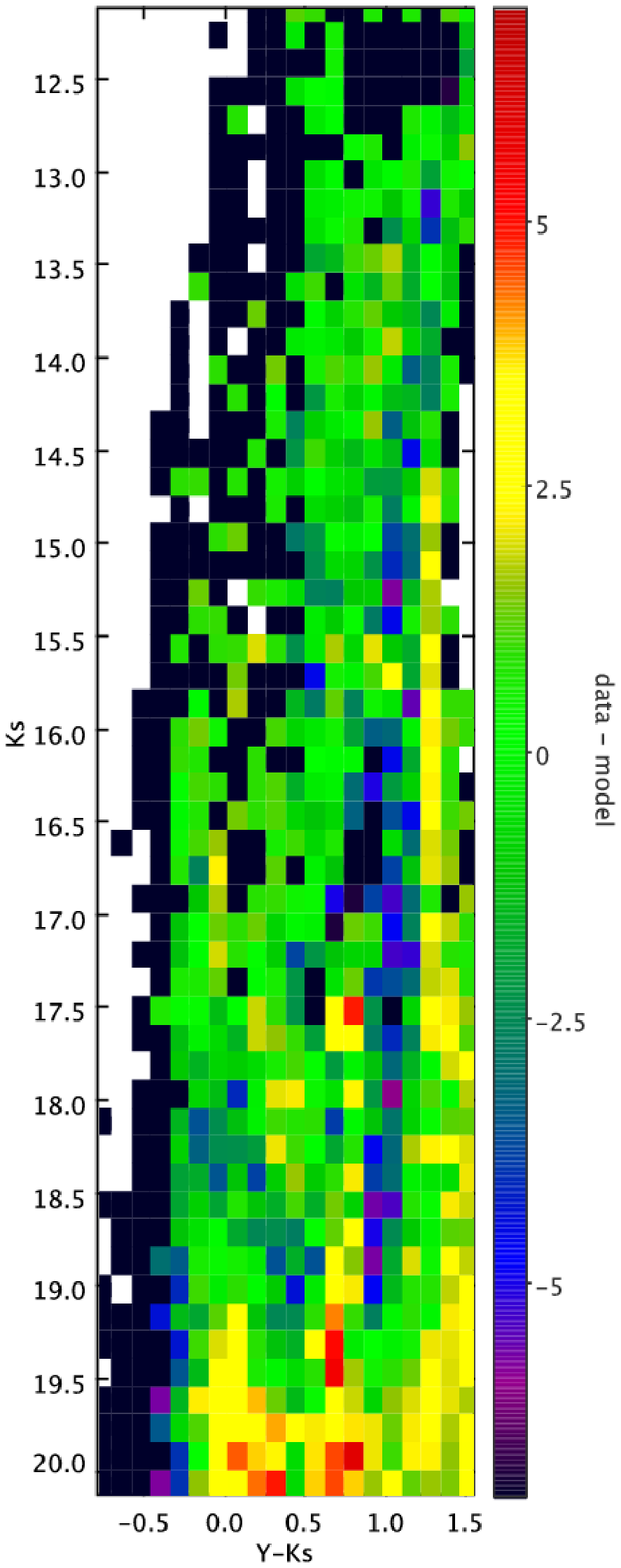}}
\caption{Examples of Hess diagrams obtained during a typical run of
  StarFISH -- in this case, for subregion G1 of tile 8\_3, with
  $\dmo=18.48$~mag and $\av=0.32$~mag. From left to right we show Hess
  diagrams for (a) the original VMC data, within the magnitude and
  colour limits used in this work, (b) the best-fitting model derived
  by StarFISH, (c) the $\chi^2$ map, that is the fractional
  contribution of each CMD bin to the total $\chi^2$, (d) the
  difference between data and best-fitting model. In panels (a), (b)
  and (d), the scale is in units of stars per CMD bin.}
\label{hess}
\end{figure*}

Using these SPMs we have recovered the SFH assuming a wide range in
the values for \av\ and \dmo. The \av\ vs.\ \dmo\ grid is regularly
spaced by 0.03~mag and 0.025~mag, respectively, with limits: from 0.06
to 0.60~mag in \av, in all three outer disk tiles; and from 18.40 to
18.55~mag in \dmo\ in tile 8\_3, 18.40 to 18.53 in 4\_3 and 18.28 to
18.50 in the 8\_8 tile.  For the 6\_6 tile we used limits from 0.30 to
0.99~mag in \av, and from 18.40 to 18.53~mag in \dmo. The different
limits reflect the fact that the different tiles are effectively found
to be at different mean values of \av\ and \dmo, as revealed by our
initial explorative work using a much coarser grid of \av\ vs.\ \dmo\
values.

The SFH was recovered simultaneously using two CMDs, $\ks$ vs.\ $\jks$
with limits $-$0.52 to 0.88~mag in colour and 12.10 to 20.45 in
magnitude, and $\ks$ vs.\ $\yks$~ with limits $-$0.82 to 1.56~mag in
colour and 12.10 to 20.15 in magnitude.  In tile 6\_6 we used
$\ks=20$~mag as the faintest limit in all CMDs.  These limits in
colour and magnitude allow us to separate the LMC stars from most
contaminating galaxies (see as an example Fig.~\ref{cmd} where most of
the galaxies are clearly well separated up to $K_{s}=19.5$~mag and
located in the faintest and redder part of the CMDs) and to derive the
SFH using CMD regions with completeness greater than 70\% in all
cases.

The StarFISH code \citep{HZ01} is used to find the combination of SPMs
that best fits the observed CMD, as illustrated in Fig.~\ref{hess}.
The parameter describing the goodness-of-fit is the \chisq-like
statistic defined by \citet{Dolphin02}. The final result of StarFISH
are the weights of each partial model, which can be directly
translated into the star formation rate (that is, the stellar mass
formed per unit time, in $\Msun\,{\rm yr}^{-1}$) as a function of
lookback time, SFR$(t)$, and into a mean AMR, $\feh(t)$
\citep[see][for details]{Kerber09, Rubele11}. We recall that StarFISH
includes a method to drift outside of local minima in the parameter
space \citep{HZ01}, which we extensively tested using simulated VMC
data \citep{Kerber09}. Thus, we are quite confident that the SFH
solutions we find are unique and the best-fitting ones.

Figures \ref{sol43} to \ref{sol88} show examples of the best-fitting
\SFRt\ and AMR (left panels), and the \chisq\ solution map as a
function of \av\ and \dmo\ (right panels), in a subregion of each
investigated tile.  In the \chisq\ solution map, the black dashed and
continuous lines illustrate the confidence error limit at $1\sigma$
and $3\sigma$, respectively.

In each one of these figures, the top left panel shows the
$\SFRt/\langle\SFRt\rangle$ (blue histograms) with stochastic errors
(blue errors bars) and the systematic \SFRt\ variations inside the
confidence level region of $1\sigma$ (gray shaded region). The middle
left panels show the best-fitting AMR recovered (red and black points)
together with its stochastic errors (red or black vertical bars). Red
points show the median metallicity for when the \SFRt\ reaches zero
values inside its $3\sigma$ limit; black points are the same as red
ones but for the cases in which the star formation is clearly
detected, i.e.\ \SFRt\ is non-zero inside the $3\sigma$ limit.  The
systematic AMR variations inside the $1\sigma$ confidence level are
shown as the shaded region. The green points indicate the centres of
the SPMs used for the recovery of the SFH (see Sect.~\ref{tab_amr}).
The bottom left panel illustrates the variation of the \SFRt\ solution
considering stochastic (dark violet line) and systematic errors
(shaded regions).

We have completed the recovery of the SFH in 12 subregions in tiles
8\_3, and 11 in 4\_3, another 11 in 8\_8, plus two small regions in
6\_6. All results are illustrated in the Appendix.  Subregions G9 in
tiles 4\_3 and 8\_8 present photometry of lower quality, caused by the
low $Y$ and $J$ sensitivity in the upper part of VIRCAM detector
number 16.  This effect is strong for these two bands for tile 4\_3,
which will not be further considered in this work.  In tile 8\_8 the
effect is weak but could influence the SFH results, so we decided to
show/use in this work only the derived parameters, distance modulus
and extinction, but not the \SFRt\ and AMR.

\begin{figure*}
\resizebox{0.35\hsize}{!}{\includegraphics{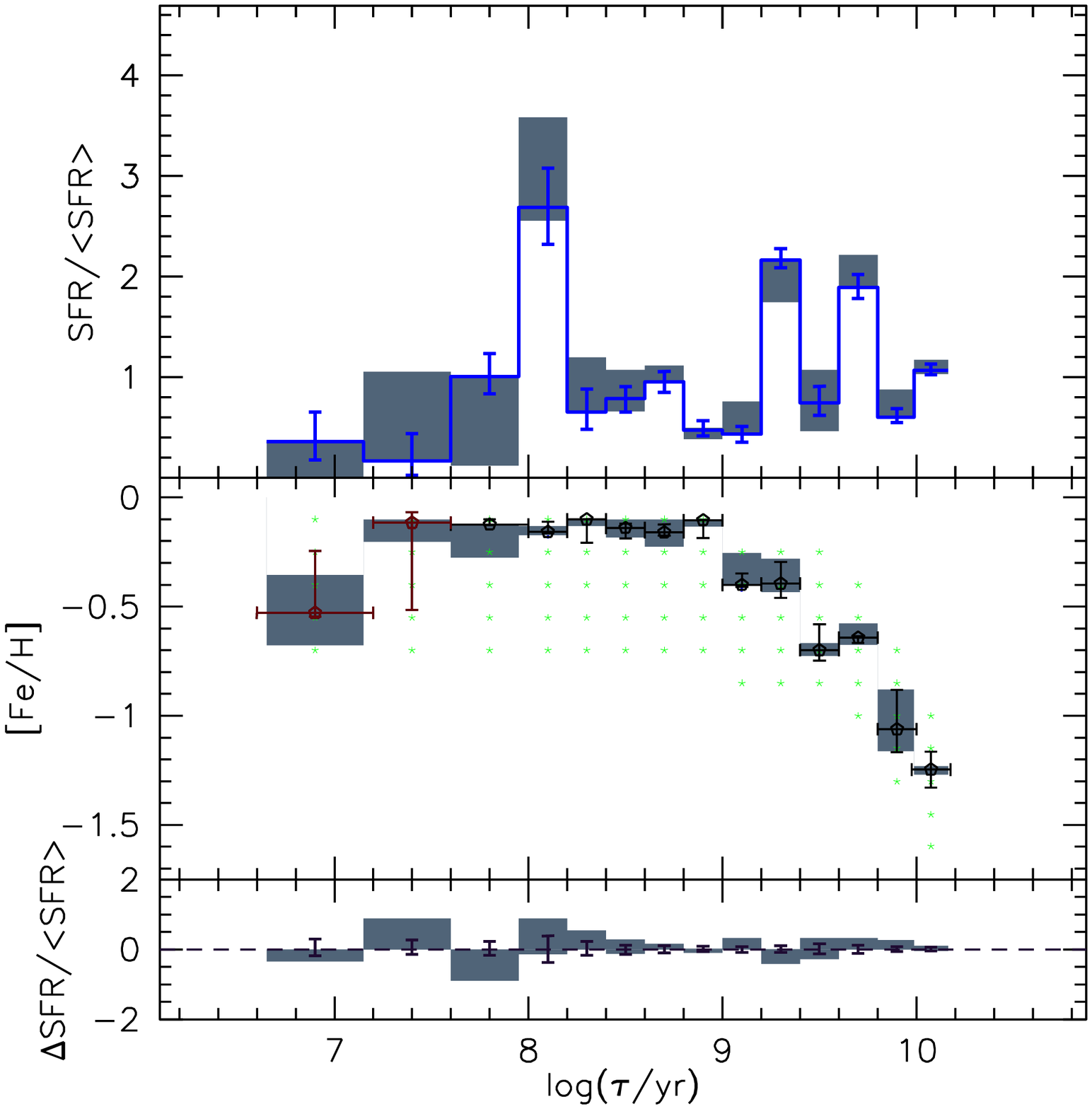}}
\resizebox{0.45\hsize}{!}{\includegraphics{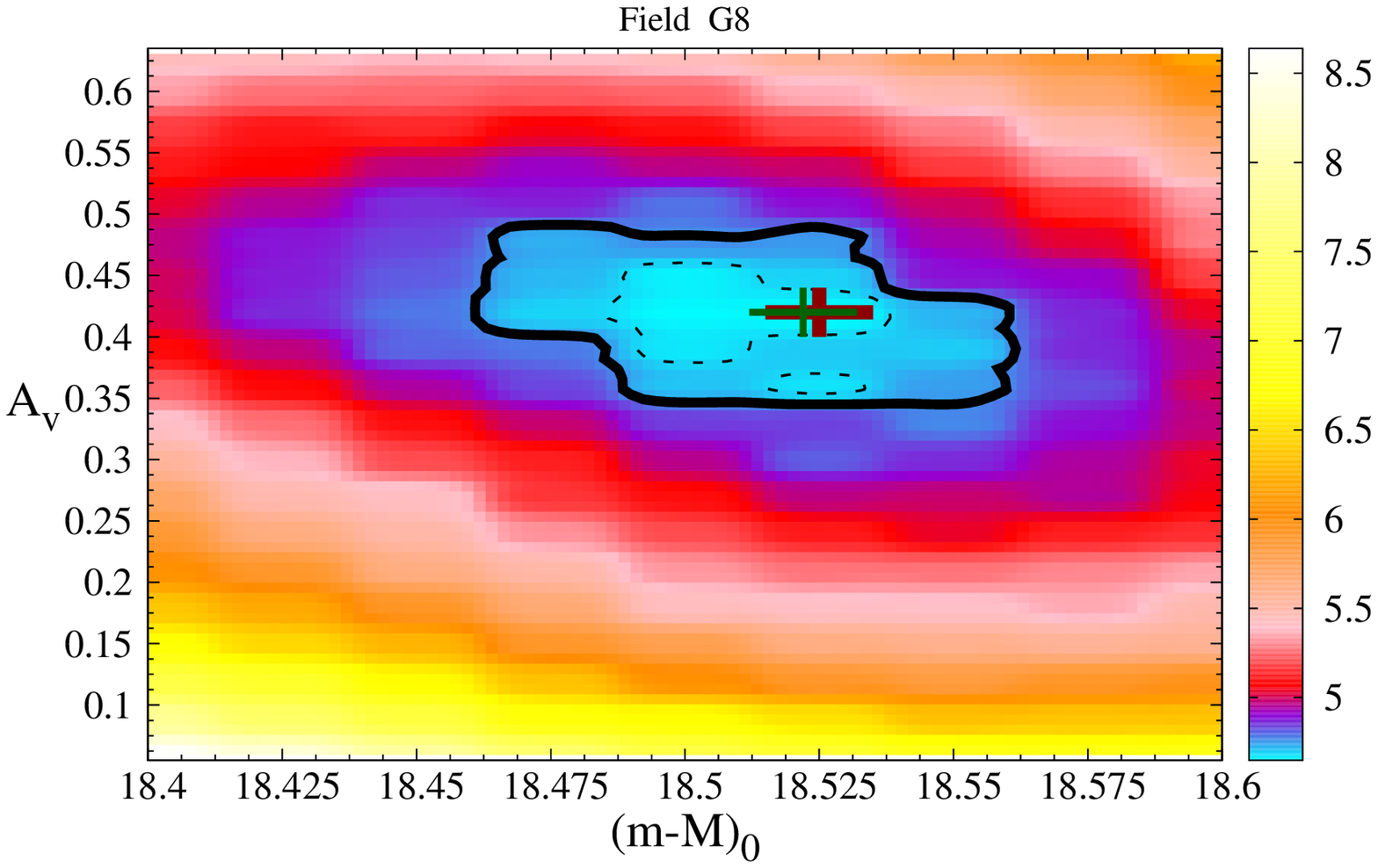}}
\caption{Example of the best-fitting solution in a subregion of tile
  4\_3.  {\bf Left panel:} The top part presents $\SFRt
  /\langle\SFRt\rangle$ (histograms) with stochastic errors (bars) and
  systematic variations (shaded regions) vs. $\logtyr$. The central
  part shows the best-fitting AMR recovered (red and black points; red
  is used in age bins for which the \SFRt\ is close to zero) with
  stochastic errors (red or black vertical bars) and systematic
  variations (shaded regions). The green dots indicate the central
  values of \feh\ and \logtyr\ of the SPMs.  The bottom part shows the
  variation of the \SFRt\ solution with stochastic errors (dashed line
  and bars) and systematic errors (shaded regions).  {\bf Right
    panel:} map of the \chisq\ values (as indicated by the colour
  scale) as a function of \av\ and \dmo\ with $1\sigma$ (dashed lines)
  and $3\sigma$ (continuous line) contours. The red cross marks the
  position of the best-fitting model. The green cross marks the
  position of the best model at the distance of the best-fitting LMC
  plane. }
\label{sol43}
\end{figure*}

\begin{figure*}[!ht]
  \resizebox{0.35\hsize}{!}{\includegraphics{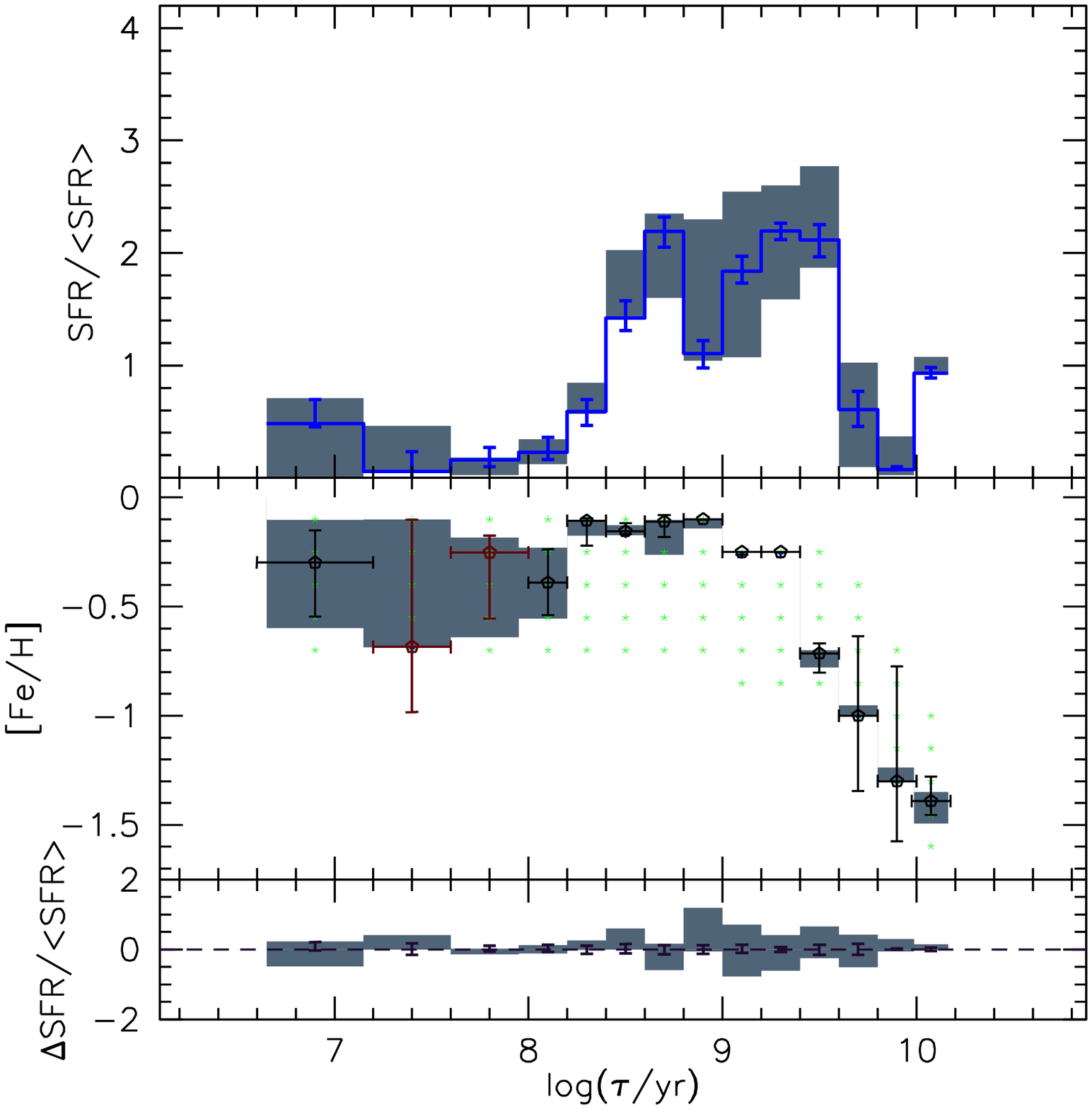}}
  \resizebox{0.45\hsize}{!}{\includegraphics{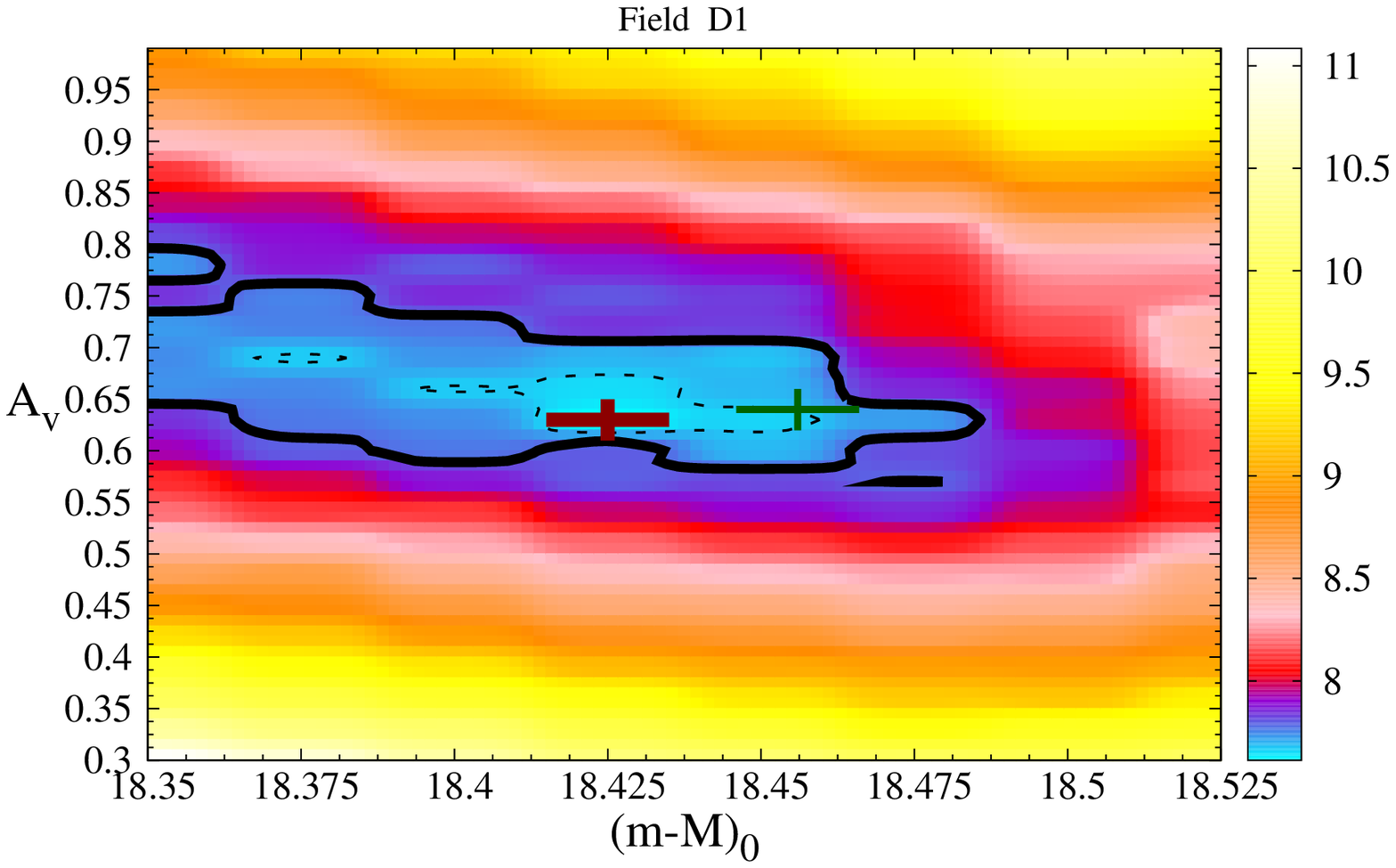}}\\
  \resizebox{0.35\hsize}{!}{\includegraphics{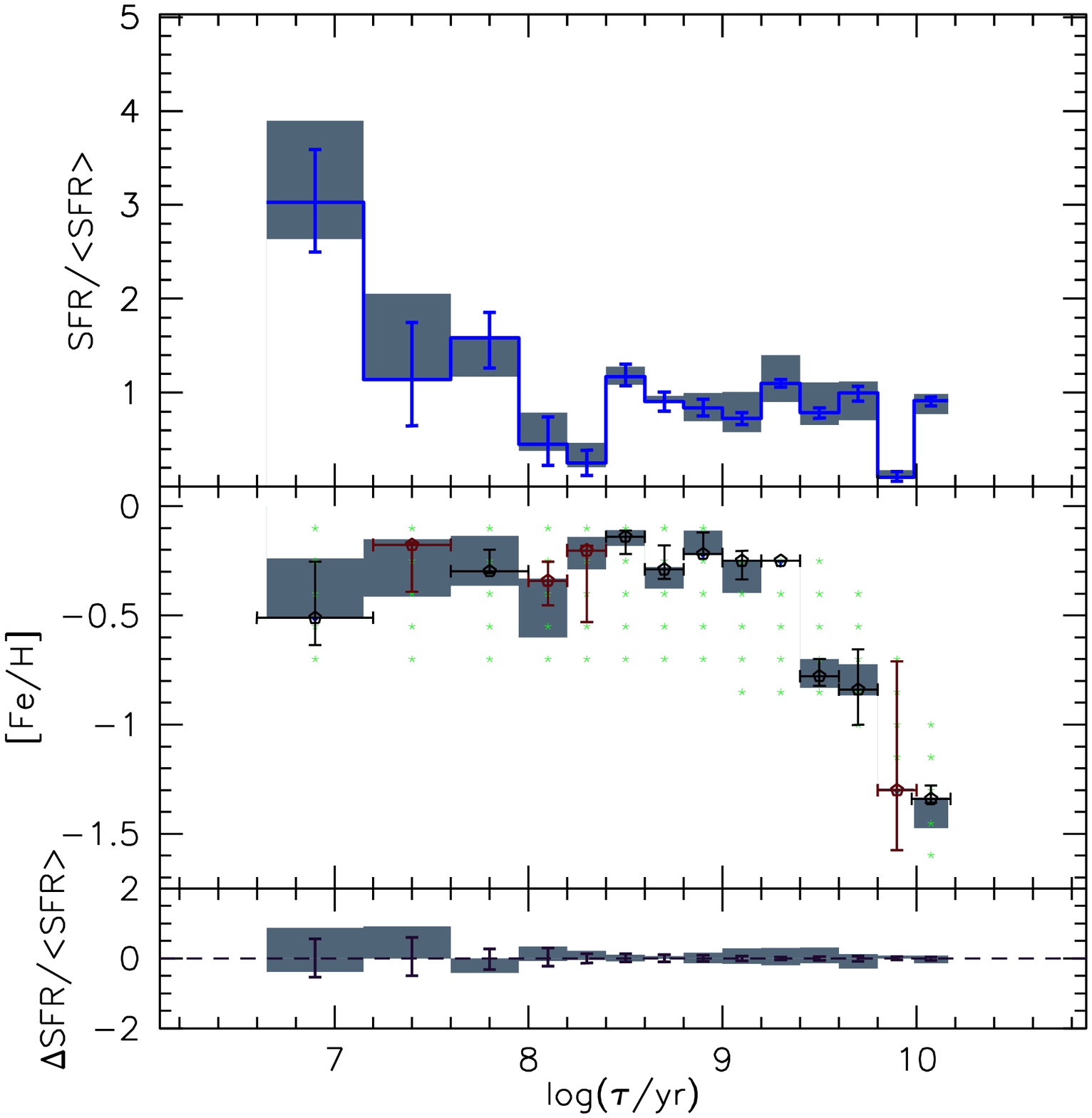}}
  \resizebox{0.45\hsize}{!}{\includegraphics{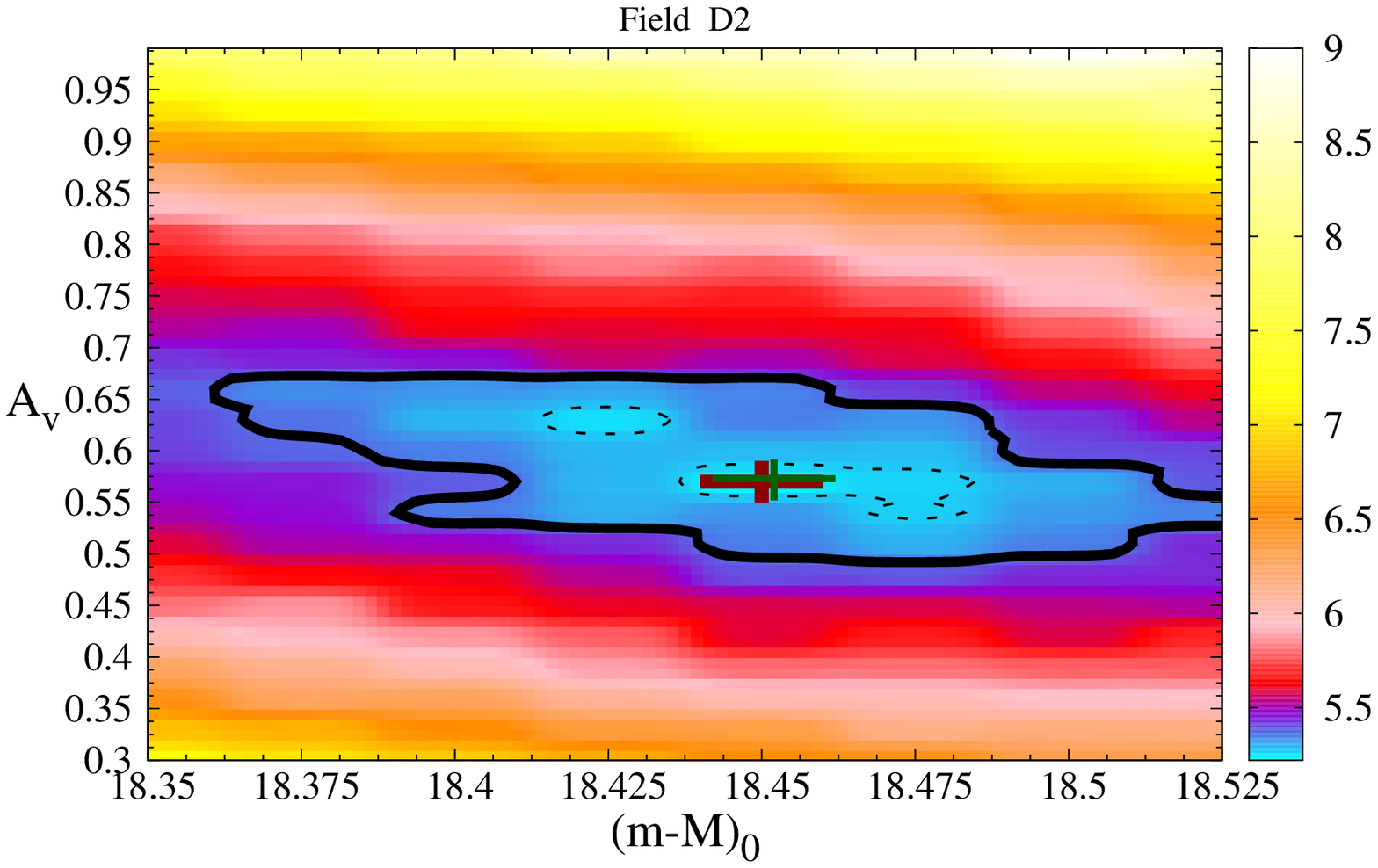}}
  \caption{Same as Fig.~\ref{sol43} but for tile 6\_6. Both
    subregions in this tile are presented.}
\label{sol66}
\end{figure*}

\begin{figure*}[!ht]
  \resizebox{0.35\hsize}{!}{\includegraphics{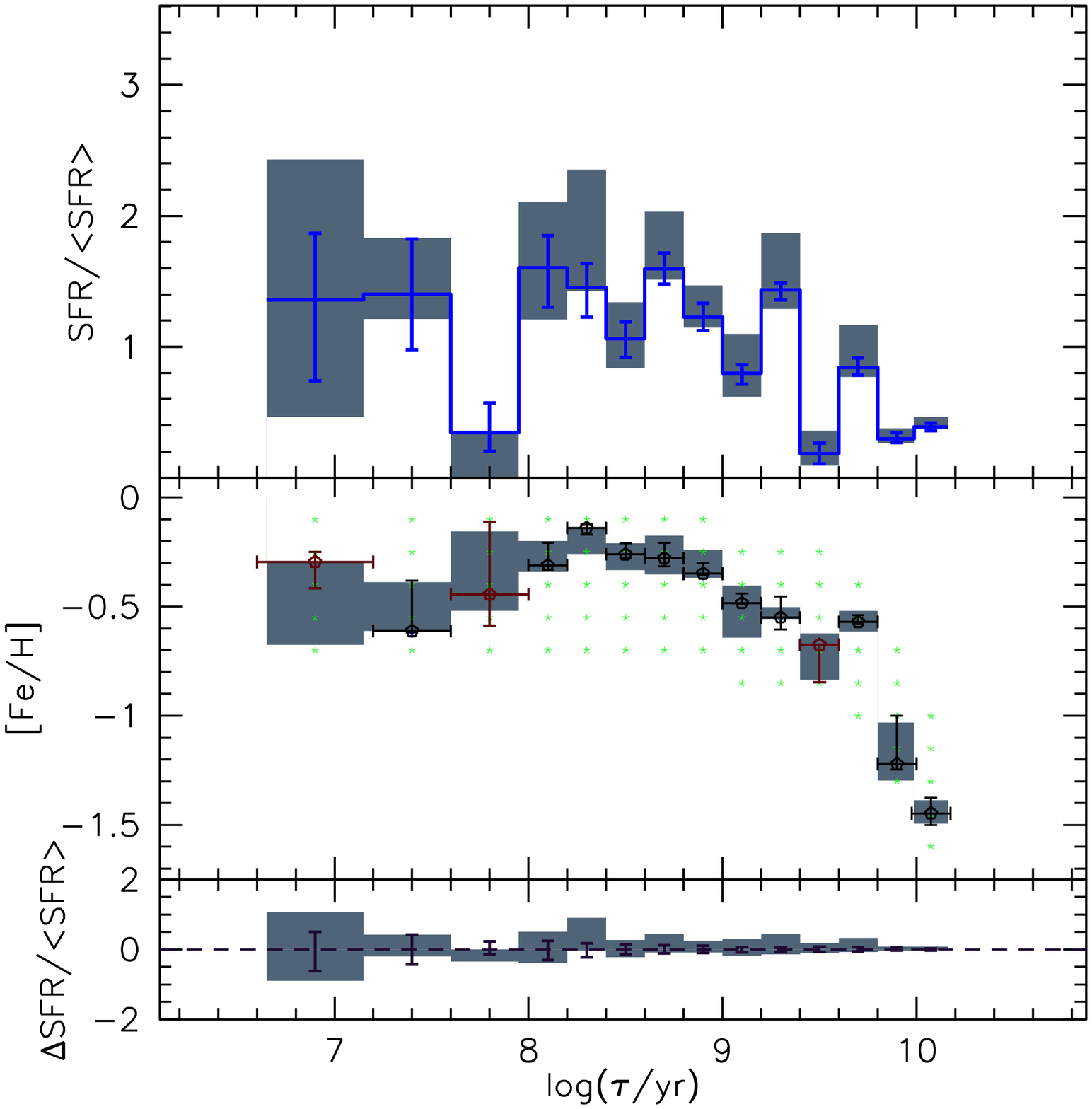}}
  \resizebox{0.45\hsize}{!}{\includegraphics{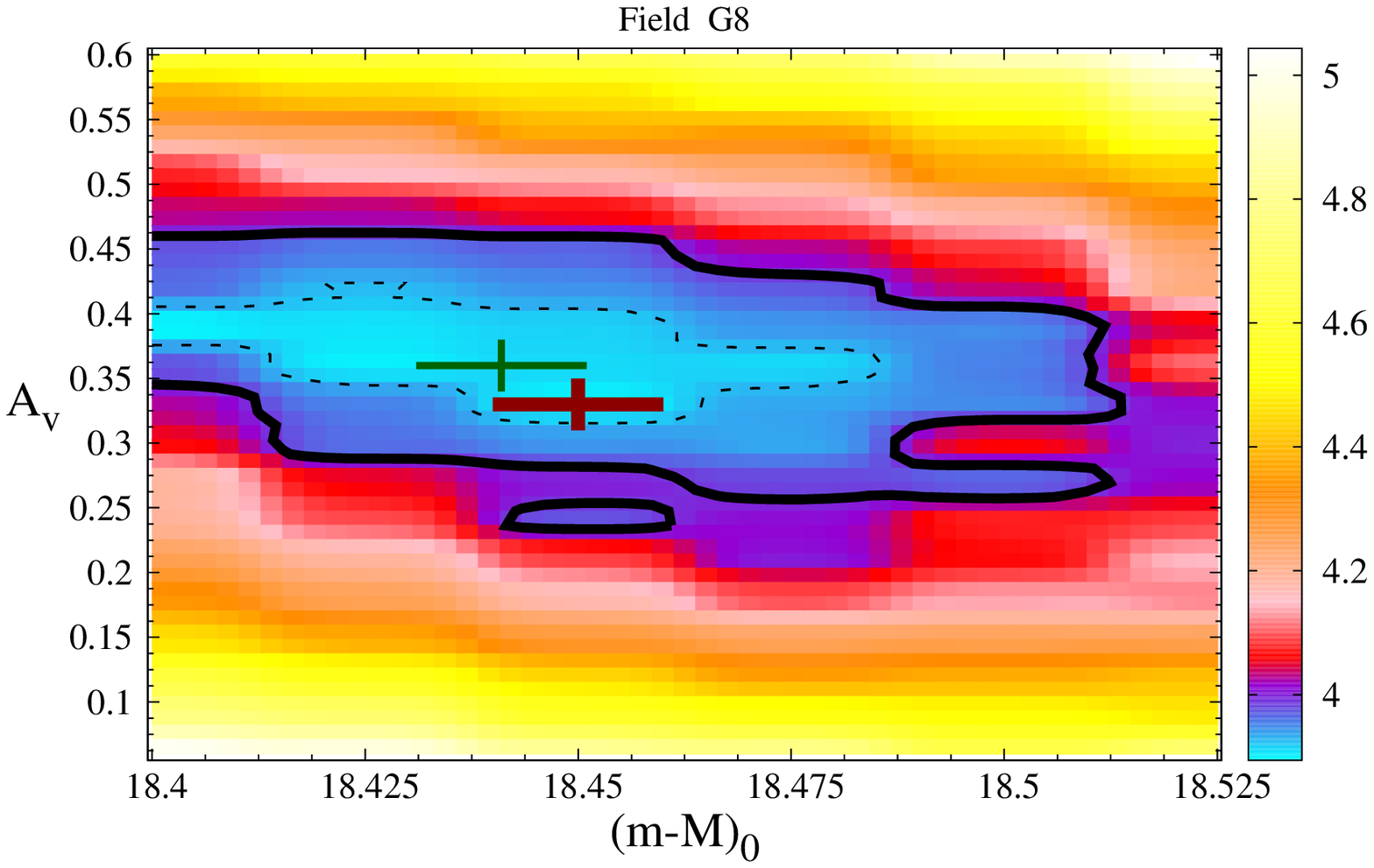}}
\caption{Same as Fig.~\ref{sol43} but for tile 8\_3.}
\label{sol83}
\end{figure*}

\begin{figure*}[!ht]
  \resizebox{0.35\hsize}{!}{\includegraphics{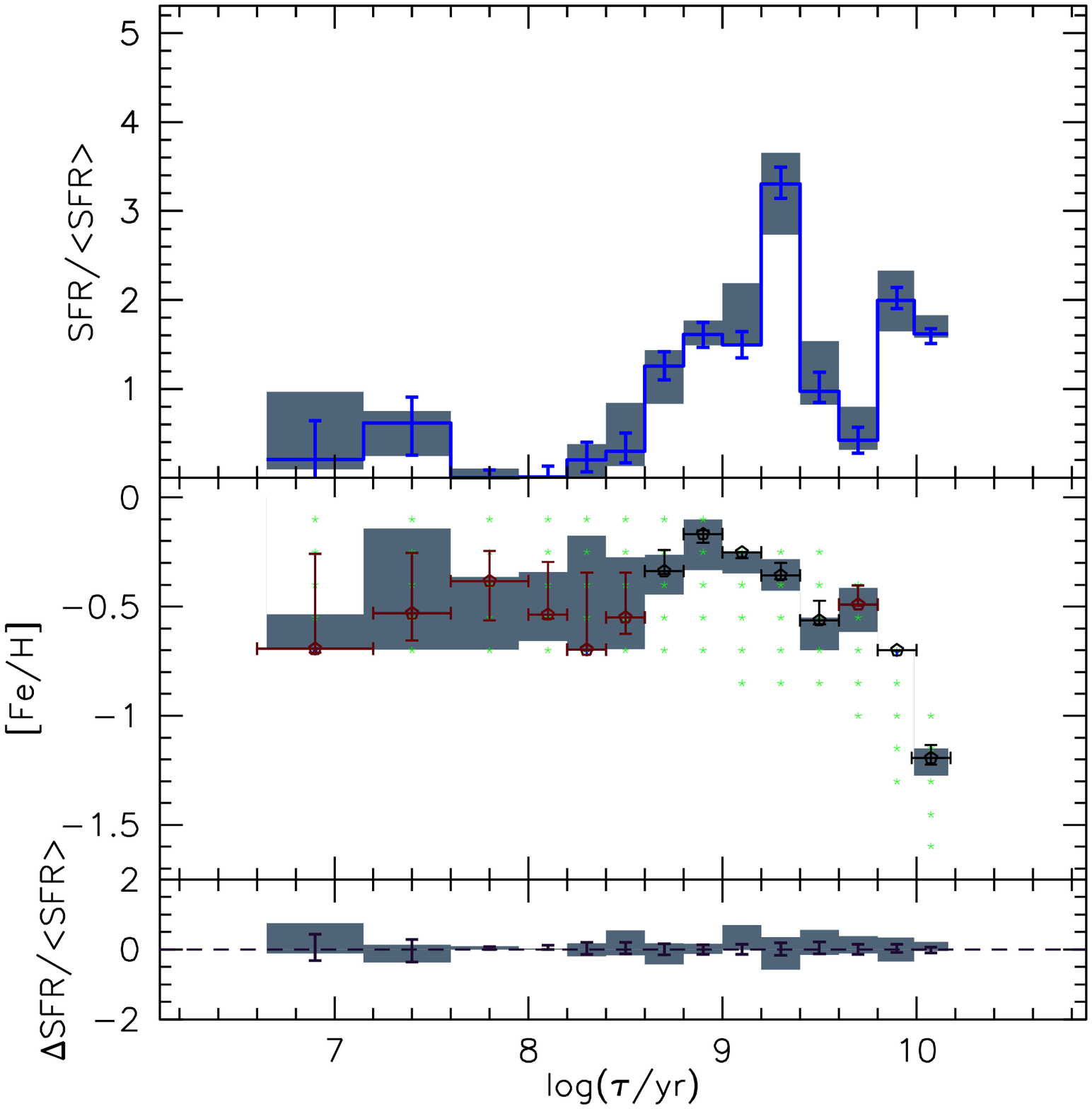}}
  \resizebox{0.45\hsize}{!}{\includegraphics{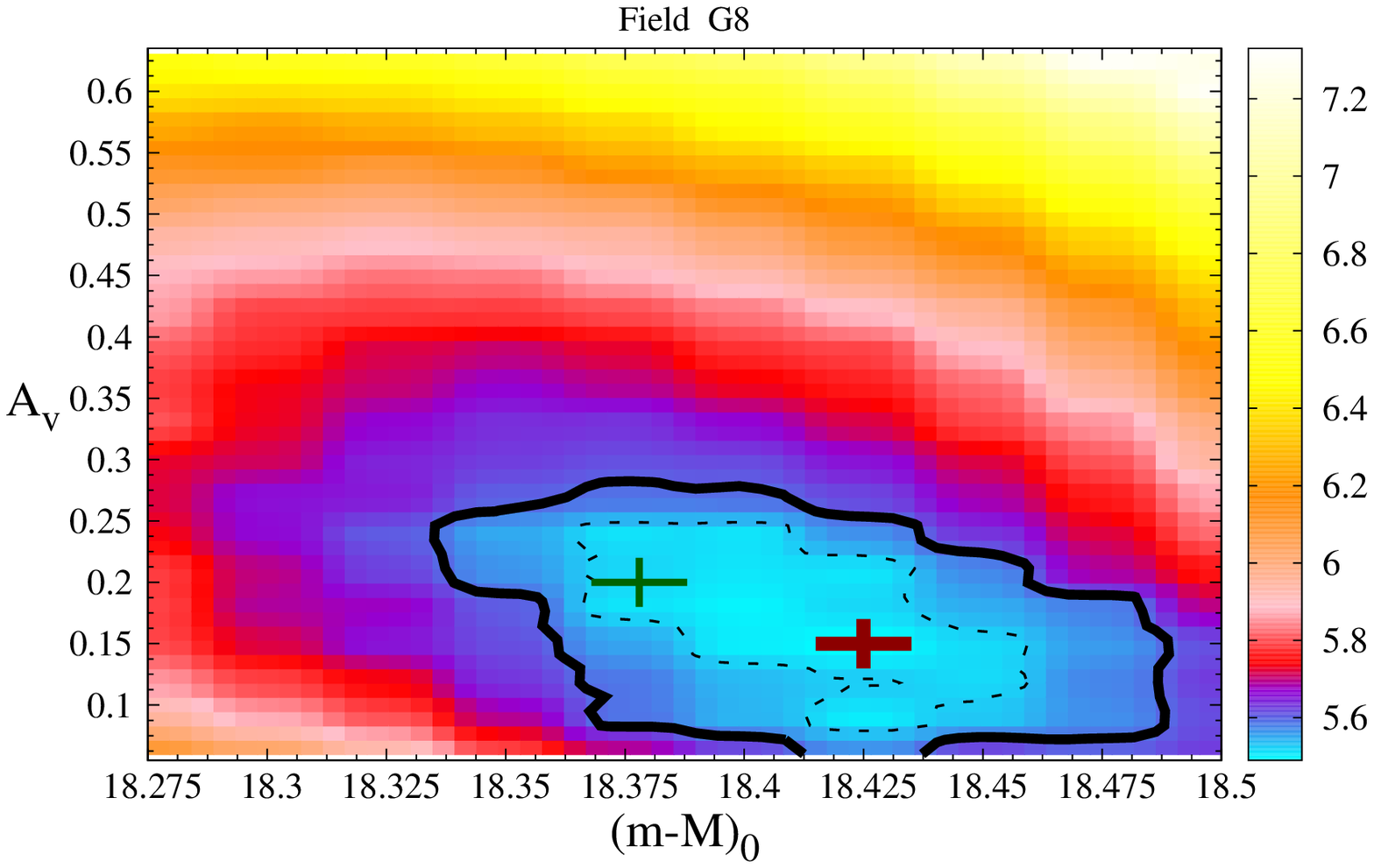}}
\caption{Same as Fig.~\ref{sol43} but for tile 8\_8.}
\label{sol88}
\end{figure*}

\section{Distance modulus and extinction}
\label{sec_dist_av}

As clearly illustrated in the right panels of Figs.~\ref{sol43} to
\ref{sol88}, the SFH recovery provides estimates of the distance
modulus and extinction, \dmo\ and \av, for each subregion.  These can
be used to probe the LMC disk geometry, as well as to build reddening
maps.

%TO BE MERGED:
%One of the remarkable improvements of the present SFH-recovery method
%in relation to ones presented by other authors is the capacity to
%recover not only the SFR(t) and AMR for each tile subregion, but also
%the distance modulus and reddening that are naturaly consistent with
%these physical information.  Therefore the best global solution for
%each tile subregion provides as a bonus the necessary results that can
%be used to probe the LMC disk geometry, as well as to build reddening
%maps.

We evaluated the \av\ and \dmo\ values in each subregion as the
average value inside the 68\% confidence level (1$\sigma$) of the
best-fitting solution, and its error from the width of this interval.
Then, we compared our results with values obtained in
\citet{Zaritsky_etal04} for the \av\ parameter, and
\citet{Nikolaev_etal04}, \citet{vdMC01}, \citet{vanderMarel_etal02},
\citet{OlsenSalyk02} and \citet{sub10} for \dmo.  Table~\ref{tab_dmav}
presents the results for all subregions considered. The parameters
\avcs\ and \avhs\ are the \av\ values from \cite{Zaritsky_etal04} in
the case of cool and hot stars, respectively, whereas \avtw\ and
\dmotw\ are the \av\ and \dmo\ values found in this work.  In the
table we also compare results for \dmo\ obtained in this work with
those derived from geometric models of the LMC, namely:
\cite{Nikolaev_etal04} in the case of $i=30.7^\circ$ and
$\theta_0=151.0^\circ$ (where $i$ is the inclination and $\theta_0$
the position angle of line of nodes), and \cite{vdMC01} and
\cite{vanderMarel_etal02} with parameters $i=34.7^\circ$,
$\theta_0=122.5^\circ$ and $\theta_0=129.9^\circ$, respectively.

\begin{table*}
  \caption{\dmo\ values obtained in this work: \dmotw\ from the 
    best-fitting SFH and \dmobftw\ from the best-fitting LMC disk. 
    Also, \av\ values obtained in this work (\avtw) compared to
    those from \cite{Zaritsky_etal04} for cool (\avcs) and hot 
    (\avhs) stars.}
\label{tab_dmav}
%\centering
\begin{tabular}{lcccccccc}
\hline
\hline  
Tile & Subregion & $\alpha$ (J2000) & $\delta$ (J2000) & \dmotw & \dmobftw  & \avtw & \avcs & \avhs\\
     &           & (deg) & (deg) & (mag) & (mag)  & (mag) & (mag) &  (mag) \\
\hline
\bf{4\_3}
& G1  &74.64&$-$72.62 & $18.55\; (-0.03,+0.02)$ &18.532& $0.219\; (-0.039,+0.021)$&  0.42$\pm$0.32&0.73$\pm$0.43\\
& G2  &74.81&$-$72.27 & $18.52\; (-0.04,+0.03)$ &18.527& $0.360\; (-0.090,+0.060)$&  0.36$\pm$0.32&0.57$\pm$0.42\\
& G3  &74.99&$-$71.92 & $18.52\; (-0.02,+0.03)$ &18.523& $0.295\; (-0.055,+0.064)$&  0.43$\pm$0.32&0.50$\pm$0.37\\
& G4  &75.16&$-$71.56 & $18.50\; (-0.03,+0.02)$ &18.518& $0.372\; (-0.042,+0.048)$&  0.46$\pm$0.30&0.54$\pm$0.36\\
& G5  &73.54&$-$72.57 & $18.55\; (-0.01,+0.01)$ &18.535& $0.372\; (-0.042,+0.048)$&  0.66$\pm$0.34&0.87$\pm$0.37\\
& G6  &73.73&$-$72.22 & $18.53\; (-0.01,+0.02)$ &18.530& $0.440\; (-0.050,+0.040)$&  0.62$\pm$0.32&0.84$\pm$0.38\\
& G7  &73.93&$-$71.87 & $18.53\; (-0.01,+0.02)$ &18.526& $0.354\; (-0.054,+0.036)$&  0.42$\pm$0.30&0.62$\pm$0.41\\
& G8  &74.11&$-$71.51 & $18.51\; (-0.01,+0.01)$ &18.521& $0.410\; (-0.050,+0.040)$&  0.48$\pm$0.29&0.52$\pm$0.38\\
& G10 &72.67&$-$72.16 & $18.54\; (-0.02,+0.03)$ &18.533& $0.240\; (-0.060,+0.060)$&  0.60$\pm$0.32&0.89$\pm$0.41\\
& G11 &72.87&$-$71.81 & $18.55\; (-0.03,+0.02)$ &18.529& $0.440\; (-0.081,+0.069)$&  0.45$\pm$0.31&0.74$\pm$0.40\\
& G12 &73.08&$-$71.45 & $18.55\; (-0.03,+0.02)$ &18.524& $0.373\; (-0.073,+0.077)$&  0.59$\pm$0.34&0.66$\pm$0.38\\
\hline
\bf{6\_6}
& D1  & 85.64  &        $-$69.92    & $18.43\; (-0.05,+0.02)$  &18.456& $0.650\; (-0.056,+0.033) $&  0.50$\pm$0.38& 0.68$\pm$0.42\\
& D2  & 83.27  &        $-$68.81    & $18.44\; (-0.01,+0.01)$  &18.450& $0.600\; (-0.040,+0.050) $&  0.44$\pm$0.34& 0.55$\pm$0.39\\
\hline
\bf{8\_3}
& G1  &76.90   & $-$66.83  &$18.48\; (-0.03,+0.05) $  &18.453& $ 0.286\; (-0.016,+0.043)$&0.45$\pm$0.32& 0.57$\pm$0.44\\
& G2  &77.00   & $-$66.48  &$18.44\; (-0.01,+0.04) $  &18.448& $ 0.321\; (-0.051,+0.038)$&0.38$\pm$0.30& 0.52$\pm$0.46\\
& G3  &77.10   & $-$66.13  &$18.44\; (-0.04,+0.03) $  &18.443& $ 0.324\; (-0.054,+0.095)$&0.35$\pm$0.26& 0.49$\pm$0.41\\
& G4  &77.20   & $-$65.77  &$18.47\; (-0.02,+0.05) $  &18.438& $ 0.237\; (-0.057,+0.032)$&0.39$\pm$0.28& 0.54$\pm$0.41\\
& G5  &76.09   & $-$66.80  &$18.44\; (-0.02,+0.03) $  &18.456& $ 0.360\; (-0.030,+0.030)$&0.41$\pm$0.32& 0.54$\pm$0.45\\
& G6  &76.19   & $-$66.44  &$18.42\; (-0.03,+0.03) $  &18.451& $ 0.372\; (-0.042,+0.048)$&0.38$\pm$0.28& 0.48$\pm$0.46\\
& G7  &76.29   & $-$66.09  &$18.44\; (-0.04,+0.04) $  &18.446& $ 0.360\; (-0.030,+0.030)$&0.33$\pm$0.26& 0.45$\pm$0.41\\
& G8  &76.40   & $-$65.74  &$18.45\; (-0.05,+0.03) $  &18.441& $ 0.366\; (-0.066,+0.053)$&0.40$\pm$0.30& 0.52$\pm$0.41\\
& G9  &75.25   & $-$66.76  &$18.45\; (-0.03,+0.02) $  &18.460& $ 0.420\; (-0.030,+0.030)$&0.37$\pm$0.26& 0.63$\pm$0.48\\
& G10 &75.38   & $-$66.40  &$18.46\; (-0.02,+0.04) $  &18.455& $ 0.276\; (-0.066,+0.054)$&0.33$\pm$0.26& 0.51$\pm$0.48\\
& G11 &75.50   & $-$66.05  &$18.42\; (-0.03,+0.05) $  &18.450& $ 0.412\; (-0.082,+0.067)$&0.34$\pm$0.28& 0.39$\pm$0.43\\
& G12 &75.62   & $-$65.70  &$18.44\; (-0.04,+0.03) $  &18.445& $ 0.303\; (-0.093,+0.057)$&0.33$\pm$0.27& 0.54$\pm$0.46\\
\hline
\bf{8\_8}  
& G1  & 90.83  &        $-$66.86    & $18.40\; (-0.05,+0.05)$  &18.389& $0.239\; (-0.118,+0.121) $&  0.25$\pm$0.20& 0.67$\pm$0.49\\
& G2  & 90.73  &        $-$66.51    & $18.38\; (-0.06,+0.04)$  &18.384& $0.238\; (-0.117,+0.122) $&  0.31$\pm$0.23& 0.83$\pm$0.45\\
& G3  & 90.64  &  $-$66.16    & $18.35\; (-0.05,+0.05)$  &18.379& $0.202\; (-0.111,+0.128) $&  0.35$\pm$0.26& 0.92$\pm$0.43\\
& G4  & 90.55  &  $-$65.81    & $18.35\; (-0.05,+0.05)$  &18.375& $0.188\; (-0.067,+0.082) $&  0.40$\pm$0.26& 0.95$\pm$0.41\\
& G5  & 89.99  &  $-$66.90          & $18.40\; (-0.02,+0.03)$  &18.393& $0.198\; (-0.077,+0.072) $&  0.27$\pm$0.22& 0.60$\pm$0.50\\
& G6  & 89.90  &  $-$66.54    & $18.38\; (-0.05,+0.03)$  &18.388& $0.213\; (-0.062,+0.087) $&  0.31$\pm$0.25& 0.77$\pm$0.47\\
& G7  & 89.81  &  $-$66.19    & $18.39\; (-0.04,+0.04)$  &18.384& $0.173\; (-0.053,+0.066) $&  0.36$\pm$0.27& 0.84$\pm$0.45\\
& G8  & 89.73  &  $-$65.84    & $18.41\; (-0.04,+0.04)$  &18.379& $0.176\; (-0.085,+0.064) $&  0.39$\pm$0.25& 0.90$\pm$0.44\\
%& G9  & 89.15  &  $-$66.92    & $18.39\; (-0.02,+0.03)$  &18.397& $0.249\; (-0.069,+0.051) $&  0.28$\pm$0.25& 0.54$\pm$0.49\\
& G10 & 89.08  &  $-$66.57          & $18.39\; (-0.02,+0.01)$  &18.393& $0.186\; (-0.036,+0.024) $&  0.36$\pm$0.29& 0.76$\pm$0.47\\
& G11 & 89.00  &  $-$66.22          & $18.41\; (-0.01,+0.01)$  &18.388& $0.146\; (-0.055,+0.034) $&  0.36$\pm$0.27& 0.77$\pm$0.41\\
& G12 & 88.93  &  $-$65.87          & $18.41\; (-0.04,+0.04)$  &18.383& $0.100\; (-0.040,+0.050) $&  0.40$\pm$0.28& 0.81$\pm$0.48\\
\hline
%\hline
\end{tabular}
\end{table*}

\subsection{Fitting the LMC disk plane}
\label{disk_fit}

For all the 36 subregions (see Table \ref{tab_dmav}) distributed in
four VMC tiles in which we have derived the SFH, we obtain {\em
  independent} determinations of the distance modulus and reddening.
The good accuracy of each distance determination allows us to fit a
disk plane geometry to the LMC disk.  To perform this fit we
considered the following five free parameters: the equatorial
coordinates for the LMC centre, \RAc\ and \DECc; the distance modulus
to the LMC centre, \dmoc; the disk inclination on the plane of the
sky, $i$ (where $i=0^\circ$ means a face-on disk); and the position
angle of the line of nodes, $\theta_{0}$.

In practice, to fit the LMC disk plane we used four different choices
for (\RAc, \DECc) in accordance with previous determinations of the
LMC disk geometry proposed by different authors
(Table~\ref{tab_disk_refs}).  This procedure simplified the search for
the best-fitting plane, and allowed us to check the dependence of the
results for the remaining three parameters with the adopted
coordinates for the LMC centre.

\begin{figure*}
%\begin{minipage}{0.85\textwidth}
\resizebox{0.50\hsize}{!}{\includegraphics{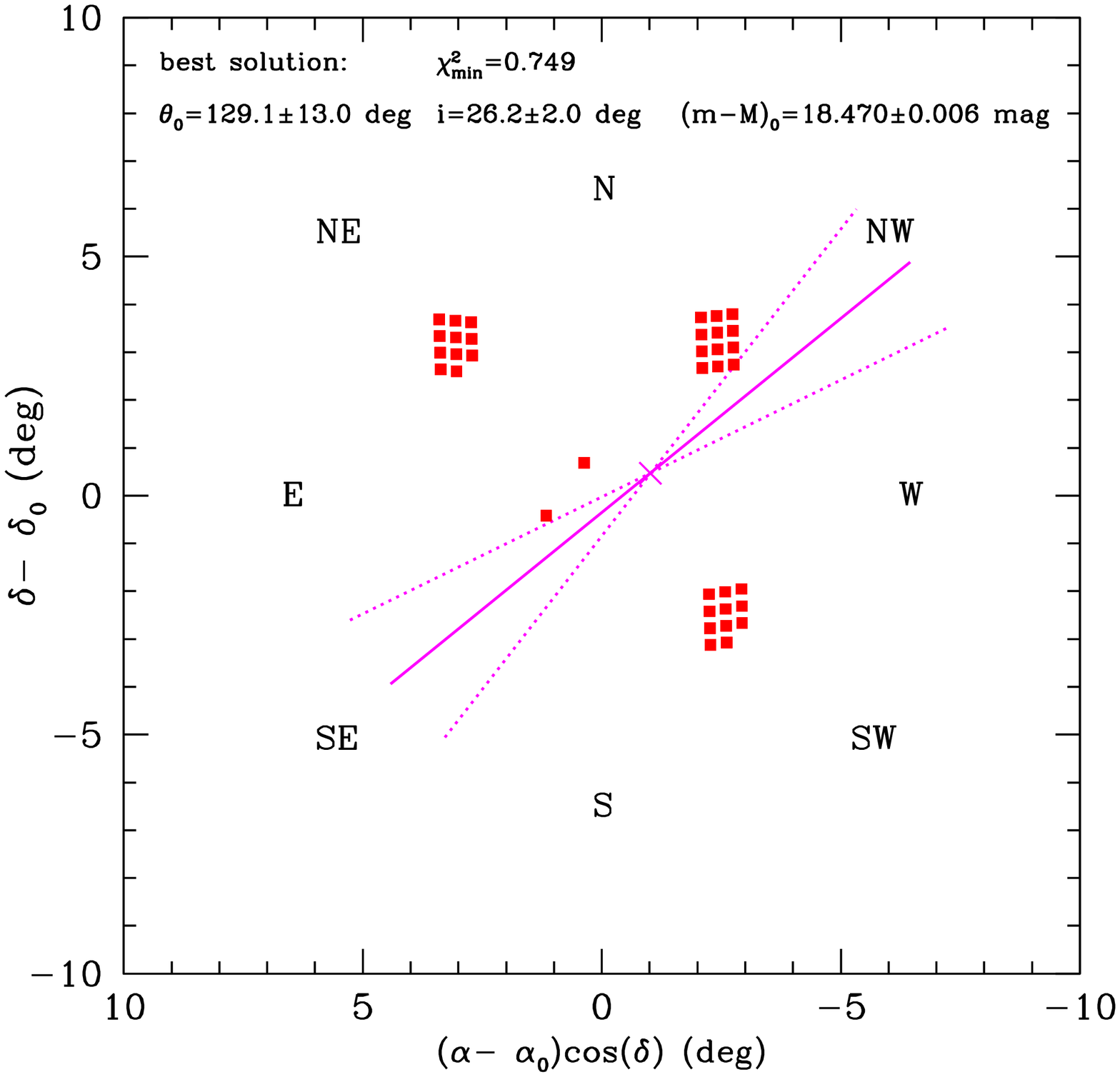}}
\resizebox{0.50\hsize}{!}{\includegraphics{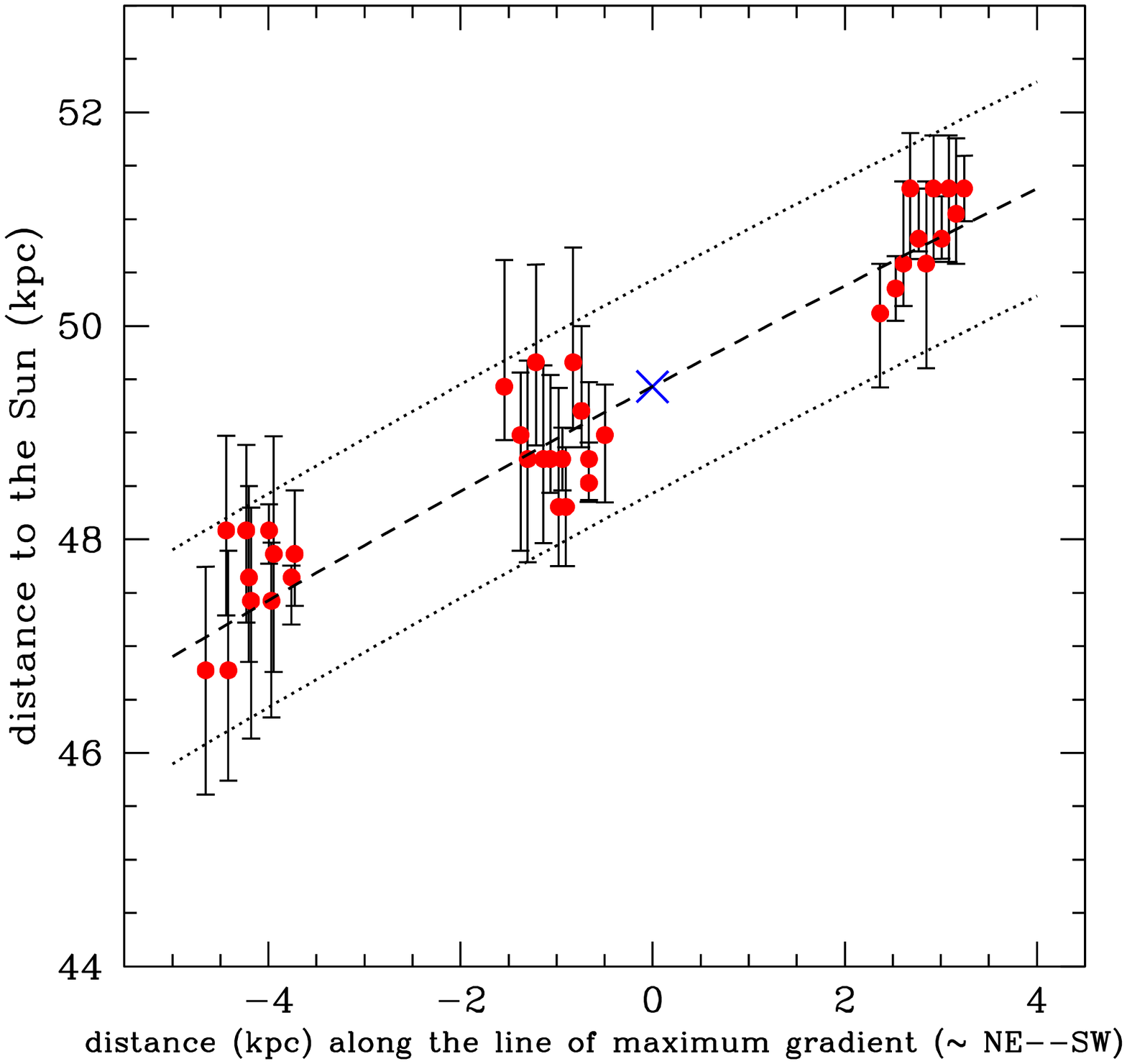}}
\caption{Best-fitting model of the LMC disk using as centre of the LMC
  the coordinates derived by \citet[][cross]{Nikolaev_etal04}.  {\bf
    Left panel:} the line of nodes (solid line) and the 36 tile
  subregions (red squares) as projected on the sky.  The disk
  parameters are shown in the figure, as well as the uncertainties in
  the position angle of the line of nodes (dotted lines).  {\bf Right
    panel:} distance to these subregions projected onto the line
  perpendicular to the line of nodes, i.e. the line of the maximum LMC
  disk gradient ($\sim$ NE--SW direction). }
\label{fig_disk_fit}
\end{figure*}

Figure \ref{fig_disk_fit} illustrates the best-fitting model for a
specific choice for (\RAc,\DECc), in this case that adopted by
\citet{Nikolaev_etal04}. The left panel shows the results of the LMC
disk as projected on the sky, whereas the right panel presents the
projection along the line perpendicular to the line of nodes, i.e.,
the line of the maximum gradient for the LMC disk.  As can be seen in
this figure, all fields studied so far are fit extremely well by a
single disk with the following parameters: $i=26.2\pm 2.0^\circ$,
$\theta_{0}=129.1\pm13.0^\circ$ and $\dmoc = 18.470\pm 0.006$~mag.
The errors in each parameter correspond to the 68\% confidence level,
and are computed using the bootstrapping technique, where 50 resamples
(for the set of \dmo\ values) are generated while accounting for the
individual errors in each measurement.  Table~\ref{tab_disk}
summarizes the final results for the LMC disk geometry, revealing that
these results are quite insensitive to the adopted (\RAc, \DECc)
values.  A comparison between Tables~\ref{tab_disk_refs} and
\ref{tab_disk} shows that we recover a disk significantly less
inclined than other authors, who tend to find $i$ values close to
$\sim35^\circ$ (with the exception of one value from \citealt{sub10}).
Regarding $\theta_{0}$, our values are well inside the wide range --
from $122.5^\circ$ to $163.7^\circ$ -- found in the literature.

The distances we recover to the LMC centre, \dmoc, are in good
agreement with most of the determinations in the recent past, which as
demonstrated by \citet{schaefer08}, tend to cluster extremely well
(and suspiciously well, from a statistical point of view) around the
value of $\dmo=18.50 \pm 0.10$~mag adopted by the HST Key Project on
the distance scale \citep{Freedman01}. Regarding our distance
determinations, we recall that they are derived in an objective way
from the global fitting of the CMD, and not from any particular set of
favoured distance indicators. Despite the good agreement with the
``standard'' distance values, we recognize that this particular result
could be affected by systematic errors in the stellar evolution models
and/or photometric ZPs, and should be verified by means of independent
methods. This is beyond the scope of the present work.  Forthcoming
papers will use different distance indicators on VMC data (in
particular the RR~Lyrae and red clump stars), to further discuss this
issue.

\begin{table*}
\caption{LMC disk parameters from the literature.}
\label{tab_disk_refs}
%\centering
\begin{tabular}{cccccc}
\hline
\hline
\RAc\ (J2000) & \DECc\ (J2000)  &  $i$ & $\theta_{0}$ &  Reference & distance indicator \\
  (deg) & (deg) & (deg) & (deg) & & \\
  \hline 
79.40 & $-$69.03 &   $30.7 \pm 1.1$ & $151.0 \pm 2.4$ & \cite{Nikolaev_etal04} & Cepheids (MACHO + 2MASS) \\
82.25 & $-$69.50 &   $34.7 \pm 6.2$ & $122.5 \pm 8.3$ & \cite{vdMC01} & AGB stars (DENIS + 2MASS) \\
81.90 & $-$69.87 &   $34.7 \pm 6.2^*$ & $129.9 \pm 6.0$ & \cite{vanderMarel_etal02} & carbon stars (kinematics) \\
79.91 & $-$69.45 &   $35.8 \pm 2.4$ & $145 \pm 4$ & \citet{OlsenSalyk02} & red clump (CTIO 0.9m $VI$ photom.)\\
79.91 & $-$69.45 &   $23.0 \pm 0.8$ & $163.7 \pm 1.5$ & \citet{sub10} & red clump (OGLE\,III $VI$ photom.) \\
79.91 & $-$69.45 &  $37.4 \pm 2.3$ & $141.2 \pm 3.7$ & \citet{sub10} & red clump (MCPS $VI$ photom.) \\
  \hline
\end{tabular}
$^*$Fixed to the \cite{vdMC01} value.
\end{table*}

\begin{table*}
\caption{Our best-fitting models for the LMC disk, for different choices of \RAc\ and \DECc.}
\label{tab_disk}
%\centering
\begin{tabular}{cccccc}
\hline
\hline
\RAc (J2000) & \DECc (J2000) & $i$ & $\theta_{0}$ & \dmoc & $\chi^{2}_{\rm{min}}$ \\
(deg) & (deg) & (deg) & (deg) & & \\
\hline
79.40 & $-$69.03 & $26.2\pm2.0$ & $129.1\pm13.0$ & $18.470\pm0.006$ & 0.749 \\
82.25 & $-$69.50 & $26.2\pm1.9$ & $126.4\pm10.1$ & $18.466\pm0.006$ & 0.785 \\
81.90 & $-$69.87 & $26.2\pm2.0$ & $130.9\pm8.9$ & $18.470\pm0.005$ & 0.750 \\
79.91 & $-$69.45 & $26.2\pm1.7$ & $129.6\pm10.1$ & $18.471\pm0.006$ & 0.769 \\
\hline
\end{tabular}
\end{table*}

\subsection{Extinction values}
\label{sect_Av}

The spatial resolution adopted in the present work is too coarse
to allow the derivation of detailed extinction maps. However, a first
comparison with other works is advisable for the sake of validating
our results, and is particularly interesting because previous
extinction maps are mainly based on optical data.
 
Table~\ref{tab_dmav} presents the \av\ values derived in this work,
\avtw, in comparison to those derived from the Magellanic Cloud
Photometric Survey (MCPS) data by \cite{Zaritsky_etal04}, in the case
of cool and hot stars (\avcs\ and \avhs, respectively). Note that our
results represent mean values for an entire subregion, in contrast to
the star-by-star values derived by \cite{Zaritsky_etal04}.  This
explains why their error bars are intrinsically much longer.

In regions with low stellar density, our \av\ values are much smaller
than those of \cite{Zaritsky_etal04}, in particular if we consider the
\avhs. Better agreement is present if we consider denser regions, in
particular in the two subregions of the 6\_6 tile. We note that
\cite{Haschke11} find a similar discrepancy between their \av\ values,
derived from both the red clump and RR~Lyrae \vi\ colours from
OGLE~III data, and those from \cite{Zaritsky_etal04}. \cite{Haschke11}
obtained, for a wide area over the LMC, a mean colour excess
$\evi=0.09\pm0.07$~mag that translates into $\av=0.22\pm0.17$~mag, in
good general agreement with our values.  The reddening issue will be
further discussed in a forthcoming VMC paper.

%%%%%%%%%%%%%%%%%%%%%%%%%%%%%%%%%%%%%%%%%%%%%%%%%%%%%%%%%%%
\section{Discussion and conclusions}
\label{sec_discussion}

\subsection{Reducing SFH errors}

We have recovered the SFH in four VMC tiles across the LMC evaluating
simultaneously their best-fitting \SFRt, AMR, \av\ and \dmo, and their
stochastic and systematic errors inside the 68\% confidence level in
each subregion for each tile. We find clear indications that these LMC
regions are distributed, to a first approximation, across a single
disk plane.

We can take advantage of this latter conclusion to further improve the
SFH results. Indeed, the uncertainties in both \dmo\ and \av\
contribute to the systematic errors in the derived \SFRt\ and AMR. If
we assume that this distance is exactly known and defined by the
best-fitting disk, only the range of \av\ values is left to be
explored, and the errors are expected to decrease. We therefore make
this assumption of a known distance, given by the plane defined in the
first row of Table~4, for each subregion in each tile. SFH-recovery
is re-done by exploring the same range of \av\ values as before, while
the errors are re-computed.

\begin{figure*}
%\resizebox{0.49\hsize}{!}{\includegraphics{sfr_all88.eps}}
%\resizebox{0.49\hsize}{!}{\includegraphics{sfr_D_all88.eps}}
\resizebox{0.25\hsize}{!}{\includegraphics{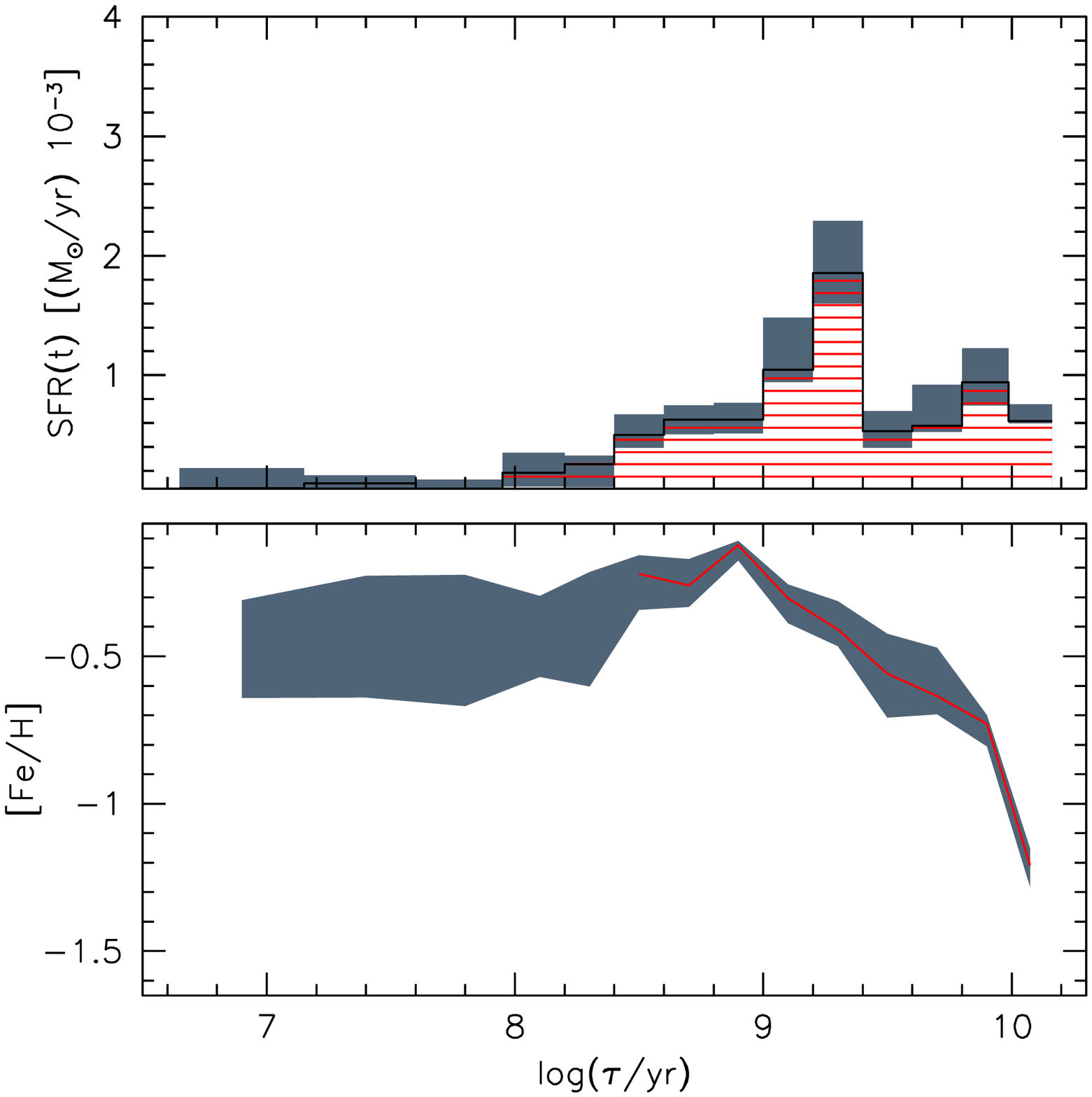}}
\resizebox{0.25\hsize}{!}{\includegraphics{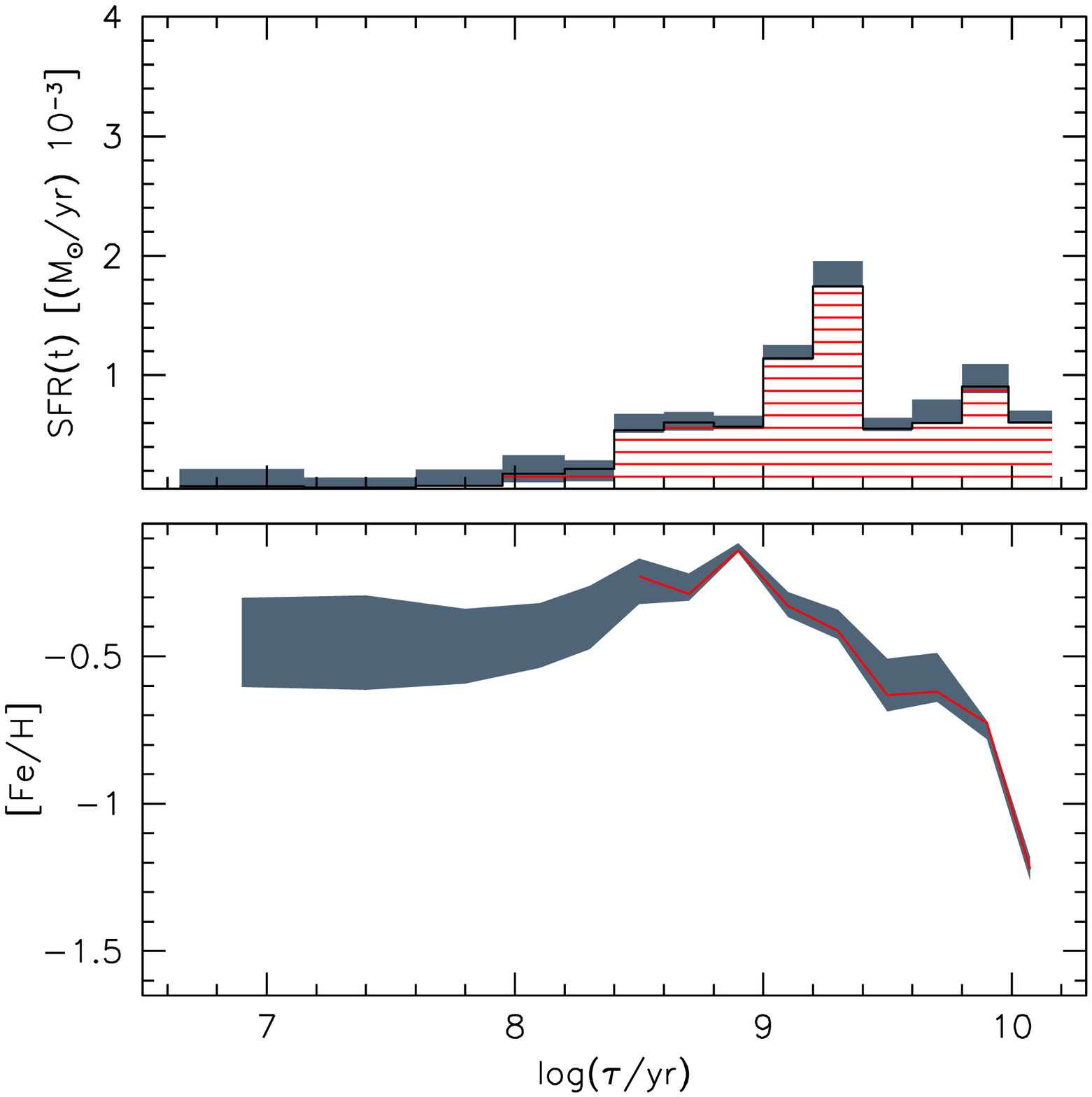}}
\resizebox{0.25\hsize}{!}{\includegraphics{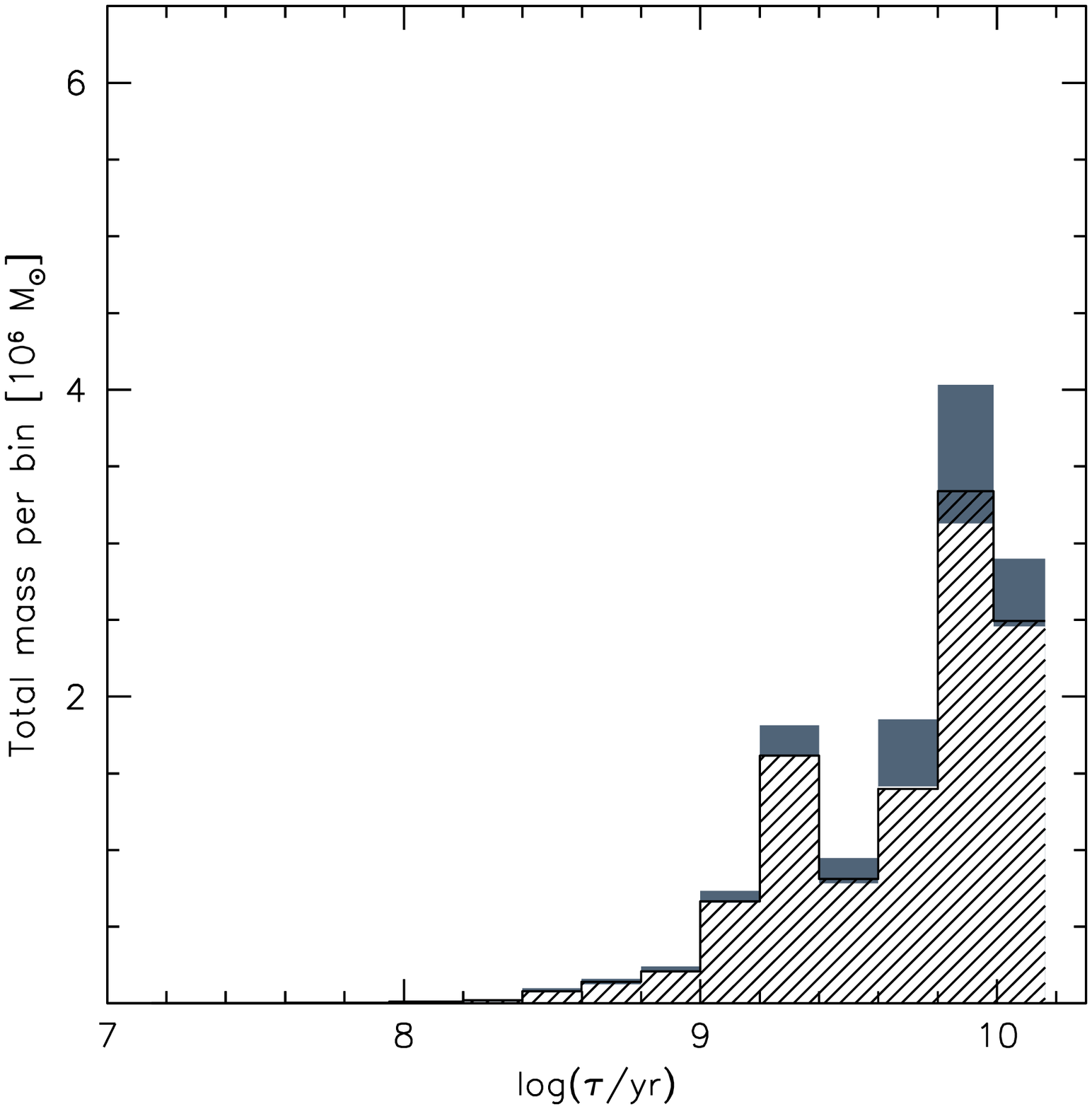}}
\caption{Total \SFRt\ and AMR derived for tile 8\_8, resulting from
  the addition of the SFH values for all subregions. {\bf Left
    panels:} The \SFRt\ as a function of \logtyr\ (red shaded
  histogram) and its systematic variations (gray), together with the
  mean AMR in all age bins in which the \SFRt\ is non-negligible (red
  line) and its systematic variations (gray). These results are
  obtained assuming that both \dmo\ and \av\ are free parameters. The
  {\bf middle panels} show the same but using a fixed \dmo\ --
  obtained from the best-fitting disk geometry -- and assuming that
  only \av\ is a free parameter. The {\bf right panel} shows the total
  stellar mass formed inside each \logtyr\ bin (shaded histogram)
  together with the systematic variations (gray area). }
\label{allsfr88}
\end{figure*}

Figure~\ref{allsfr88} illustrates the typical results of this
exercise.  Note that, in addition to illustrating the effect of
assuming a known distance, the figure also presents the total results
added over all subregions in a tile -- in this case, the 8\_8 tile.
The same general features are observed in every single subregion we
examined.

It is striking in Fig.~\ref{allsfr88} that in both cases (with
unknown/known distance) we derive about the same mean \SFRt\ and AMR,
but the error bars are significantly reduced when the distance is
known.  For the oldest age bins, the error bars are reduced by a
factor of about 2. This exercise clearly indicates the potential of
exploring the SFH over wide areas in the LMC: we can take advantage of
known correlations between parameters over large scales (in this case,
\dmo), to improve the SFH obtained in small regions of the galaxy.

\subsection{The stellar mass formation history} 
\label{ssec_stmass}

In the following subsections we comment on the global results for each
tile, starting from those with the best photometry. Before doing that,
we call attention to the rightmost panels in
Figs.~\ref{allsfr88}--\ref{sol661}, which show the distribution of
{\em total mass of formed stars} along the age bins defined in this
work. These plots indicate that the star formation at young ages
represents, as a rule, just a minor fraction of the total stellar mass
formed in the LMC. For all tiles studied so far, most of the star
formation has occurred at ages $\logtyr\!>\!9.5$ ($t\!>\!3$~Gyr),
whereas just $\la\!9$~\% has occurred at ages $\logtyr<9.0$.
Therefore, even strong peaks detected with high significance in the
\SFRt\ at young ages, may not represent major events in the formation
of stellar mass in the LMC.

Another remarkable feature in these panels is the modest size of the
systematic errors, especially at the oldest age bins where one could
have expected them to be significant because of the incompleteness and
larger photometric errors at the level of the oldest MSTOs. Instead,
both random and systematic errors keep modest because of the large
numbers of stars sampled by VMC, as anticipated by \citet{Kerber09}.
Summing the values in the right panels in
Fig.~\ref{allsfr88}--\ref{sol661}, one can estimate the total stellar
mass formed for every analised region of the LMC, with errors of just
$\la\!20$~\%. The main limitation of these estimates are ptobably in
the uncertainties in the initial mass function, which determines the
fraction of total mass going into faint, undetectable main sequence
stars.
 
\subsection{Tile 8\_8} 
\label{ssec_88}

This tile represents on average $\av\!\sim\!0.2$~mag and
$\dmo\!\sim\!18.39$~mag.  
Considering old stellar populations, the \SFRt\ (Fig.~\ref{allsfr88})
is similar from subregion to subregion. This is more evident comparing
the different panels in Fig.~\ref{sfr88}.  In addition to the oldest
star formation detected at $\logtyr>10.0$, which appears with
$\feh\!\sim\!-1.0$~dex, it is possible to see evidence for two other
main star formation (SF) episodes:
\begin{enumerate}
\item The first happened at $\logtyr=9.9$ and has
  $\feh\!\sim\!-0.70$~dex. It may have formed up to 31~\% of the total
  stellar mass in this tile.
\item The second is more recent and occurred between $\logtyr=9.1$ and
  9.3, with an average $\feh\!\sim\!-0.42$~dex. This metallicity
  coincides with the values found in LMC intermediate age clusters of
  similar age \citep{Kerber_etal07, Olszewski_etal91, MG03,
    Grocholski_etal06}. Although it appears as a very prominent peak
  in the \SFRt\ plot (left and middle panels), it represents just
  21~\% of the total mass formed in this tile (right panel).
\end{enumerate}
In addition, examination of Fig.~\ref{sfr88} reveals the presence of a
third peak in the \SFRt\ at $\logtyr=8.5-8.7$ , which is well evident
only in subregions G1, G5, G6, and marginally seen also in G7, G10,
G11 and G12. These are also the innermost subregions of this tile.
Therefore, there is a clear indication that this more recent SF
episode did not occur outside of a given radius in the LMC disk.

The SF for ages less than $\logtyr=8.3$ seems to be negligible in this
tile.  It was not possible to evaluate the AMR of the youngest stellar
populations, because of the very low \SFRt\ at all ages $\logtyr<8.3$.

\subsection{Tile 8\_3} 
\label{ssec_83}

\begin{figure*}
\resizebox{0.25\hsize}{!}{\includegraphics{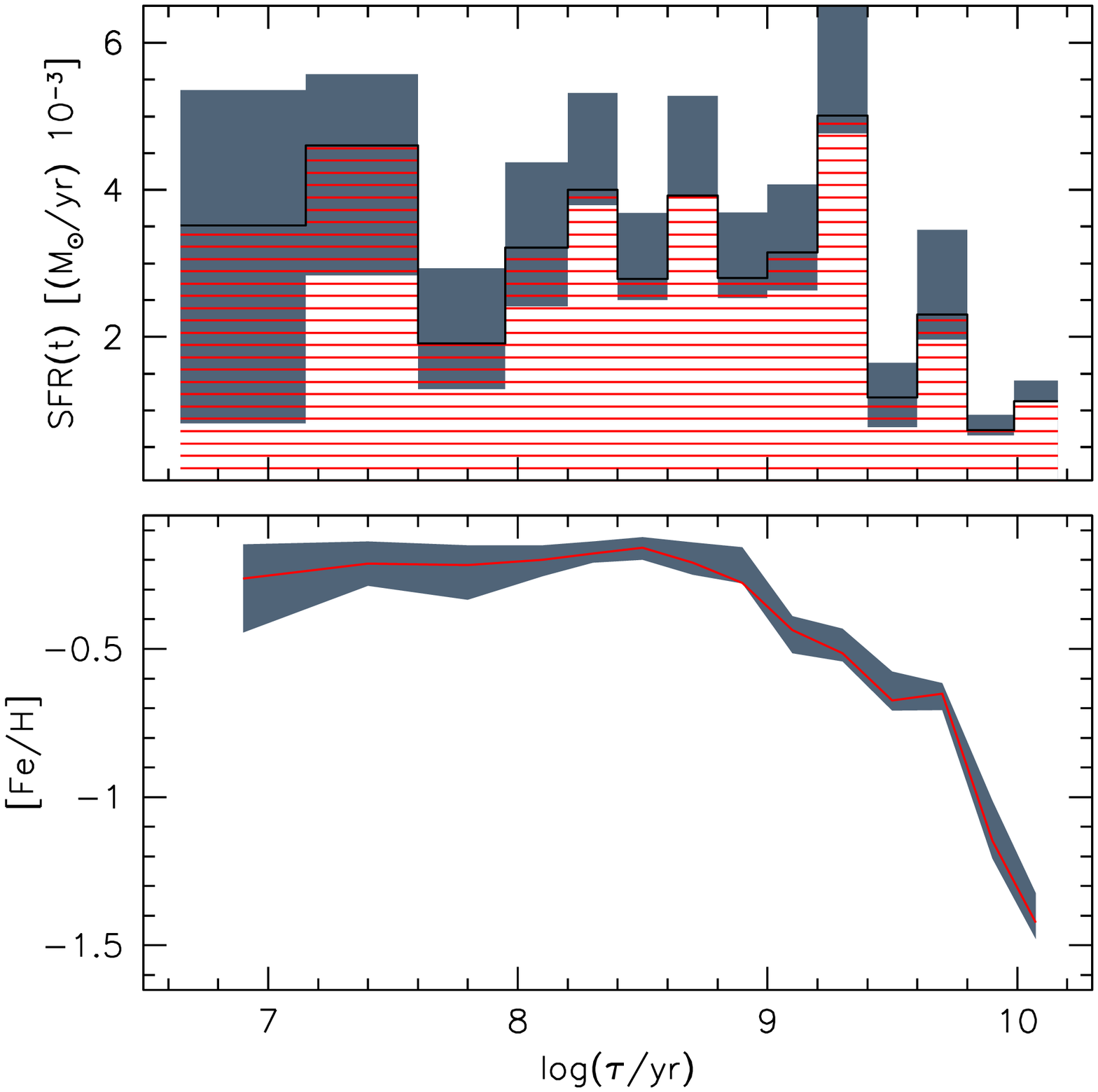}}
\resizebox{0.25\hsize}{!}{\includegraphics{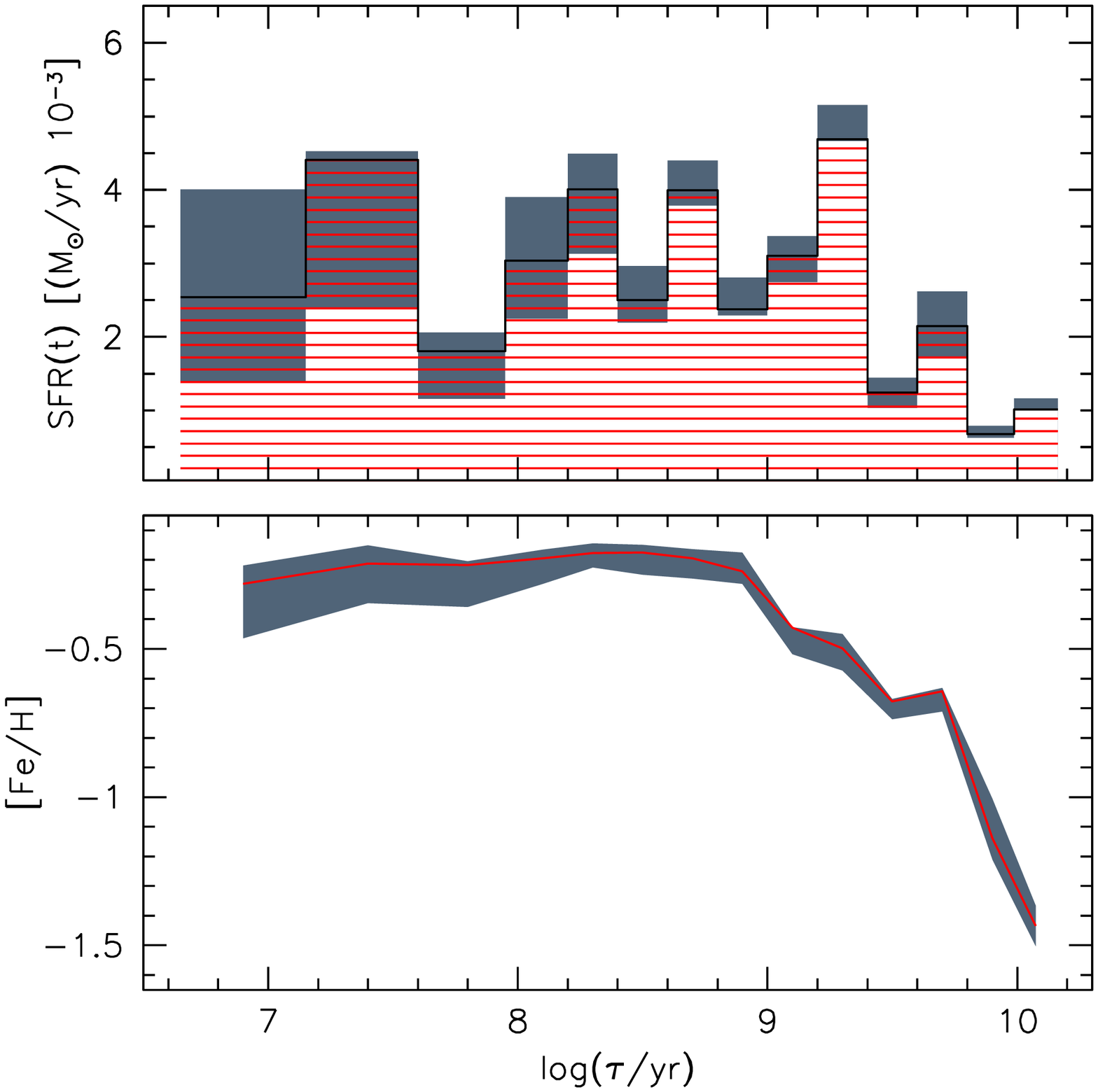}}
\resizebox{0.25\hsize}{!}{\includegraphics{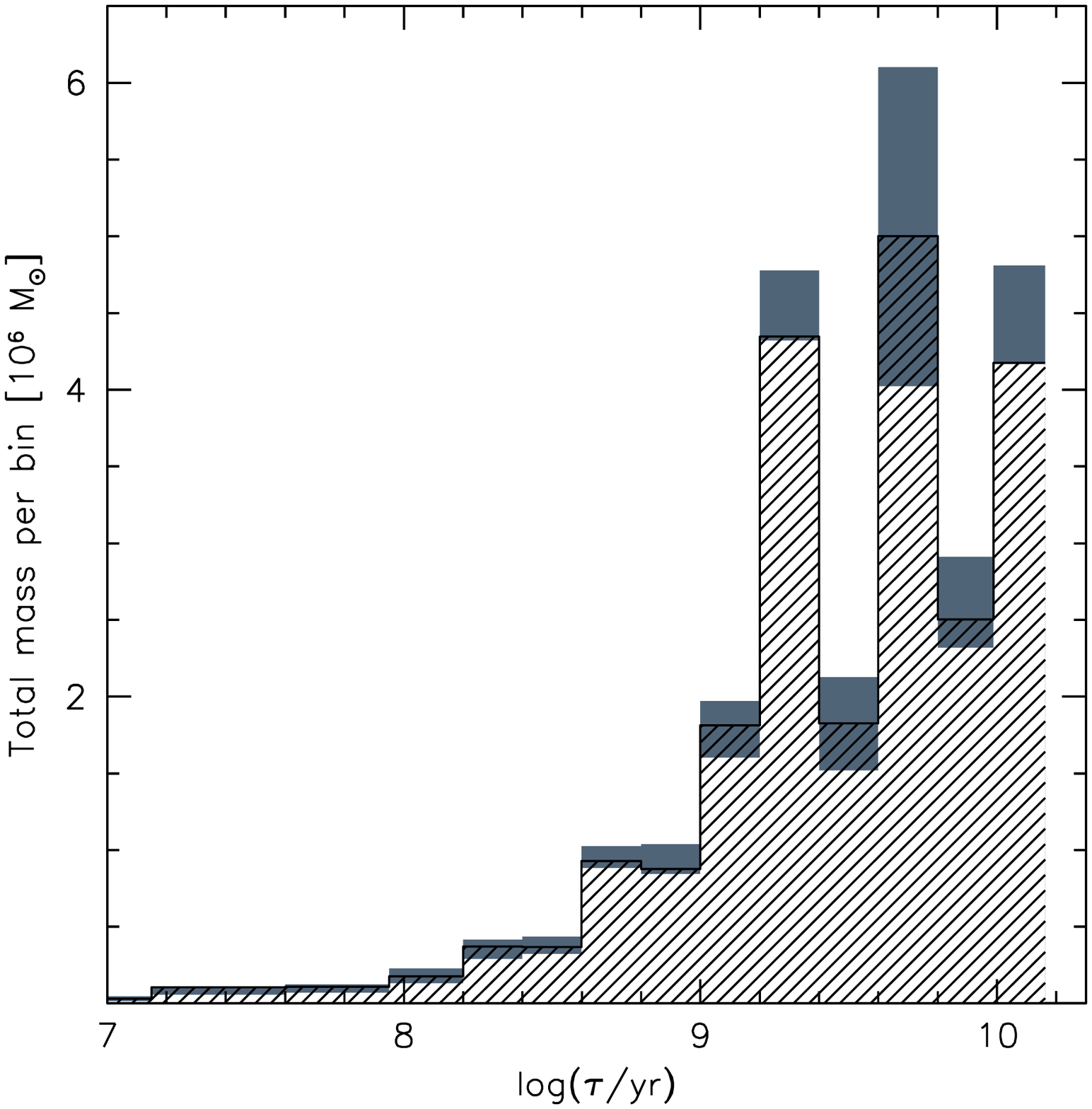}}
\caption{Total \SFRt\ and AMR in tile 8\_3. Colours and lines are as in
  Fig.~\ref{allsfr88}.}
\label{allsfr83}
\end{figure*}

In this tile the typical \av\ and \dmo\ values are $\sim\!0.33$ and
$\sim\!18.45$~mag, respectively.  Fig.~\ref{allsfr83} shows the total
\SFRt\ and AMR in the same way as for tile $8\_8$. Note the prominent,
and almost constant \SFRt\ over the most recent couple of Gyr.
This young SF changes significantly from subregion to subregion, as
revealed by Fig.~\ref{sfr83}. SF for ages $\logtyr<8$ is present in
most of the subregions in the West part of this tile (from G5 to G12),
although with large error bars. This recent \SFRt\ can be associated
to the presence of gas in the proximity of N11, the second-most most
proficient star-forming region in the LMC, which falls next to the
Western border of the 8\_3 tile.

As for the tile 8\_8, in addition to the oldest star formation
detected at $\logtyr>10.0$ with $\feh\!\sim\!-1.0$~dex, it is also
possible to identify two main SF episodes:
\begin{enumerate}
\item The first at $\logtyr=9.7$ and with a $\feh\!\sim\!-0.65$~dex,
  which formed 22~\% of the stellar mass.
\item The second between $\logtyr=9$ and 9.4 and with an average
  $\feh\!\sim\!-0.47$~dex, forming about 27~\% of the stellar mass.
\end{enumerate} 

The total AMR seems to change little from subregion to subregion (see
Fig.~\ref{sfr83}), and shows small variations also for the young
stellar population.
 
\subsection{Tile 4\_3} 
\label{ssec_43}

\begin{figure*}
\resizebox{0.25\hsize}{!}{\includegraphics{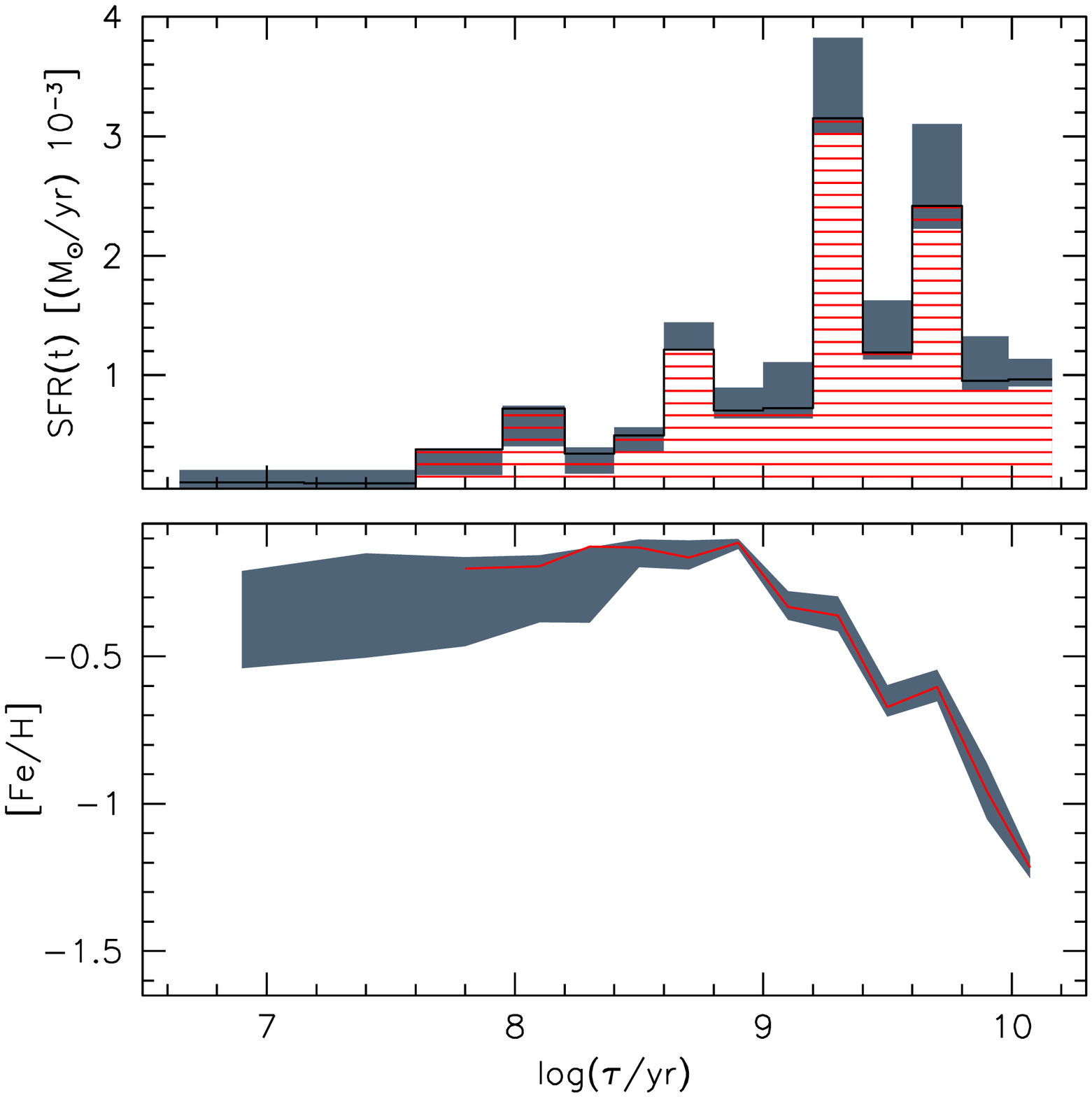}}
\resizebox{0.25\hsize}{!}{\includegraphics{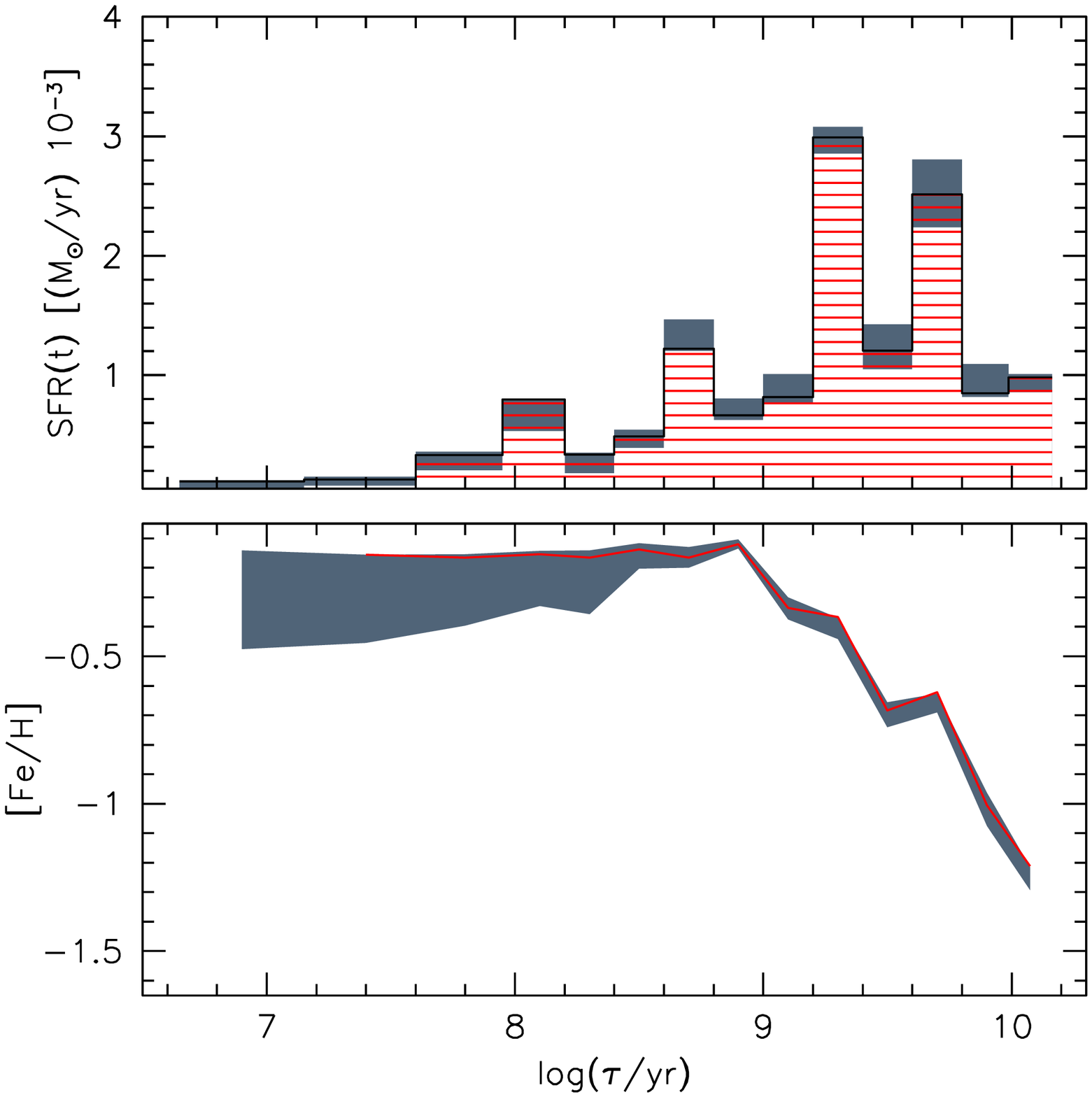}}
\resizebox{0.25\hsize}{!}{\includegraphics{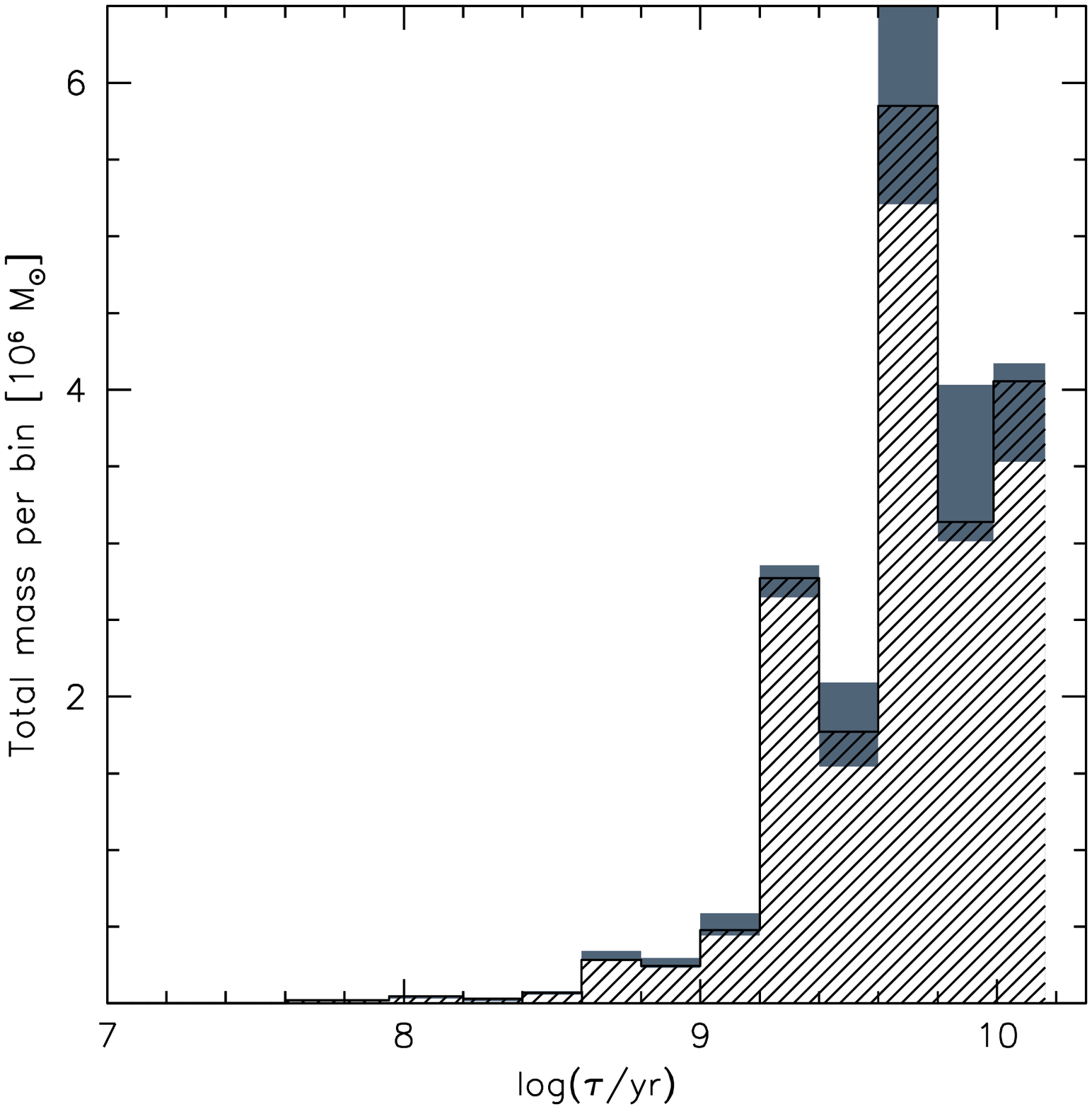}}
\caption{Total \SFRt\ and AMR in tile 4\_3. Colours and lines are as in
  figure \ref{allsfr88}.}
\label{allsfr43}
\end{figure*}

This tile presents average \av\ and \dmo\ values close to $\sim\!0.33$
and $\sim\!18.54$~mag, respectively.  Fig.~\ref{allsfr43} shows the
total \SFRt\ and AMR. The SF of the young stellar population
($\logtyr<8$) is as weak as in the 8\_8 tile, whereas the SF in the
older stellar population appears similar in most subregions (see
Fig.~\ref{sfr43}). In addition to the oldest period of star formation
at $\logtyr\!>\!10.0$ with $\feh\!\sim\!-1.0$~dex, it is possible to
deduce three main peaks in the \SFRt:
\begin{enumerate}
\item The first at $\logtyr=9.7$ and with $\feh\!\sim\!-0.62$~dex,
  forming 31~\% of the stellar mass in the tile.
\item The second at $\logtyr=9.3$ and with an average
  $\feh\!\sim\!-0.35$~dex, forming 15~\% of the stellar mass.
\item The youngest at $\logtyr=8.7$ with average
  $\feh\!\sim\!-0.18$~dex. Although this peak appears evident in the
  \SFRt\ plot, it represents just a modest 1.5~\% of the stellar mass.
\end{enumerate} 
Note that the young SFR, with $\logtyr\la9$, is concentrated in
subregions G4 and G8 (see Fig.~\ref{sfr43}), which are those more
centrally located over the LMC disk. Regarding the two oldest peaks in
the \SFRt\, they are very similar to those derived for the 8\_3 tile,
supporting a good degree of mixing among older stellar populations
across the LMC disk \citep[see also][]{HZ99, HZ09, Cioni_etal00,
  Cioni09, NW00, blu06, Carrera_etal11}.

The AMR is well evaluated for all ages $\logtyr>7.6$, where the \SFRt\
is non negligible.

\subsection{Tile 6\_6} 
\label{ssec_66}

\begin{figure}
\resizebox{0.49\hsize}{!}{\includegraphics{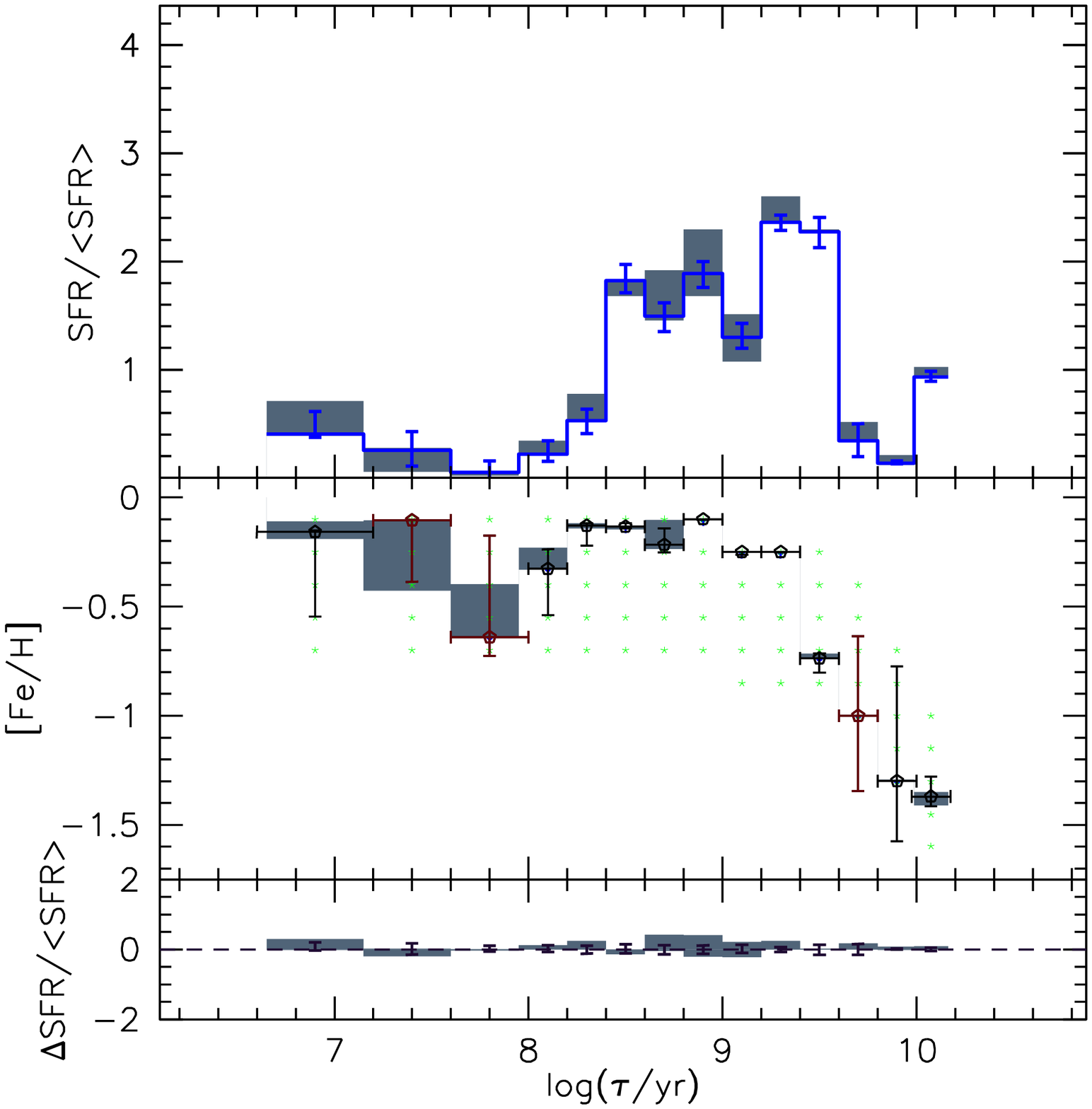}}
\resizebox{0.49\hsize}{!}{\includegraphics{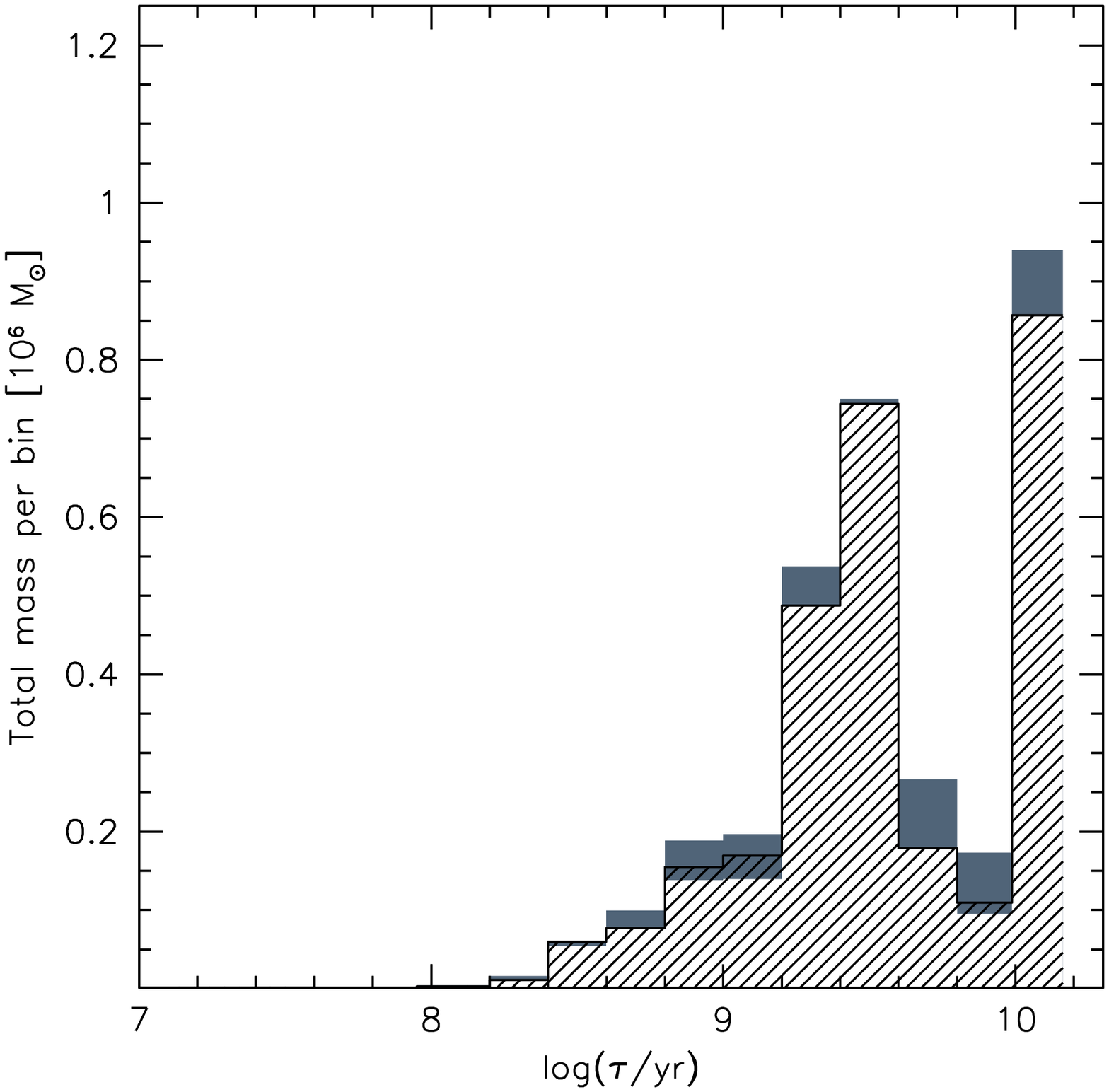}}
\\
\resizebox{0.49\hsize}{!}{\includegraphics{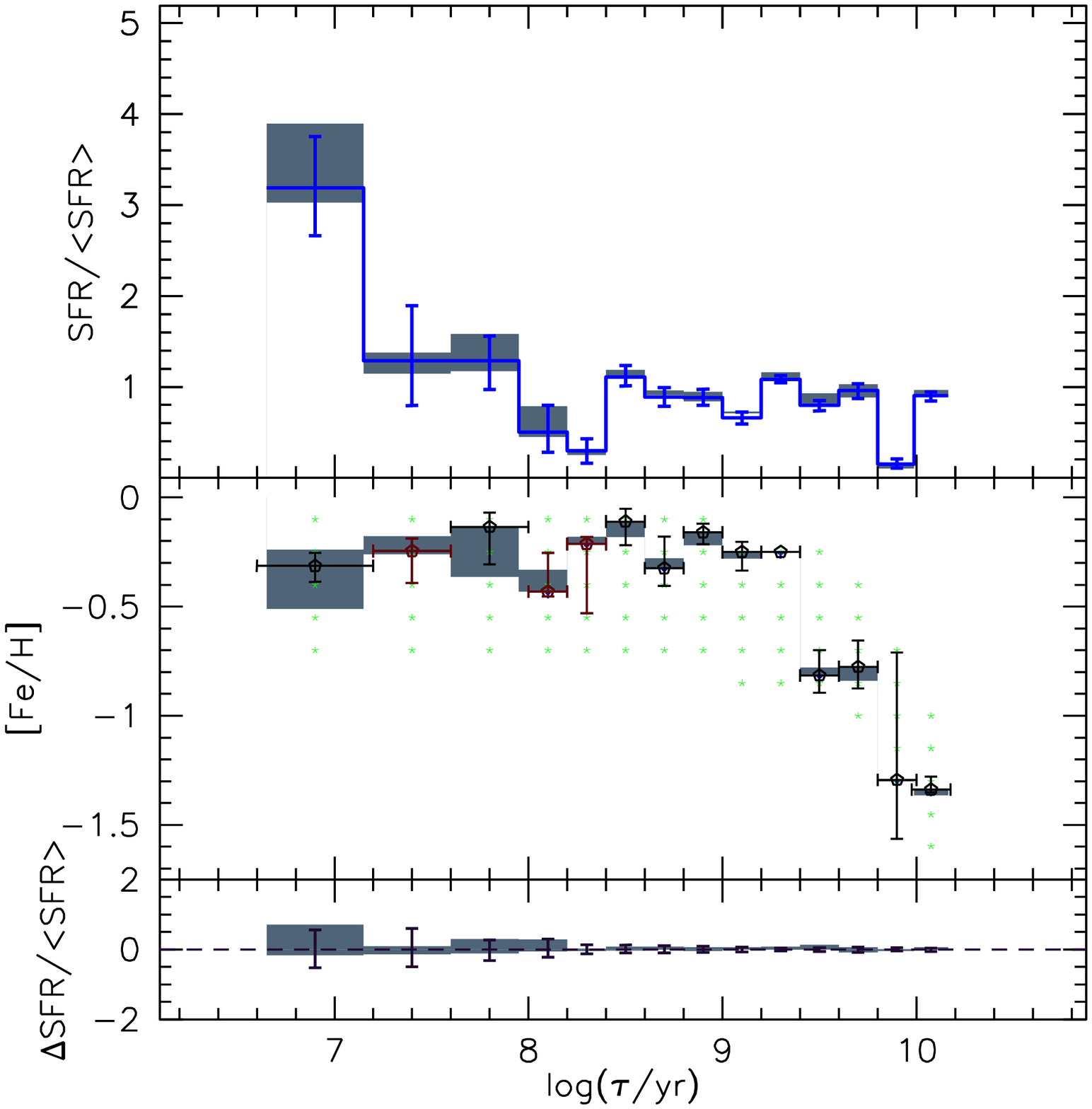}}
\resizebox{0.49\hsize}{!}{\includegraphics{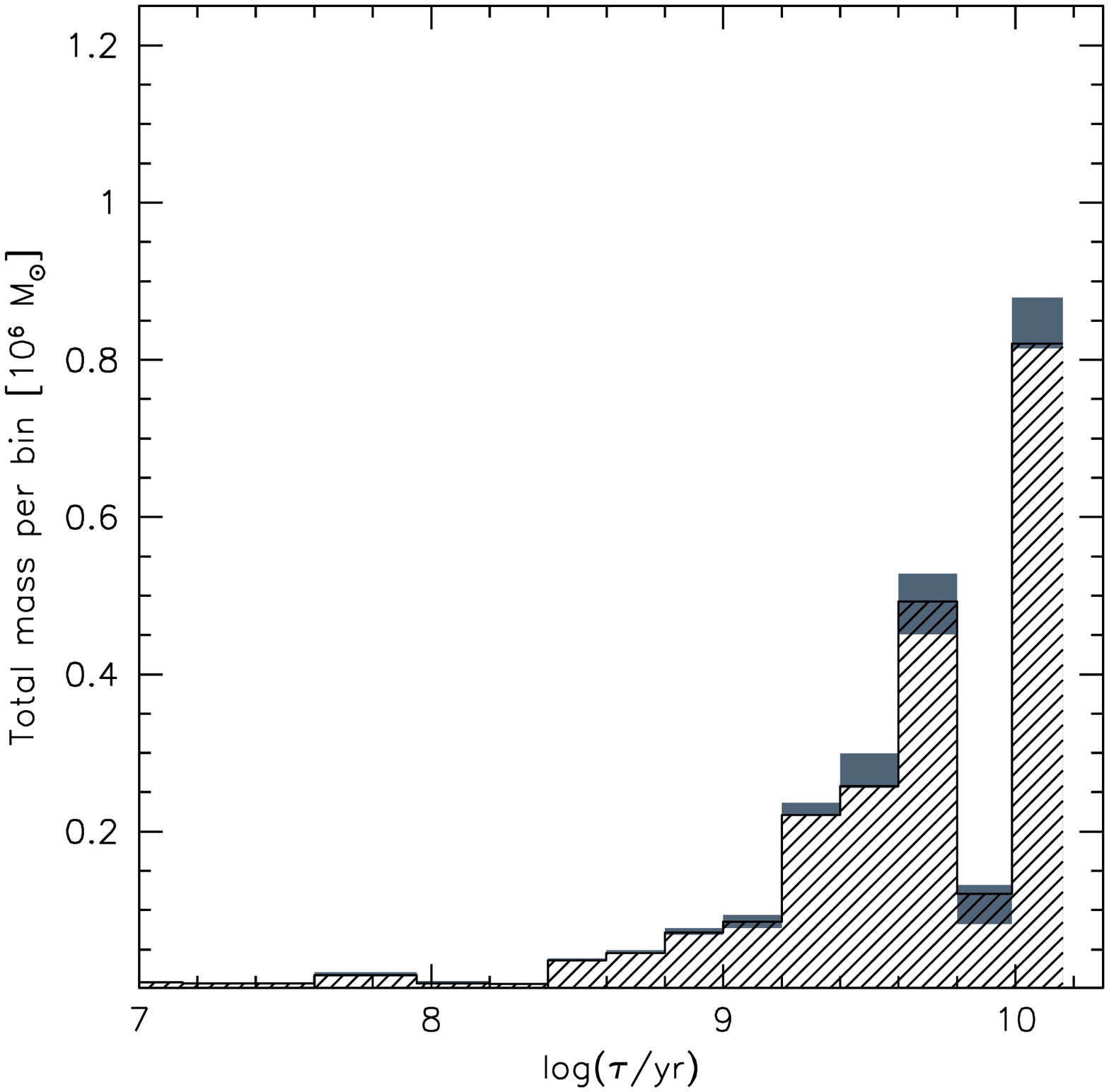}}
%\resizebox{0.45\hsize}{!}{\includegraphics{sfr66_G1_18.45_0.63.eps}}
%\resizebox{0.45\hsize}{!}{\includegraphics{sfr66_G2_18.45_0.57.eps}}
%\resizebox{0.495\hsize}{!}{\includegraphics{66_chisur_G1.eps}}
%\resizebox{0.495\hsize}{!}{\includegraphics{66_chisur_G2.eps}}
\caption{SFH results for subregions D1 (top panels) and D2 (bottom
  panels) in tile 6\_6, assuming a fixed distance. The left panels
  must be compared to the left panels in Fig.~\ref{sol66}, to
  illustrate the effect of fixing the distance.  The right panels show
  the total mass formed per age bin.}
\label{sol661}
\end{figure}

In this tile we have recovered the SFH for the two small subregions D1
and D2, located on opposite sides of the tile and away from the
30~Doradus regions.  Fig.~\ref{sol661} shows their SFHs as rederived
after assuming a known distance. Since both regions are substantially
different, their results have not been added as for the other tiles.

Subregion D1 is the most crowded and has some superposition with the
LMC bar. Its best-fitting SFH solution presents a $\chi^2$ larger than
those typically found in any other subregion analysed in this work,
probably because of the larger photometric errors and/or higher
differential extinction. Note also the larger \av\ values we find in
both D1 and D2, with respect to other tiles: $\avtw=0.65 \;
(-0.06,+0.03)$~mag for D1 and $\avtw=0.60\pm0.03$~mag for D2. For
comparison, the \av\ values obtained by \citet{Zaritsky_etal04} in D1
and D2 are $\avcs=0.50 \pm 0.38$~mag (cool stars) and $\avhs=0.68 \pm
0.42$~mag (hot stars), $\avcs=0.44 \pm 0.34$~mag (cool stars) and
$\avhs=0.55 \pm 0.39$~mag (hot stars), respectively, in good agreement
with values derived here.

Despite the larger degree of crowding in this tile, the number of
stars available in each subregion is very high and allows derivation
of a very accurate SFH, as shown by the small error bars in
Fig.~\ref{sol661}, especially at the oldest ages. If we compare the
\SFRt\ of the two subregions for ages older than $\logtyr=8.0$ the
results are similar, with the oldest SF peak at $\logtyr=10.1$, an age
gap at $\logtyr=9.9$, and a remarkably continuous \SFRt\ between
$\logtyr\simeq9.7$ and $\logtyr=8.4$. According to the right panels in
Fig.~\ref{sol661}, 64 and 53~\% of the total mass has been formed in
this latter age interval, for subregions D1 and D2 respectively.  It
is tempting to associate this prolonged period of SF with the
formation of the LMC bar.  The gap in the \SFRt\ for ages
$\logtyr=9.9$ has already been noticed before, and is extensively
commented on \citet{HZ09} as a main period of quiescent SF between
$\sim5$ and $\sim12$~Gyr ago.  Previous results regarding this feature
\citep[including][]{Olsen99, Holtzman_etal99, Smecker-Hane_etal02,
  HZ09} were based on deep HST data covering very small regions of the
LMC bar. It is remarkable, and very encouraging, that the same feature
can now be detected in ground-based data.

The \SFRt\ for ages younger than $\logtyr=8.0$ is more intense in D2,
which is the subregion closer to both the 30~Dor complex and to the
LMC centre.

%\section{Conclusion}
%\label{sec_conclusion}

\subsection{The Magellanic interaction history and SFH peaks} 
\label{ssec_kenji}

We here discuss the present observational results on the two oldest
peaks in the \SFRt\ of the LMC in the context of the past interaction
history of the LMC with the SMC and the Galaxy. These peaks are
remarkable in the \SFRt\ plots of outer disk tiles (4\_3, 8\_3 and
8\_8), and correspond to the formation of $\sim15$\,\% to 30\,\% of
the total stellar mass each.  A previous numerical model on chemical
and dynamical evolution of the LMC showed that the star formation
history of the LMC strongly depends on the LMC--SMC--Galaxy
interaction history \citep[e.g.,][]{BekkiChiba05}.  The model showed
that the LMC--SMC tidal interaction for the last few Gyr can
significantly enhance the SF in the LMC owing to the stronger tidal
interaction between the LMC and the SMC (see their Fig.~9).  The
observed second peak at $\logtyr=9.3$ ($\sim 2$ Gyr ago) can therefore
correspond to the epoch when the mutual distance between the LMC and
the SMC becomes significantly smaller so as to interact rather
violently.  The model also showed that the SFR of the LMC has peaks at
5.5 Gyr and 6.5 Gyr ago, which correspond to the epochs when the LMC
strongly interacted with the Galaxy.  Therefore, the observed first
peak in the SFR at $\logtyr=9.7$ ($\sim 5$ Gyr ago) might well
correspond to the epoch when the LMC started its strong tidal
interaction with the Galaxy.  Given that there is no SFR peak for
$\logtyr>9.7$, it would be possible that the first SFR peak can
contain fossil information as to when the LMC was accreted onto the
Galaxy and commenced its tidal interaction with the Galaxy.

Some regions in tiles 4\_3 and 8\_3 show significant \SFRt\ peaks at
$\logtyr=8.0 \sim 8.2$, which however represent the formation of just
a tiny fraction of their total stellar mass ($\sim 0.25$~\% in 4\_3
and 1.6~\% in 8\_3; see Figs.~\ref{allsfr43} and \ref{allsfr83}). It
is possible that these episodes of enhanced \SFRt\ are triggered by
the last strong LMC--SMC interaction, which occurred roughly at this
epoch according to previous numerical simulations
\citep[e.g.,][]{GardinerNoguchi96}. The present study has also
revealed that \feh\ appears to decrease after $\logtyr\sim 8.1$
\citep[see also ][]{vanLoon_etal05} in spite of enhanced SFRs in some
regions (e.g., tile $6\_6$). This result is intriguing, because
canonical closed-box chemical evolution models predict increase in
\feh\ of stars with time.  If the observed apparent decrease in \feh\
is real (e.g., $\Delta\feh\sim 0.2$ between $\logtyr=8.2$ and
$\logtyr=7.0$ in the tile $6\_6$), then this means the following two
possibilities. One is that the metal-poor gas, initially in the outer
part of the LMC, was transferred to the inner regions to be converted
into stars more metal-poor than the existing ones; the other is that
the metal-poor gas was accreted by the LMC from outside (e.g., the
Galactic Halo or the SMC) and then converted into metal-poor stars.

A number of observations showed that stellar populations in the LMC
have a shallow radial metallicity gradient. For the $\sim
0.047$~dex~kpc$^{-1}$ gradient derived from AGB stars
(\citealt{Cioni09}, but see also \citealt{Feast_etal10}), two
different regions with a difference in radial distances of $\sim
4$~kpc can have $\Delta\feh\sim 0.2$~dex.  Therefore it is not
unlikely that star formation from gas initially in the outer part of
the LMC can be responsible for the decrease in [Fe/H] in the AMR for
some regions of the LMC.  The second possibility ( metal-poor gas
being accreted from outside the LMC), which is more intriguing than
the first, is strongly supported by a previous dynamical model that
showed gas-transfer between the LMC and the SMC $\sim 0.2$~Gyr ago
\citep{BekkiChiba07}.  The model clearly showed that the outer part of
the SMC's gas disk can be stripped by the LMC--SMC tidal interaction
and finally be accreted efficiently onto the LMC.  Given that the SMC
has a metallicity that is significantly smaller than that of the LMC,
new stars formed from gas transferred from the SMC inevitably have
smaller metallicities and thus explain the observed young and
metal-poor stars in the LMC.

Recently \citet{Olsen_etal11} found that about 5\% of the LMC AGB
stars have line-of-sight velocities that appear to oppose the sense of
rotation of the LMC disk.  They have also found an association of the
kinematically distinct population with the peculiar gaseous arm in the
LMC and accordingly claimed that the stars and the peculiar arm can
originate from the SMC.  These accreted SMC populations in the LMC
clearly support the second possibility.  However, it is still unclear
how much of the SMC gas needs to be accreted onto the LMC so as to
{\it quantitatively} explain the observed metallicities of young
metal-poor stars in the LMC. It is doubtlessly worthwhile for future
theoretical studies to investigate this issue by using sophisticated
chemodynamical simulations of the LMC and the SMC for the most recent
0.2~Gyr.

So far we have considered only a scenario in which the LMC and the SMC
have had bound orbits around the Galaxy for the last $\sim6$~Gyr.  It
is however observationally unclear whether these bound orbits are
consistent with recent proper motion results of the MCs.  Recent HST
proper motion studies of the LMC \citep[e.g.,][]{KvdMA06a} have shown
that the LMC has a high velocity ($\sim380$~km\,s$^{-1}$), which
suggests that the LMC passed by the Galaxy for the first time about
0.2~Gyr ago \citep[e.g.,][]{Besla_etal07}. On the other hand, the
latest ground-based observational studies of the proper motions of the
MCs have shown that the LMC has a lower velocity
(300--340~km\,s$^{-1}$; e.g. \citealt{Costa_etal09, Vieira_etal10}).
Recent numerical simulations of the past orbit of the LMC (and the
SMC) demonstrated that the LMC is bound to the Galaxy for at least
$\sim5$~Gyr if the LMC's velocity is less than 360~km\,s$^{-1}$
\citep{Bekki11}. The observational results by \citet{KvdMA06a} and
\citet{Vieira_etal10} are around this limit and are barely consistent
with each other, implying that more observational data sets are
required to provide strong constraints on the orbital history of the
LMC.

\subsection{Concluding remarks}
This work demonstrates that SFH-recovery in VMC data works as well as
expected \citep{Kerber09}, and produces sensible results for the
values of distance and reddening in every $\sim\!0.12$~\sqdeg\
subregion analysed. Moreover, we show that we can take advantage of
the correlation between the parameters of different subregions -- in
our case, the distances -- to improve the SFH results, notably
reducing the final error bars. It is obvious that the same techniques
can be further improved by taking advantage of other correlations --
e.g., of the smoothness of the old SFH over large scales, which is
also evident in this work.

These aspects will be fully explored as more data from the survey is
considered.  We note that the present results refer to just a small
subset of the VMC: four tiles covering less than 5.6~\sqdeg, with some
subregions deliberately left out of the analysis because of either
problems with the photometry (in VIRCAM detector 16 of each tile), or
of the large and variable extinction (as for most of the 30~Dor
region).  The VMC \ks-band imaging has been 100\,\% completed in only
one of these tiles.  The final survey area as planned will include
116~\sqdeg\ covering the entire LMC, and additional 45~\sqdeg\ over
the SMC, 20~\sqdeg\ over the Bridge, and a small subset (3~\sqdeg) of
the Stream. The perspectives for deriving detailed and reliable
spatially resolved SFHs all across the Magellanic system, seem very
promising.

%%%%%%%%%%%%%%%%%%%%%%%%%%%%%%%%%%%%%%%%%%%%%%%%%%%%%%%%%%%%%%%%%%%%%%%%%

\begin{acknowledgements}
  We thank the UK team responsible for the realisation of VISTA, and
  the ESO team who have been operating and maintaining this facility.
  The UK's VISTA Data Flow System comprising the VISTA pipeline at the
  Cambridge Astronomy Survey Unit (CASU) and the VISTA Science Archive
  at Wide Field Astronomy Unit (Edinburgh) (WFAU) provided calibrated
  data products, and is supported by STFC.

  L.K. acknowledges support from the Brazilian funding
  agency CNPq.  RdG acknowledges partial research support through
  grant 11073001 from the National Natural Science Foundation of
  China. MG and MATG acknowledge support from the Belgian Federal
  Science Policy (project MO/33/026).

  This publication makes use of data products from the Two Micron All
  Sky Survey, which is a joint project of the University of
  Massachusetts and the Infrared Processing and Analysis
  Center/California Institute of Technology, funded by the National
  Aeronautics and Space Administration and the National Science
  Foundation.
\end{acknowledgements}

% for the bibliography, at the end
\bibliographystyle{aa} %style aa.bst
\bibliography{sfh} % your references file.bib

\begin{thebibliography}{66}
\expandafter\ifx\csname natexlab\endcsname\relax\def\natexlab#1{#1}\fi

\bibitem[{{Bekki}(2011)}]{Bekki11}
{Bekki}, K. 2011, \mnras, 1126

\bibitem[{{Bekki} \& {Chiba}(2005)}]{BekkiChiba05}
{Bekki}, K. \& {Chiba}, M. 2005, \mnras, 356, 680

\bibitem[{{Bekki} \& {Chiba}(2007)}]{BekkiChiba07}
{Bekki}, K. \& {Chiba}, M. 2007, \mnras, 381, L16

\bibitem[{{Bertin} {et~al.}(2002){Bertin}, {Mellier}, {Radovich}, {Missonnier},
  {Didelon}, \& {Morin}}]{Bertin02}
{Bertin}, E., {Mellier}, Y., {Radovich}, M., {et~al.} 2002, in Astronomical
  Society of the Pacific Conference Series, Vol. 281, Astronomical Data
  Analysis Software and Systems XI, ed. {D.~A.~Bohlender, D.~Durand, \&
  T.~H.~Handley}, 228--+

\bibitem[{{Besla} {et~al.}(2007){Besla}, {Kallivayalil}, {Hernquist},
  {Robertson}, {Cox}, {van der Marel}, \& {Alcock}}]{Besla_etal07}
{Besla}, G., {Kallivayalil}, N., {Hernquist}, L., {et~al.} 2007, \apj, 668, 949

\bibitem[{{Blum} {et~al.}(2006){Blum}, {Mould}, {Olsen}, {Frogel}, {Werner},
  {Meixner}, {Markwick-Kemper}, {Indebetouw}, {Whitney}, {Meade}, {Babler},
  {Churchwell}, {Gordon}, {Engelbracht}, {For}, {Misselt}, {Vijh}, {Leitherer},
  {Volk}, {Points}, {Reach}, {Hora}, {Bernard}, {Boulanger}, {Bracker},
  {Cohen}, {Fukui}, {Gallagher}, {Gorjian}, {Harris}, {Kelly}, {Kawamura},
  {Latter}, {Madden}, {Mizuno}, {Mizuno}, {Nota}, {Oey}, {Onishi}, {Paladini},
  {Panagia}, {Perez-Gonzalez}, {Shibai}, {Sato}, {Smith}, {Staveley-Smith},
  {Tielens}, {Ueta}, {Van Dyk}, \& {Zaritsky}}]{blu06}
{Blum}, R.~D., {Mould}, J.~R., {Olsen}, K.~A., {et~al.} 2006, \aj, 132, 2034

\bibitem[{{Bonatto} {et~al.}(2004){Bonatto}, {Bica}, \& {Girardi}}]{bon04}
{Bonatto}, C., {Bica}, E., \& {Girardi}, L. 2004, \aap, 415, 571

\bibitem[{{Carrera} {et~al.}(2011){Carrera}, {Gallart}, {Aparicio}, \&
  {Hardy}}]{Carrera_etal11}
{Carrera}, R., {Gallart}, C., {Aparicio}, A., \& {Hardy}, E. 2011, \aj, 142, 61

\bibitem[{{Carrera} {et~al.}(2008){Carrera}, {Gallart}, {Hardy}, {Aparicio}, \&
  {Zinn}}]{Carrera_etal08}
{Carrera}, R., {Gallart}, C., {Hardy}, E., {Aparicio}, A., \& {Zinn}, R. 2008,
  \aj, 135, 836

\bibitem[{{Chabrier}(2001)}]{Chabrier01}
{Chabrier}, G. 2001, \apj, 554, 1274

\bibitem[{{Cioni}(2009)}]{Cioni09}
{Cioni}, M.-R.~L. 2009, \aap, 506, 1137

\bibitem[{{Cioni} {et~al.}(2011){Cioni}, {Clementini}, {Girardi}, {Guandalini},
  {Gullieuszik}, {Miszalski}, {Moretti}, {Ripepi}, {Rubele}, {Bagheri},
  {Bekki}, {Cross}, {de Blok}, {de Grijs}, {Emerson}, {Evans}, {Gibson},
  {Gonzales-Solares}, {Groenewegen}, {Irwin}, {Ivanov}, {Lewis}, {Marconi},
  {Marquette}, {Mastropietro}, {Moore}, {Napiwotzki}, {Naylor}, {Oliveira},
  {Read}, {Sutorius}, {van Loon}, {Wilkinson}, \& {Wood}}]{Cioni11}
{Cioni}, M.-R.~L., {Clementini}, G., {Girardi}, L., {et~al.} 2011, \aap, 527,
  A116+

\bibitem[{{Cioni} {et~al.}(2000){Cioni}, {Habing}, \& {Israel}}]{Cioni_etal00}
{Cioni}, M.-R.~L., {Habing}, H.~J., \& {Israel}, F.~P. 2000, \aap, 358, L9

\bibitem[{{Cole} {et~al.}(2005){Cole}, {Tolstoy}, {Gallagher}, \&
  {Smecker-Hane}}]{Cole_etal05}
{Cole}, A.~A., {Tolstoy}, E., {Gallagher}, III, J.~S., \& {Smecker-Hane}, T.~A.
  2005, \aj, 129, 1465

\bibitem[{{Costa} {et~al.}(2009){Costa}, {M{\'e}ndez}, {Pedreros}, {Moyano},
  {Gallart}, {No{\"e}l}, {Baume}, \& {Carraro}}]{Costa_etal09}
{Costa}, E., {M{\'e}ndez}, R.~A., {Pedreros}, M.~H., {et~al.} 2009, \aj, 137,
  4339

\bibitem[{{Dalton} {et~al.}(2006){Dalton}, {Caldwell}, {Ward}, {Whalley},
  {Woodhouse}, {Edeson}, {Clark}, {Beard}, {Gallie}, {Todd}, {Strachan},
  {Bezawada}, {Sutherland}, \& {Emerson}}]{Dalton_etal06}
{Dalton}, G.~B., {Caldwell}, M., {Ward}, A.~K., {et~al.} 2006, in Society of
  Photo-Optical Instrumentation Engineers (SPIE) Conference Series, Vol. 6269,
  Society of Photo-Optical Instrumentation Engineers (SPIE) Conference Series

\bibitem[{{Dolphin}(2002)}]{Dolphin02}
{Dolphin}, A.~E. 2002, \mnras, 332, 91

\bibitem[{{Emerson} {et~al.}(2006){Emerson}, {McPherson}, \&
  {Sutherland}}]{Emerson_etal06}
{Emerson}, J., {McPherson}, A., \& {Sutherland}, W. 2006, The Messenger, 126,
  41

\bibitem[{{Emerson} {et~al.}(2004){Emerson}, {Irwin}, {Lewis}, {Hodgkin},
  {Evans}, {Bunclark}, {McMahon}, {Hambly}, {Mann}, {Bond}, {Sutorius}, {Read},
  {Williams}, {Lawrence}, \& {Stewart}}]{Emerson_etal04}
{Emerson}, J.~P., {Irwin}, M.~J., {Lewis}, J., {et~al.} 2004, in Society of
  Photo-Optical Instrumentation Engineers (SPIE) Conference Series, Vol. 5493,
  Society of Photo-Optical Instrumentation Engineers (SPIE) Conference Series,
  ed. {P.~J.~Quinn \& A.~Bridger}, 401--410

\bibitem[{{Feast} {et~al.}(2010){Feast}, {Abedigamba}, \&
  {Whitelock}}]{Feast_etal10}
{Feast}, M.~W., {Abedigamba}, O.~P., \& {Whitelock}, P.~A. 2010, \mnras, 408,
  L76

\bibitem[{{Freedman} {et~al.}(2001){Freedman}, {Madore}, {Gibson}, {Ferrarese},
  {Kelson}, {Sakai}, {Mould}, {Kennicutt}, {Ford}, {Graham}, {Huchra},
  {Hughes}, {Illingworth}, {Macri}, \& {Stetson}}]{Freedman01}
{Freedman}, W.~L., {Madore}, B.~F., {Gibson}, B.~K., {et~al.} 2001, \apj, 553,
  47

\bibitem[{{Gardiner} \& {Noguchi}(1996)}]{GardinerNoguchi96}
{Gardiner}, L.~T. \& {Noguchi}, M. 1996, \mnras, 278, 191

\bibitem[{{Gaustad} {et~al.}(2001){Gaustad}, {McCullough}, {Rosing}, \& {Van
  Buren}}]{shassa}
{Gaustad}, J.~E., {McCullough}, P.~R., {Rosing}, W., \& {Van Buren}, D. 2001,
  \pasp, 113, 1326

\bibitem[{{Girardi} {et~al.}(2002){Girardi}, {Bertelli}, {Bressan}, {Chiosi},
  {Groenewegen}, {Marigo}, {Salasnich}, \& {Weiss}}]{Girardi_etal02}
{Girardi}, L., {Bertelli}, G., {Bressan}, A., {et~al.} 2002, \aap, 391, 195

\bibitem[{{Girardi} {et~al.}(2008){Girardi}, {Dalcanton}, {Williams}, {de
  Jong}, {Gallart}, {Monelli}, {Groenewegen}, {Holtzman}, {Olsen}, {Seth},
  {Weisz}, \& {the ANGST/ANGRRR Collaboration}}]{Girardi_etal08}
{Girardi}, L., {Dalcanton}, J., {Williams}, B., {et~al.} 2008, \pasp, 120, 583

\bibitem[{{Girardi} {et~al.}(2005){Girardi}, {Groenewegen}, {Hatziminaoglou},
  \& {da Costa}}]{Girardi_etal05}
{Girardi}, L., {Groenewegen}, M.~A.~T., {Hatziminaoglou}, E., \& {da Costa}, L.
  2005, \aap, 436, 895

\bibitem[{{Grocholski} {et~al.}(2006){Grocholski}, {Cole}, {Sarajedini},
  {Geisler}, \& {Smith}}]{Grocholski_etal06}
{Grocholski}, A.~J., {Cole}, A.~A., {Sarajedini}, A., {Geisler}, D., \&
  {Smith}, V.~V. 2006, \aj, 132, 1630

\bibitem[{{Gullieuszik} {et~al.}(2011){Gullieuszik}, {Groenewegen}, {Cioni},
  {de Grijs}, {van Loon}, {Girardi}, {Ivanov}, {Oliveira}, {Emerson}, \&
  {Guandalini}}]{Gullieuszik_etal11}
{Gullieuszik}, M., {Groenewegen}, M.~A.~T., {Cioni}, M.~.~L., {et~al.} 2011,
  A\&A in press, arXiv 1110.4497

\bibitem[{{Hambly} {et~al.}(2004){Hambly}, {Mann}, {Bond}, {Sutorius}, {Read},
  {Williams}, {Lawrence}, \& {Emerson}}]{Hambly_etal04}
{Hambly}, N.~C., {Mann}, R.~G., {Bond}, I., {et~al.} 2004, in Society of
  Photo-Optical Instrumentation Engineers (SPIE) Conference Series, Vol. 5493,
  Society of Photo-Optical Instrumentation Engineers (SPIE) Conference Series,
  ed. {P.~J.~Quinn \& A.~Bridger}, 423--431

\bibitem[{{Harris} \& {Zaritsky}(1999)}]{HZ99}
{Harris}, J. \& {Zaritsky}, D. 1999, \aj, 117, 2831

\bibitem[{{Harris} \& {Zaritsky}(2001)}]{HZ01}
{Harris}, J. \& {Zaritsky}, D. 2001, \apjs, 136, 25

\bibitem[{{Harris} \& {Zaritsky}(2004)}]{HZ04}
{Harris}, J. \& {Zaritsky}, D. 2004, \aj, 127, 1531

\bibitem[{{Harris} \& {Zaritsky}(2009)}]{HZ09}
{Harris}, J. \& {Zaritsky}, D. 2009, \aj, 138, 1243

\bibitem[{{Haschke} {et~al.}(2011){Haschke}, {Grebel}, \& {Duffau}}]{Haschke11}
{Haschke}, R., {Grebel}, E.~K., \& {Duffau}, S. 2011, \aj, 141, 158

\bibitem[{{Hidalgo} {et~al.}(2011){Hidalgo}, {Aparicio}, {Skillman}, {Monelli},
  {Gallart}, {Cole}, {Dolphin}, {Weisz}, {Bernard}, {Cassisi}, {Mayer},
  {Stetson}, {Tolstoy}, \& {Ferguson}}]{Hidalgo11}
{Hidalgo}, S.~L., {Aparicio}, A., {Skillman}, E., {et~al.} 2011, \apj, 730, 14

\bibitem[{{Hodgkin} {et~al.}(2009){Hodgkin}, {Irwin}, {Hewett}, \&
  {Warren}}]{Hodgkin_etal10}
{Hodgkin}, S.~T., {Irwin}, M.~J., {Hewett}, P.~C., \& {Warren}, S.~J. 2009,
  \mnras, 394, 675

\bibitem[{{Holtzman} {et~al.}(1999){Holtzman}, {Gallagher}, {Cole}, {Mould},
  {Grillmair}, {Ballester}, {Burrows}, {Clarke}, {Crisp}, {Evans}, {Griffiths},
  {Hester}, {Hoessel}, {Scowen}, {Stapelfeldt}, {Trauger}, \&
  {Watson}}]{Holtzman_etal99}
{Holtzman}, J.~A., {Gallagher}, III, J.~S., {Cole}, A.~A., {et~al.} 1999, \aj,
  118, 2262

\bibitem[{{Irwin} {et~al.}(2004){Irwin}, {Lewis}, {Hodgkin}, {Bunclark},
  {Evans}, {McMahon}, {Emerson}, {Stewart}, \& {Beard}}]{Irwin_etal04}
{Irwin}, M.~J., {Lewis}, J., {Hodgkin}, S., {et~al.} 2004, in Society of
  Photo-Optical Instrumentation Engineers (SPIE) Conference Series, Vol. 5493,
  Society of Photo-Optical Instrumentation Engineers (SPIE) Conference Series,
  ed. {P.~J.~Quinn \& A.~Bridger}, 411--422

\bibitem[{{Kallivayalil} {et~al.}(2006){Kallivayalil}, {van der Marel},
  {Alcock}, {Axelrod}, {Cook}, {Drake}, \& {Geha}}]{KvdMA06a}
{Kallivayalil}, N., {van der Marel}, R.~P., {Alcock}, C., {et~al.} 2006, \apj,
  638, 772

\bibitem[{{Kerber} {et~al.}(2009){Kerber}, {Girardi}, {Rubele}, \&
  {Cioni}}]{Kerber09}
{Kerber}, L.~O., {Girardi}, L., {Rubele}, S., \& {Cioni}, M. 2009, \aap, 499,
  697

\bibitem[{{Kerber} {et~al.}(2007){Kerber}, {Santiago}, \&
  {Brocato}}]{Kerber_etal07}
{Kerber}, L.~O., {Santiago}, B.~X., \& {Brocato}, E. 2007, \aap, 462, 139

\bibitem[{{Mackey} \& {Gilmore}(2003)}]{MG03}
{Mackey}, A.~D. \& {Gilmore}, G.~F. 2003, \mnras, 338, 85

\bibitem[{{Ma{\'{\i}}z-Apell{\'a}niz}(2007)}]{MaizApellaniz07}
{Ma{\'{\i}}z-Apell{\'a}niz}, J. 2007, in Astronomical Society of the Pacific
  Conference Series, Vol. 364, The Future of Photometric, Spectrophotometric
  and Polarimetric Standardization, ed. {C.~Sterken}, 227--+

\bibitem[{{Marigo} {et~al.}(2008){Marigo}, {Girardi}, {Bressan}, {Groenewegen},
  {Silva}, \& {Granato}}]{Marigo_etal08}
{Marigo}, P., {Girardi}, L., {Bressan}, A., {et~al.} 2008, \aap, 482, 883

\bibitem[{{Miszalski} {et~al.}(2011{\natexlab{a}}){Miszalski}, {Napiwotzki},
  {Cioni}, {Groenewegen}, {Oliveira}, \& {Udalski}}]{Miszalski_etal11}
{Miszalski}, B., {Napiwotzki}, R., {Cioni}, M.-R.~L., {et~al.}
  2011{\natexlab{a}}, \aap, 531, A157+

\bibitem[{{Miszalski} {et~al.}(2011{\natexlab{b}}){Miszalski}, {Napiwotzki},
  {Cioni}, \& {Nie}}]{Miszalski_etal11a}
{Miszalski}, B., {Napiwotzki}, R., {Cioni}, M.-R.~L., \& {Nie}, J.
  2011{\natexlab{b}}, \aap, 529, A77+

\bibitem[{{Nikolaev} {et~al.}(2004){Nikolaev}, {Drake}, {Keller}, {Cook},
  {Dalal}, {Griest}, {Welch}, \& {Kanbur}}]{Nikolaev_etal04}
{Nikolaev}, S., {Drake}, A.~J., {Keller}, S.~C., {et~al.} 2004, \apj, 601, 260

\bibitem[{{Nikolaev} \& {Weinberg}(2000)}]{NW00}
{Nikolaev}, S. \& {Weinberg}, M.~D. 2000, \apj, 542, 804

\bibitem[{{No{\"e}l} {et~al.}(2009){No{\"e}l}, {Aparicio}, {Gallart},
  {Hidalgo}, {Costa}, \& {M{\'e}ndez}}]{Noel09}
{No{\"e}l}, N.~E.~D., {Aparicio}, A., {Gallart}, C., {et~al.} 2009, \apj, 705,
  1260

\bibitem[{{Olsen}(1999)}]{Olsen99}
{Olsen}, K.~A.~G. 1999, \aj, 117, 2244

\bibitem[{{Olsen} \& {Salyk}(2002)}]{OlsenSalyk02}
{Olsen}, K.~A.~G. \& {Salyk}, C. 2002, \aj, 124, 2045

\bibitem[{{Olsen} {et~al.}(2011){Olsen}, {Zaritsky}, {Blum}, {Boyer}, \&
  {Gordon}}]{Olsen_etal11}
{Olsen}, K.~A.~G., {Zaritsky}, D., {Blum}, R.~D., {Boyer}, M.~L., \& {Gordon},
  K.~D. 2011, \apj, 737, 29

\bibitem[{{Olszewski} {et~al.}(1991){Olszewski}, {Schommer}, {Suntzeff}, \&
  {Harris}}]{Olszewski_etal91}
{Olszewski}, E.~W., {Schommer}, R.~A., {Suntzeff}, N.~B., \& {Harris}, H.~C.
  1991, \aj, 101, 515

\bibitem[{{Orban} {et~al.}(2008){Orban}, {Gnedin}, {Weisz}, {Skillman},
  {Dolphin}, \& {Holtzman}}]{Orban08}
{Orban}, C., {Gnedin}, O.~Y., {Weisz}, D.~R., {et~al.} 2008, \apj, 686, 1030

\bibitem[{{Rubele} {et~al.}(2011){Rubele}, {Girardi}, {Kozhurina-Platais},
  {Goudfrooij}, \& {Kerber}}]{Rubele11}
{Rubele}, S., {Girardi}, L., {Kozhurina-Platais}, V., {Goudfrooij}, P., \&
  {Kerber}, L. 2011, \mnras, 550

\bibitem[{{Schaefer}(2008)}]{schaefer08}
{Schaefer}, B.~E. 2008, \aj, 135, 112

\bibitem[{{Smecker-Hane} {et~al.}(2002){Smecker-Hane}, {Cole}, {Gallagher}, \&
  {Stetson}}]{Smecker-Hane_etal02}
{Smecker-Hane}, T.~A., {Cole}, A.~A., {Gallagher}, III, J.~S., \& {Stetson},
  P.~B. 2002, \apj, 566, 239

\bibitem[{{Stetson}(1987)}]{daophot}
{Stetson}, P.~B. 1987, \pasp, 99, 191

\bibitem[{{Subramanian} \& {Subramaniam}(2010)}]{sub10}
{Subramanian}, S. \& {Subramaniam}, A. 2010, \aap, 520, A24+

\bibitem[{{van der Marel} {et~al.}(2002){van der Marel}, {Alves}, {Hardy}, \&
  {Suntzeff}}]{vanderMarel_etal02}
{van der Marel}, R.~P., {Alves}, D.~R., {Hardy}, E., \& {Suntzeff}, N.~B. 2002,
  \aj, 124, 2639

\bibitem[{{van der Marel} \& {Cioni}(2001)}]{vdMC01}
{van der Marel}, R.~P. \& {Cioni}, M.-R.~L. 2001, \aj, 122, 1807

\bibitem[{{van Loon} {et~al.}(2005){van Loon}, {Marshall}, \&
  {Zijlstra}}]{vanLoon_etal05}
{van Loon}, J.~T., {Marshall}, J.~R., \& {Zijlstra}, A.~A. 2005, \aap, 442, 597

\bibitem[{{Vieira} {et~al.}(2010){Vieira}, {Girard}, {van Altena}, {Zacharias},
  {Casetti-Dinescu}, {Korchagin}, {Platais}, {Monet}, {L{\'o}pez}, {Herrera},
  \& {Castillo}}]{Vieira_etal10}
{Vieira}, K., {Girard}, T.~M., {van Altena}, W.~F., {et~al.} 2010, \aj, 140,
  1934

\bibitem[{{Weisz} {et~al.}(2011){Weisz}, {Dolphin}, {Dalcanton}, {Skillman},
  {Holtzman}, {Williams}, {Gilbert}, {Seth}, {Cole}, {Gogarten}, {Rosema},
  {Karachentsev}, {McQuinn}, \& {Zaritsky}}]{Weisz11}
{Weisz}, D.~R., {Dolphin}, A.~E., {Dalcanton}, J.~J., {et~al.} 2011, ArXiv
  e-prints

\bibitem[{{Williams} {et~al.}(2009){Williams}, {Dalcanton}, {Dolphin},
  {Holtzman}, \& {Sarajedini}}]{Williams09}
{Williams}, B.~F., {Dalcanton}, J.~J., {Dolphin}, A.~E., {Holtzman}, J., \&
  {Sarajedini}, A. 2009, \apjl, 695, L15

\bibitem[{{Zaritsky} {et~al.}(2004){Zaritsky}, {Harris}, {Thompson}, \&
  {Grebel}}]{Zaritsky_etal04}
{Zaritsky}, D., {Harris}, J., {Thompson}, I.~B., \& {Grebel}, E.~K. 2004, \aj,
  128, 1606

\end{thebibliography}
%\
\appendix

\section{SFH results for each subregion}
\label{app}

For the sake of completeness, this appendix presents the SFH results
for all subregions of all tiles. All SFH data, including tables and
figures, are available on the VMC main site,
http://star.herts.ac.uk/~mcioni/vmc/, and are regularly updated as the
survey and analysis proceed.

\centering
\begin{figure*}[p]
\begin{minipage}{\textwidth}
\subfigure[\SFRt\ and AMR of subregions G4, G8, and G12]{
\resizebox{0.30\hsize}{!}{\includegraphics{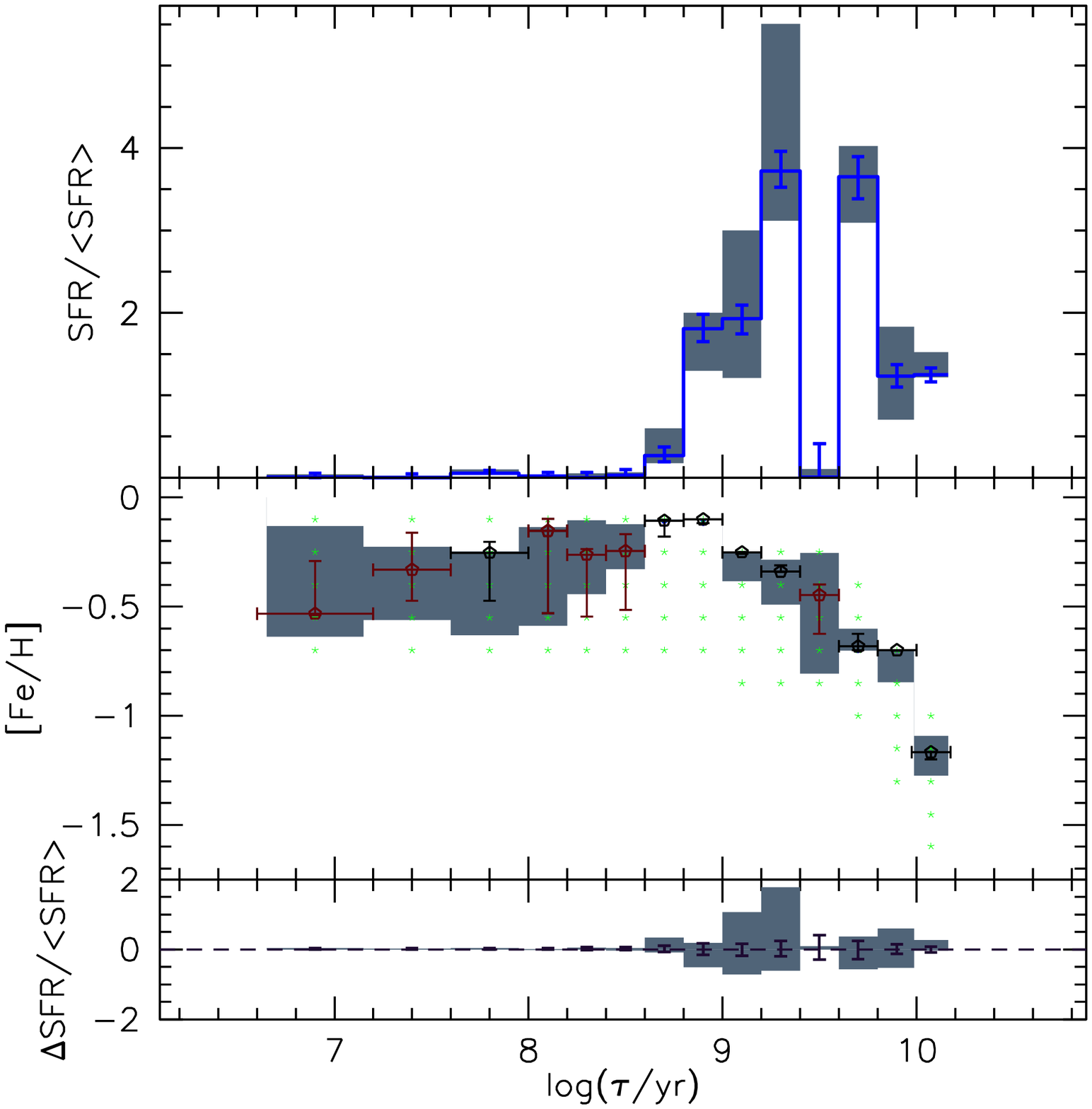}}
\resizebox{0.30\hsize}{!}{\includegraphics{sfr88_G8_18.425_0.15.eps}}
\resizebox{0.30\hsize}{!}{\includegraphics{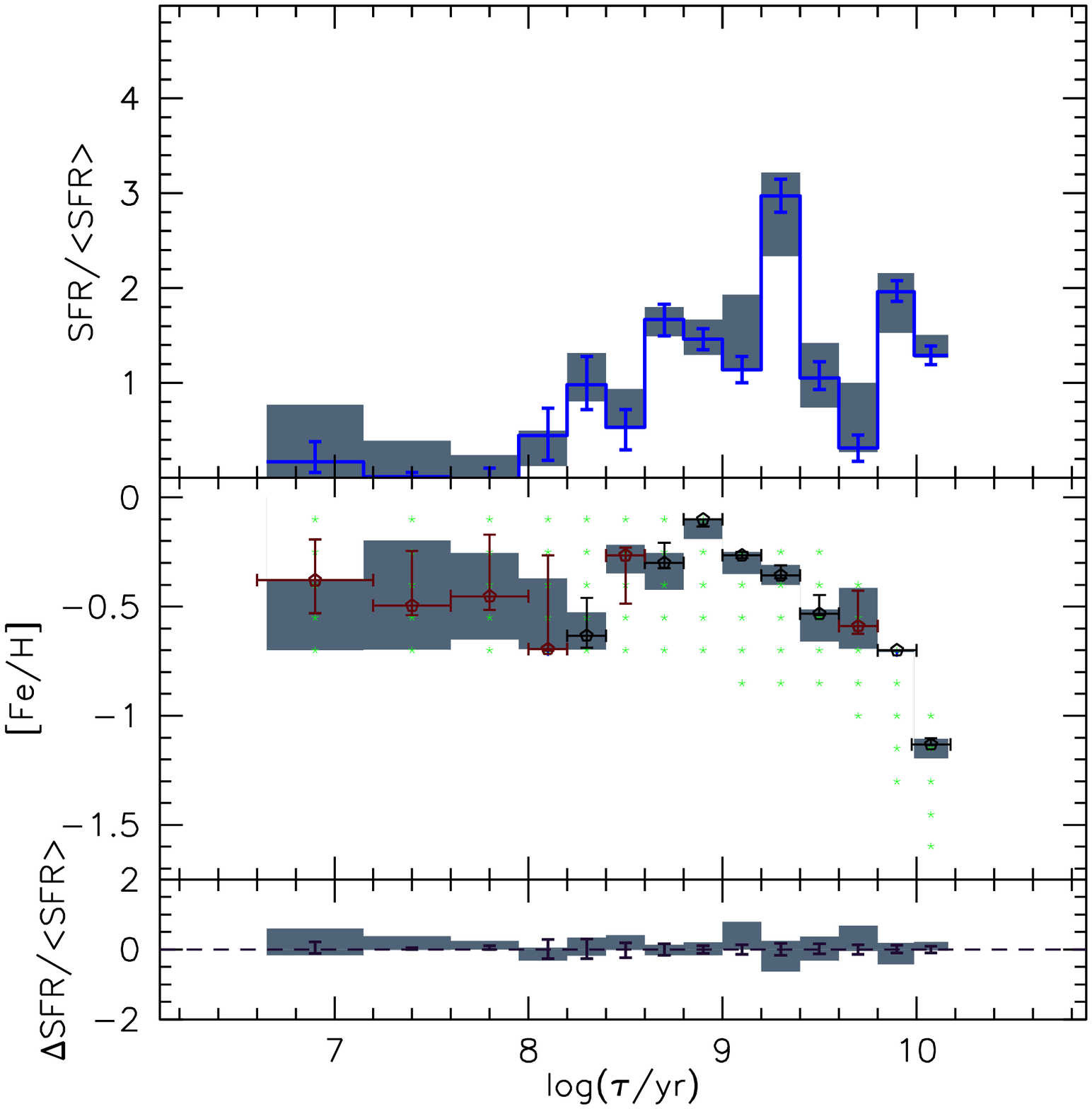}}
}
\end{minipage}
\hfill
\begin{minipage}{\textwidth}
\subfigure[\SFRt\ and AMR of subregions G3, G7 and G11]{
\resizebox{0.30\hsize}{!}{\includegraphics{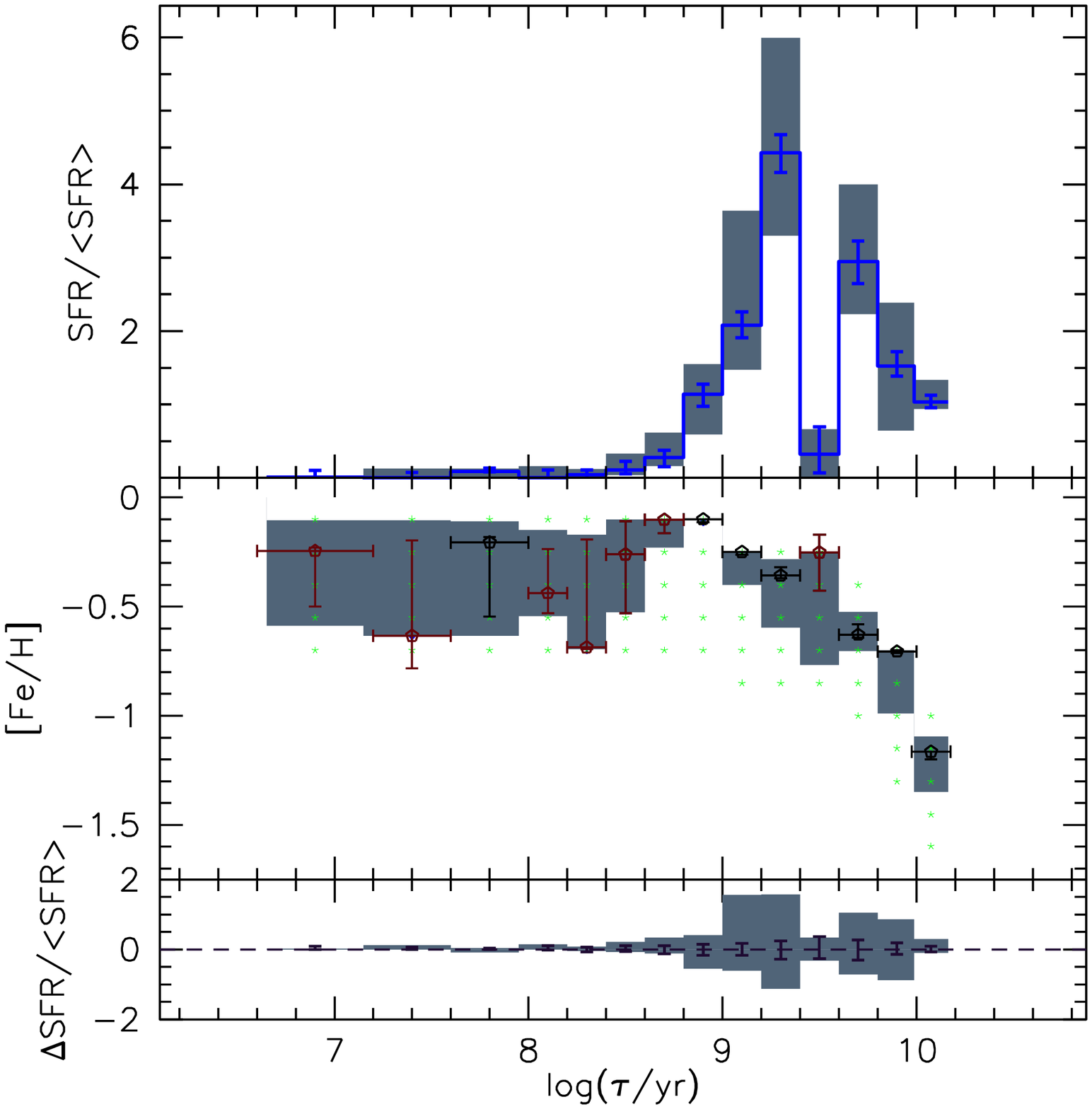}}
\resizebox{0.30\hsize}{!}{\includegraphics{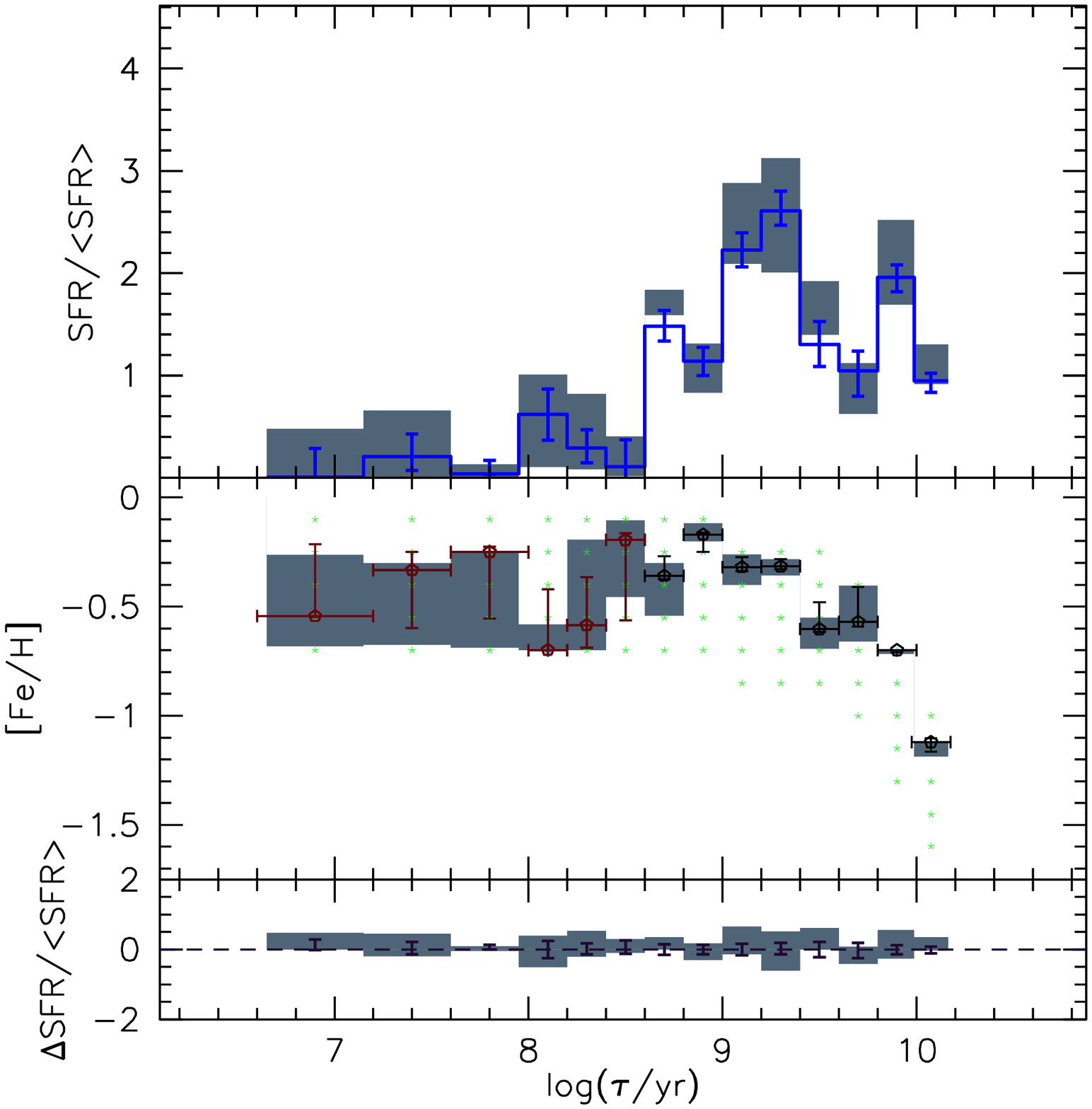}}
\resizebox{0.30\hsize}{!}{\includegraphics{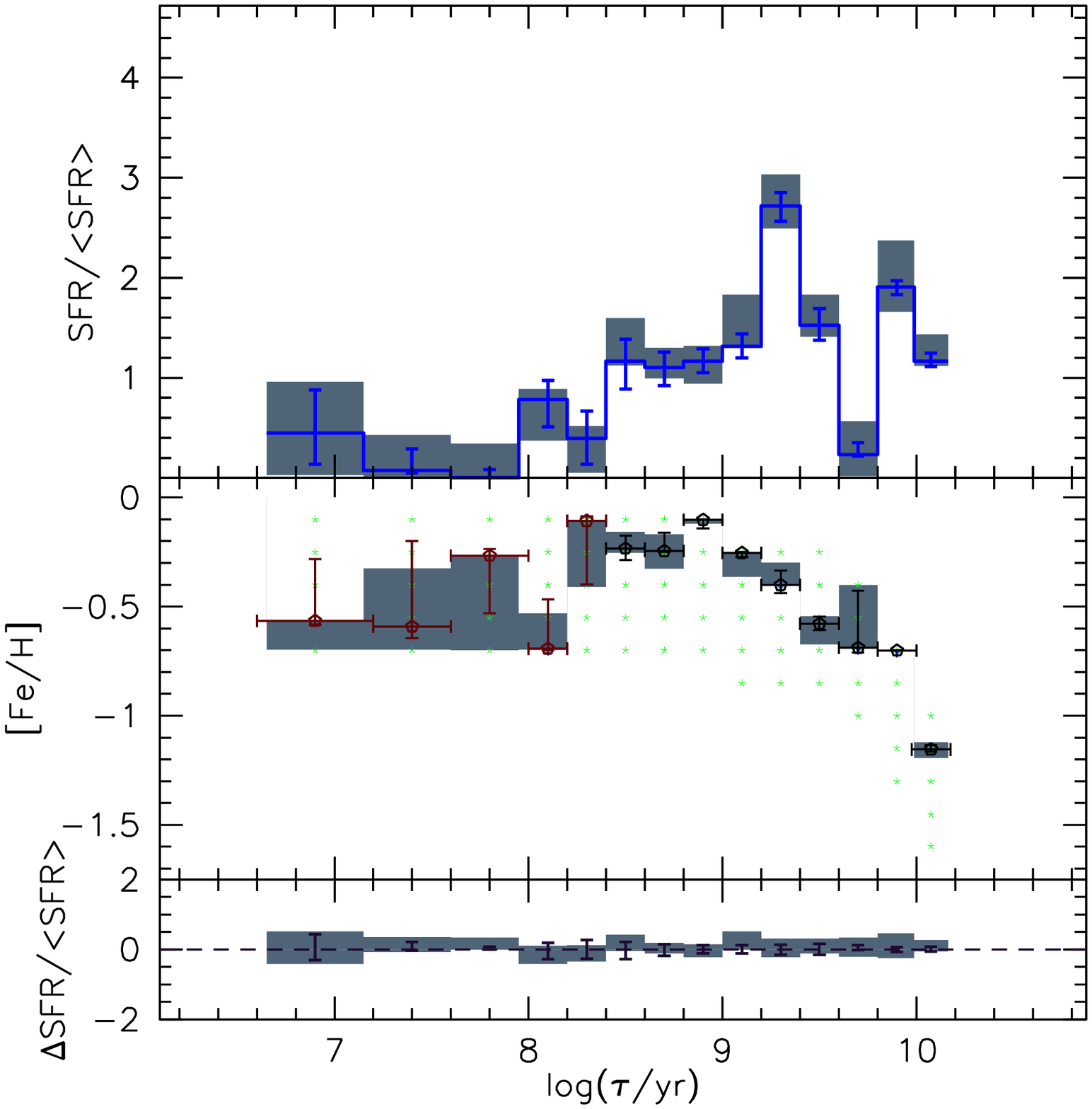}}
}
\end{minipage}
\hfill
\begin{minipage}{\textwidth}
\subfigure[\SFRt\ and AMR of subregions G2, G6 and G10]{
\resizebox{0.30\hsize}{!}{\includegraphics{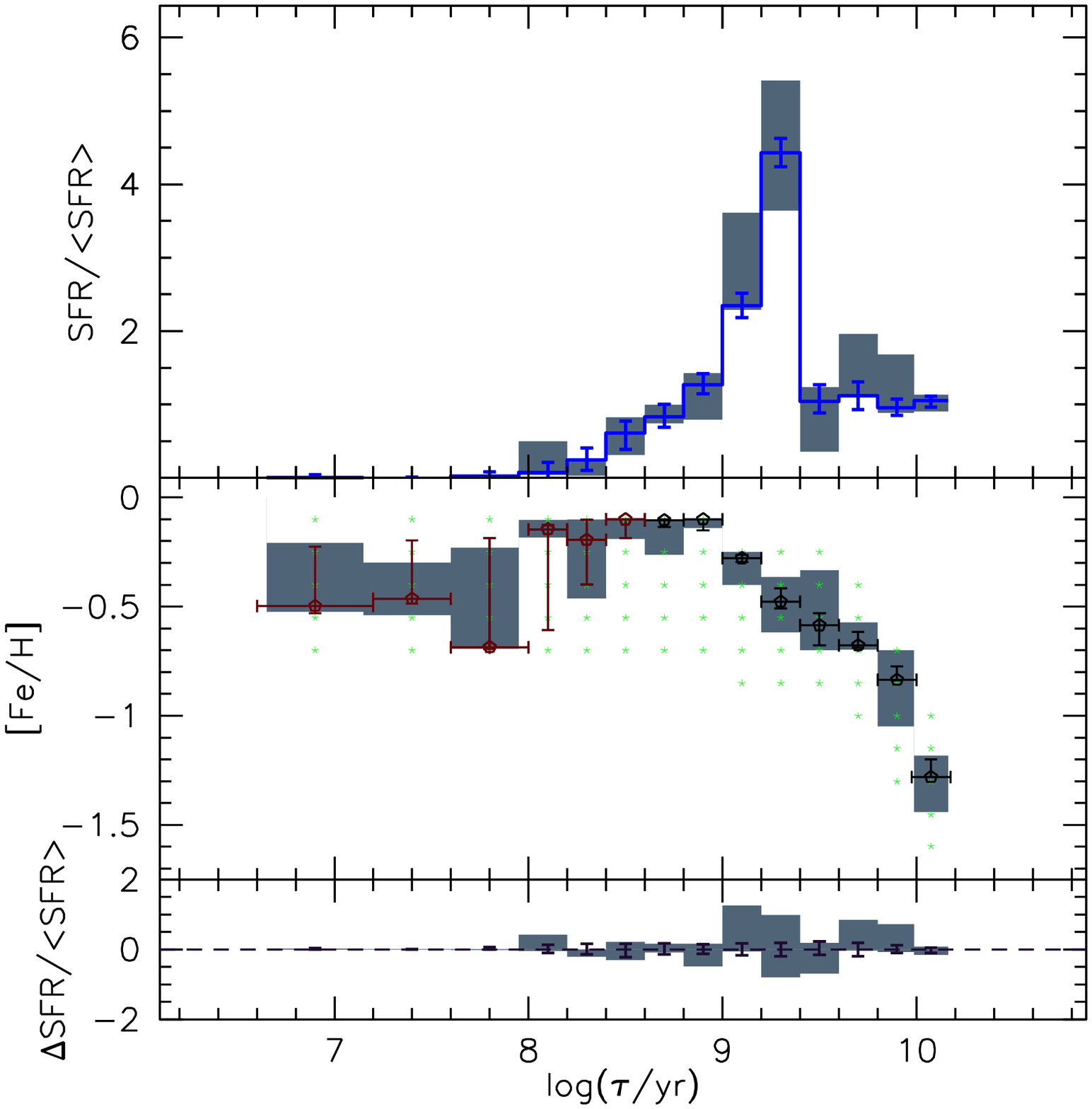}}
\resizebox{0.30\hsize}{!}{\includegraphics{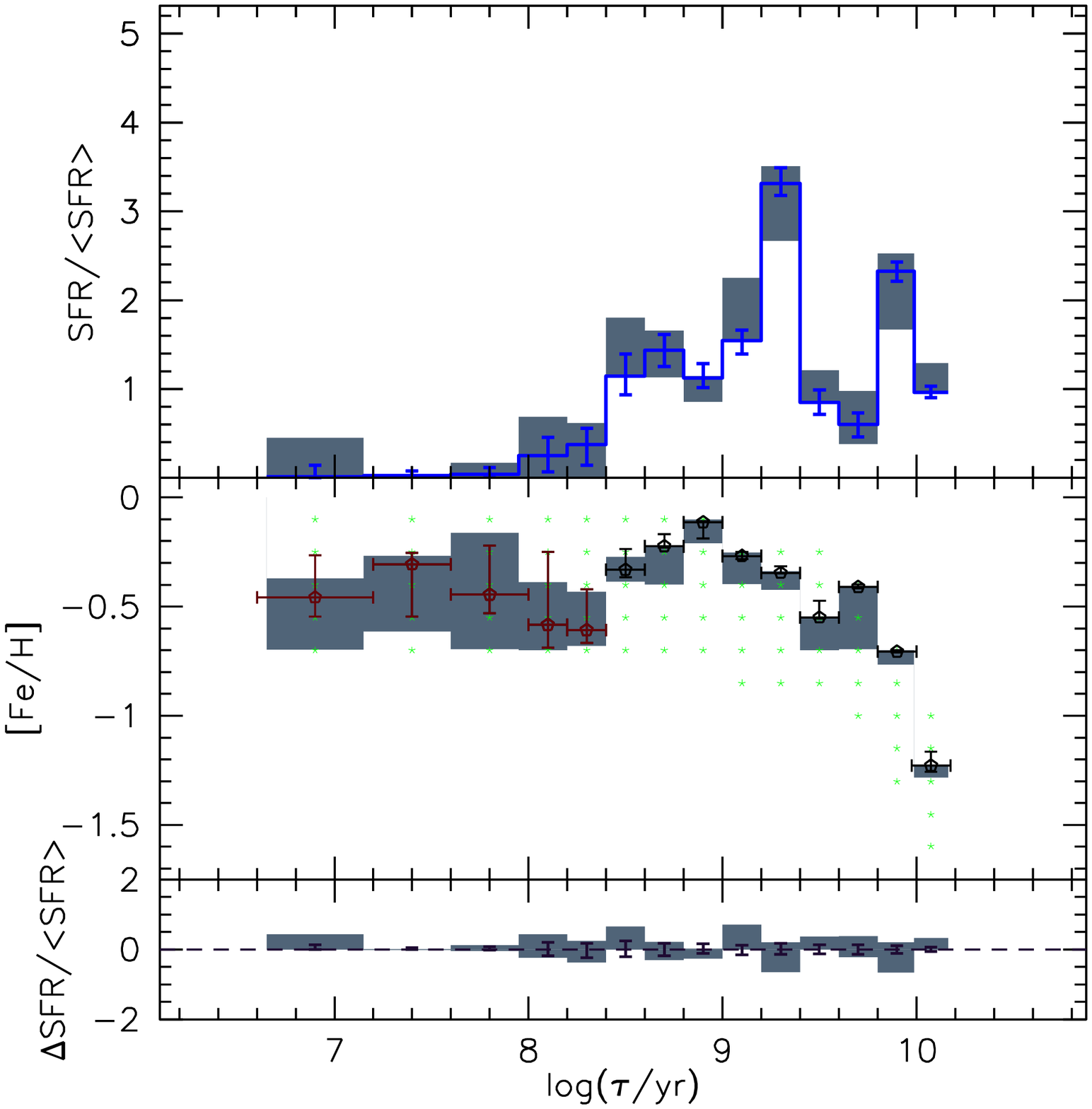}}
\resizebox{0.30\hsize}{!}{\includegraphics{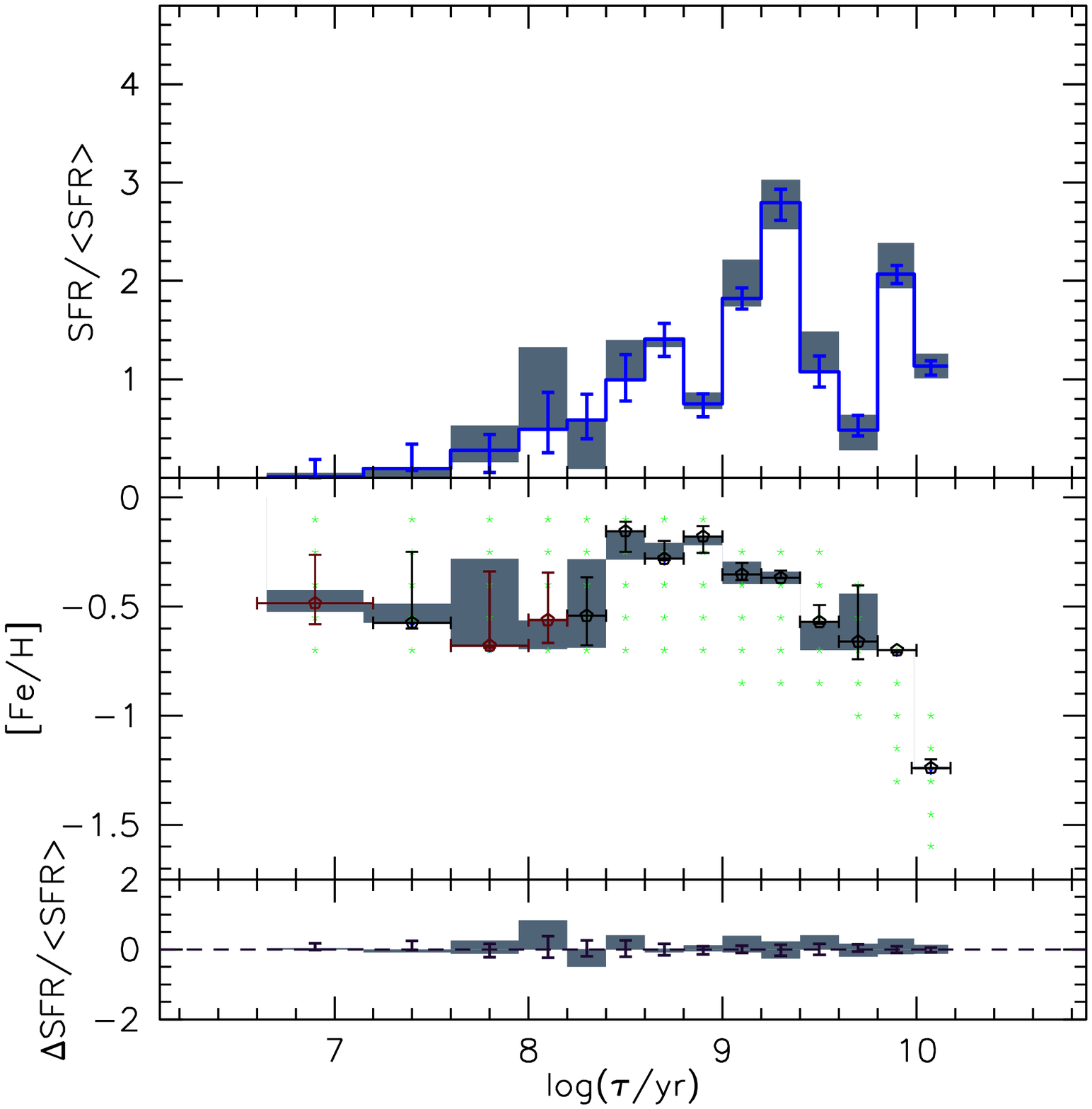}}
}
\end{minipage}
\hfill
\begin{minipage}{\textwidth}
\subfigure[\SFRt\ and AMR of subregions G1 and G5]{
\resizebox{0.30\hsize}{!}{\includegraphics{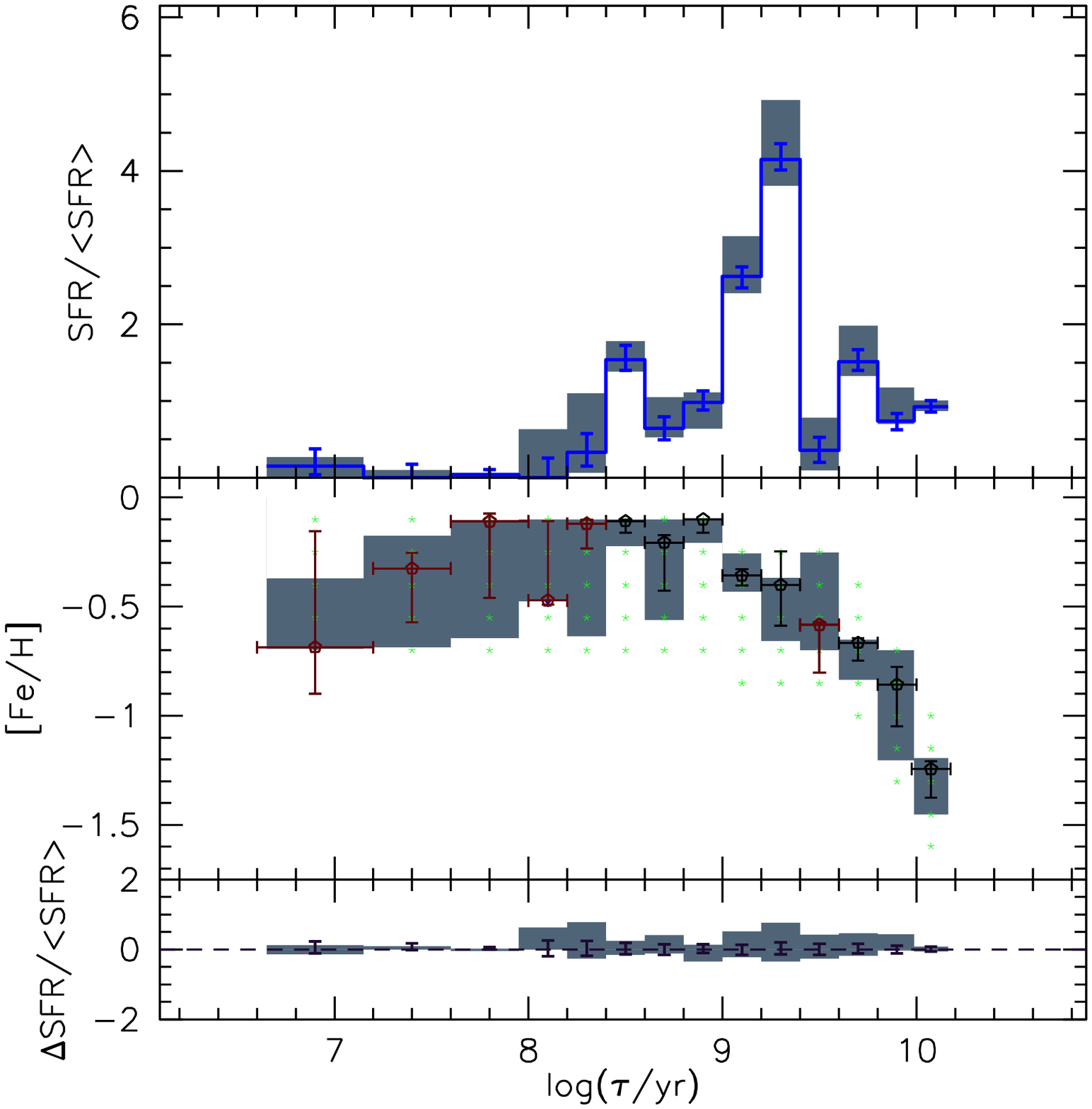}}
\resizebox{0.30\hsize}{!}{\includegraphics{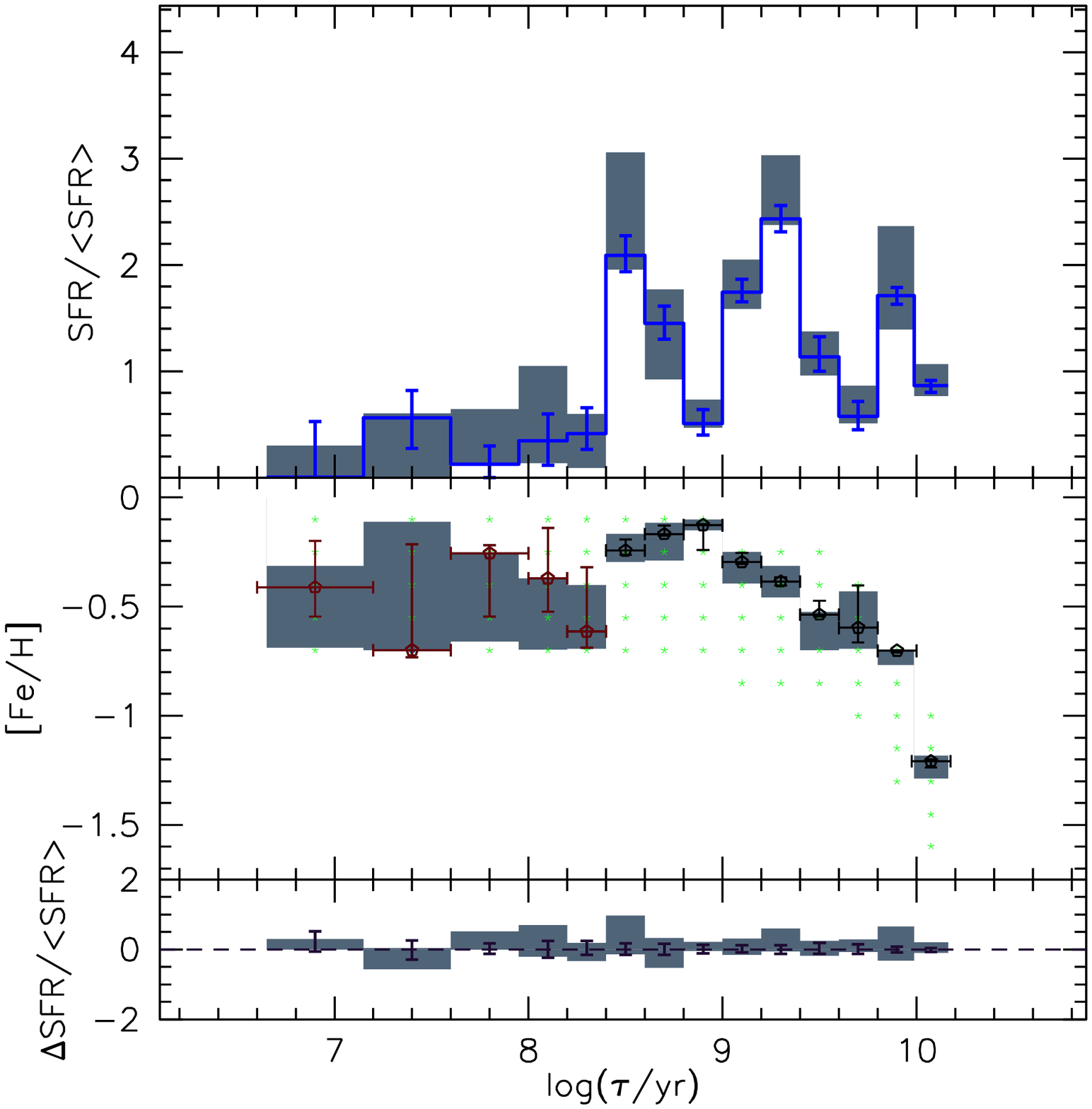}}
}
\end{minipage}
\caption{SFH for the subregions of the 8\_8 tile. The panel are
  disposed in the same way as seen in the sky, with North to the top
  and West to the right (see Fig.~\ref{field}). The bottom right
  panels represent subregions closer to the LMC centre.}
\label{sfr88}
\end{figure*}

\begin{figure*}[p]
\begin{minipage}{\textwidth}
\subfigure[\SFRt\ and AMR of subregions G4, G8 and G12]{
%\subfigure[SFR of sub region G1, G2, and G3]{
\resizebox{0.30\hsize}{!}{\includegraphics{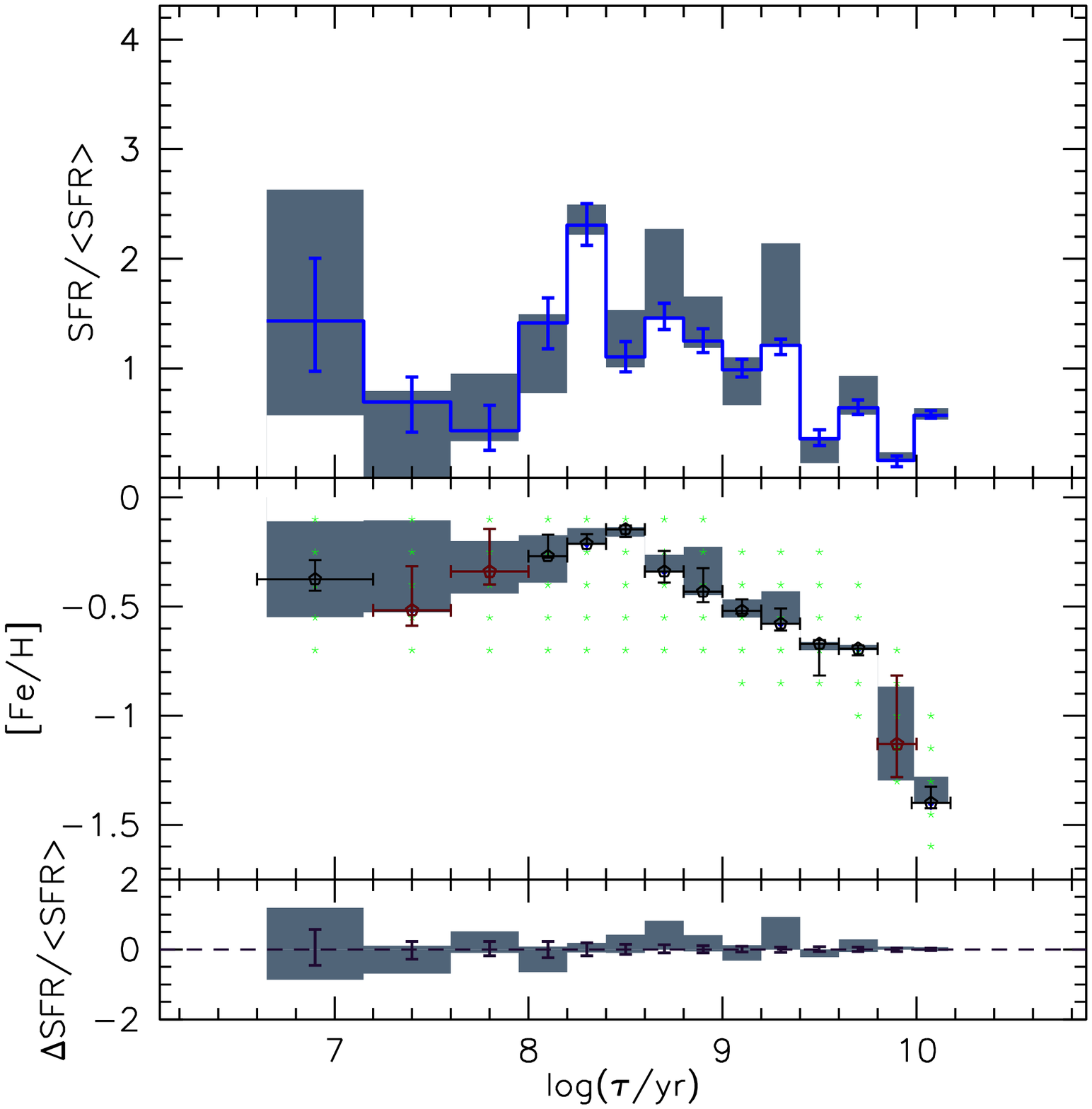}}
\resizebox{0.30\hsize}{!}{\includegraphics{sfr83_G8_18.45_0.39.eps}}
\resizebox{0.30\hsize}{!}{\includegraphics{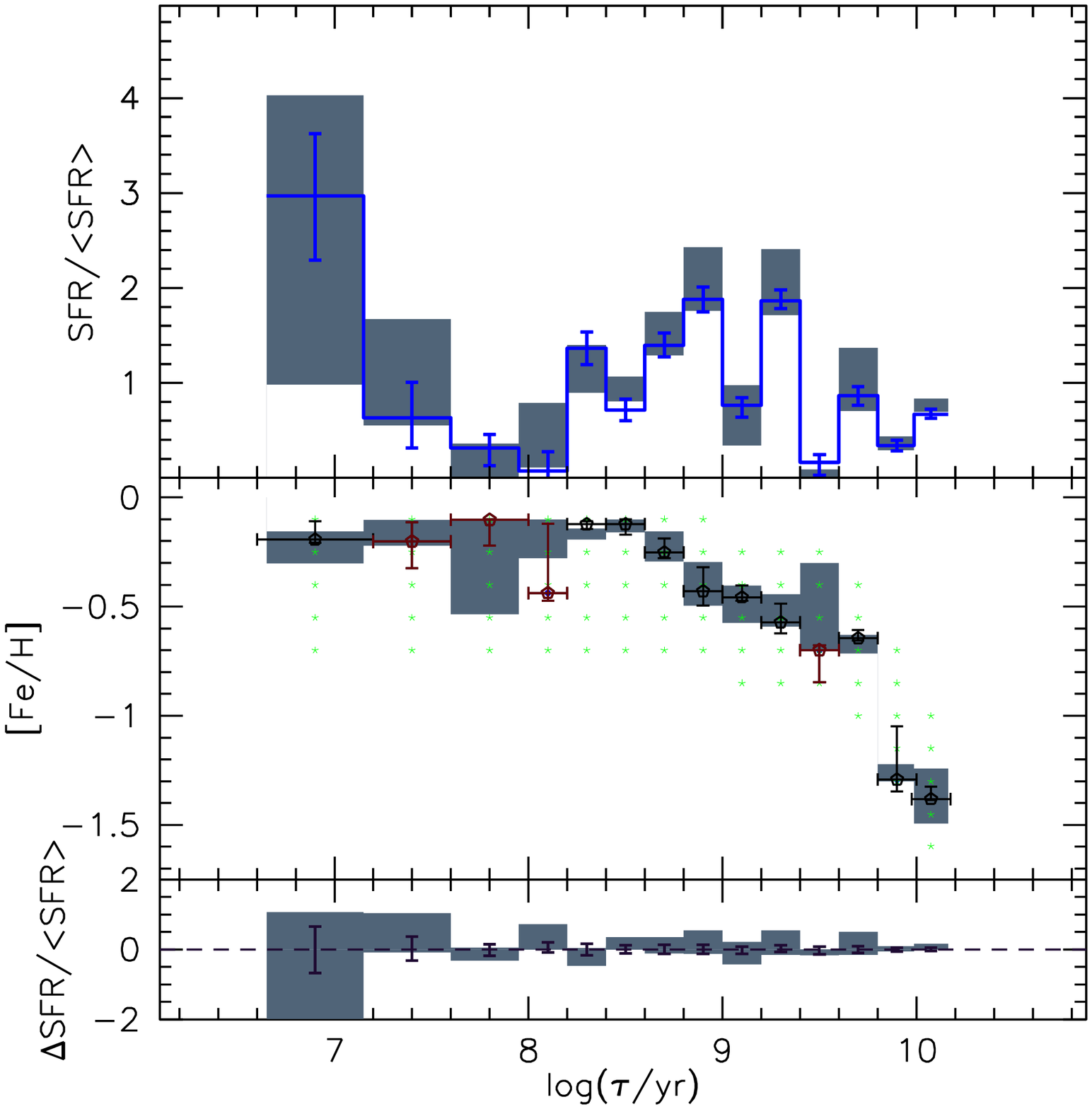}}
}
\end{minipage}
\hfill
\begin{minipage}{\textwidth}
\subfigure[\SFRt\ and AMR of subregions G3, G7 and G11]{
%\subfigure[SFR of sub region G4, G5 and G6]{
\resizebox{0.30\hsize}{!}{\includegraphics{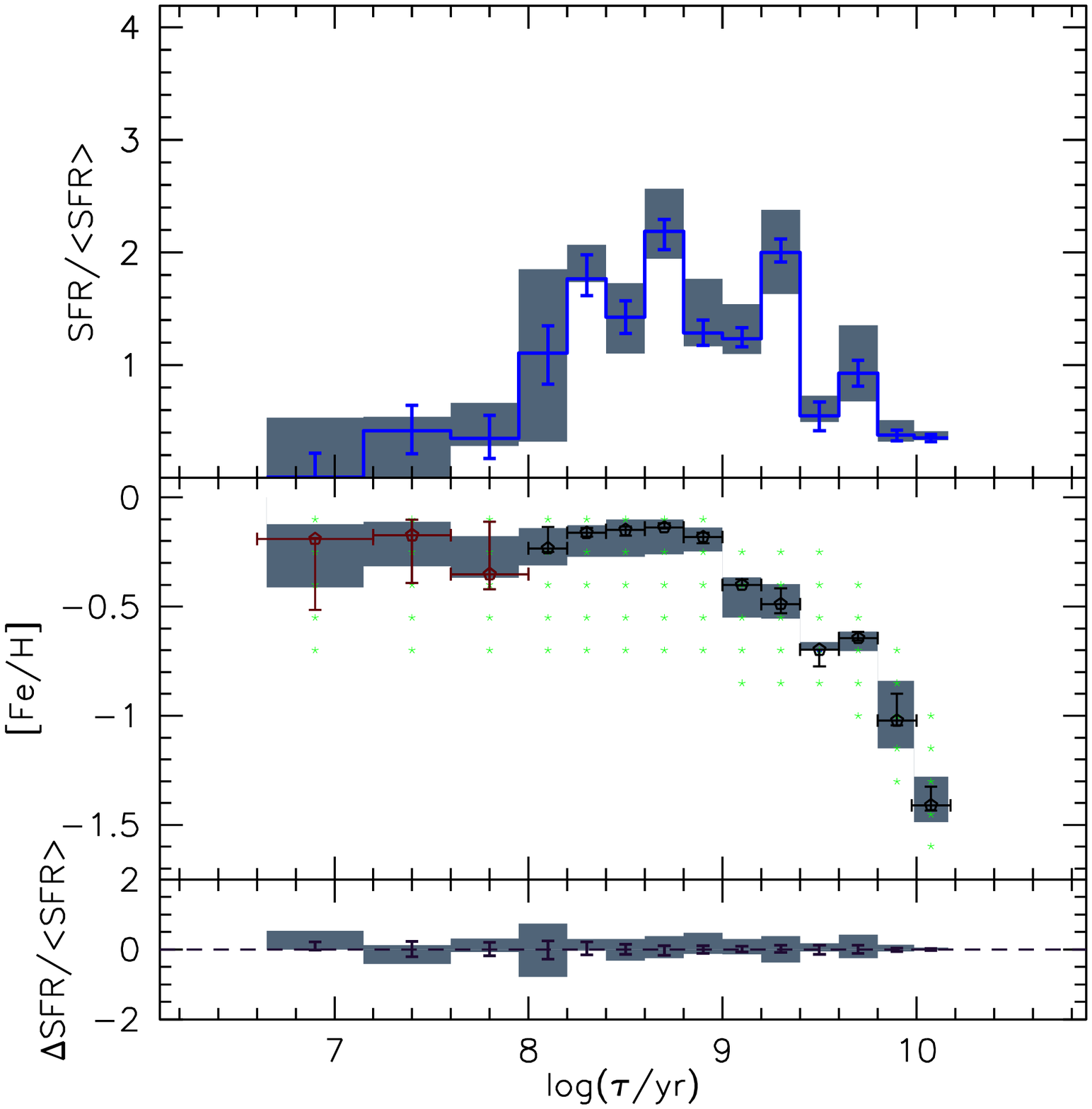}}
\resizebox{0.30\hsize}{!}{\includegraphics{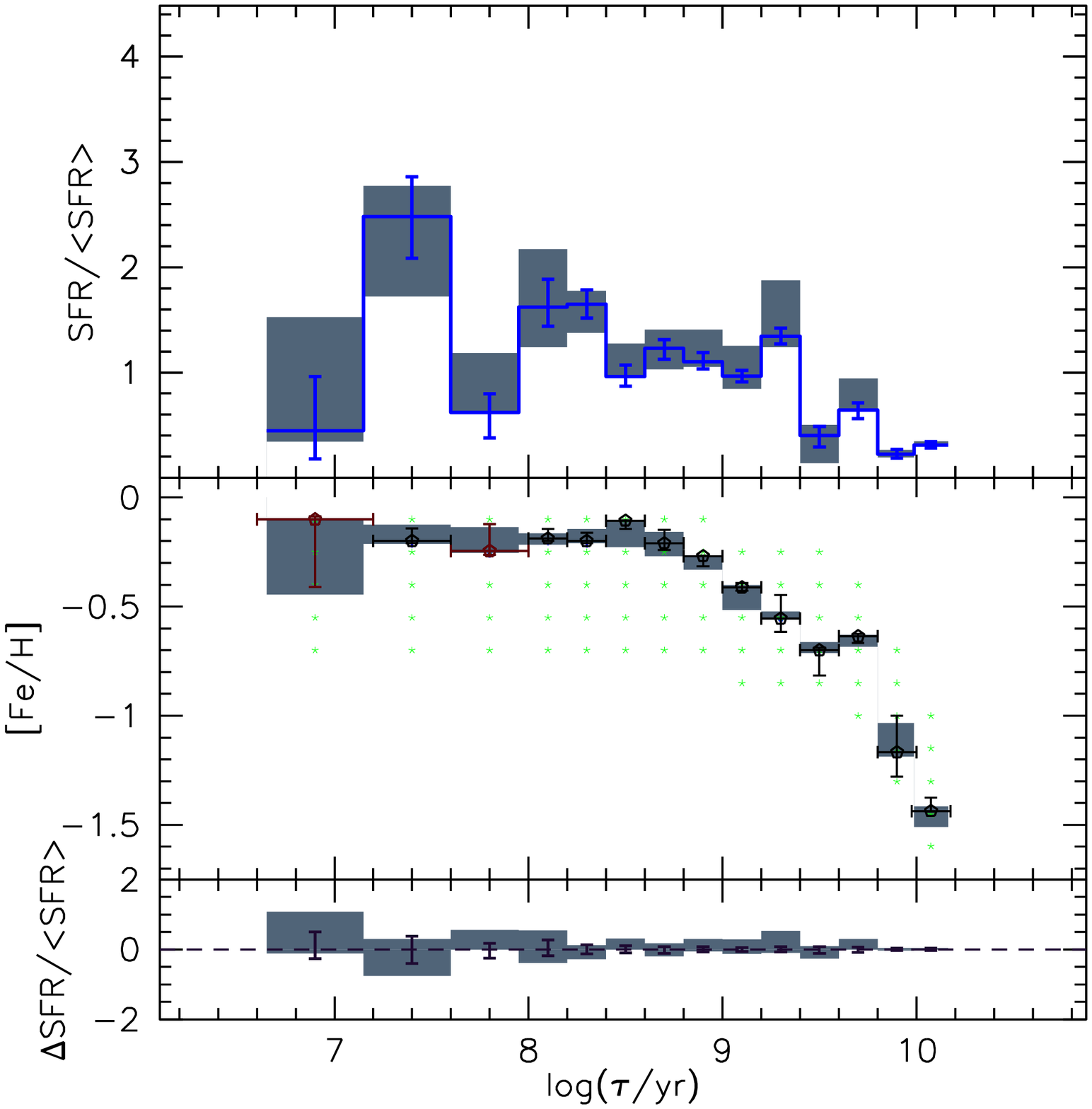}}
\resizebox{0.30\hsize}{!}{\includegraphics{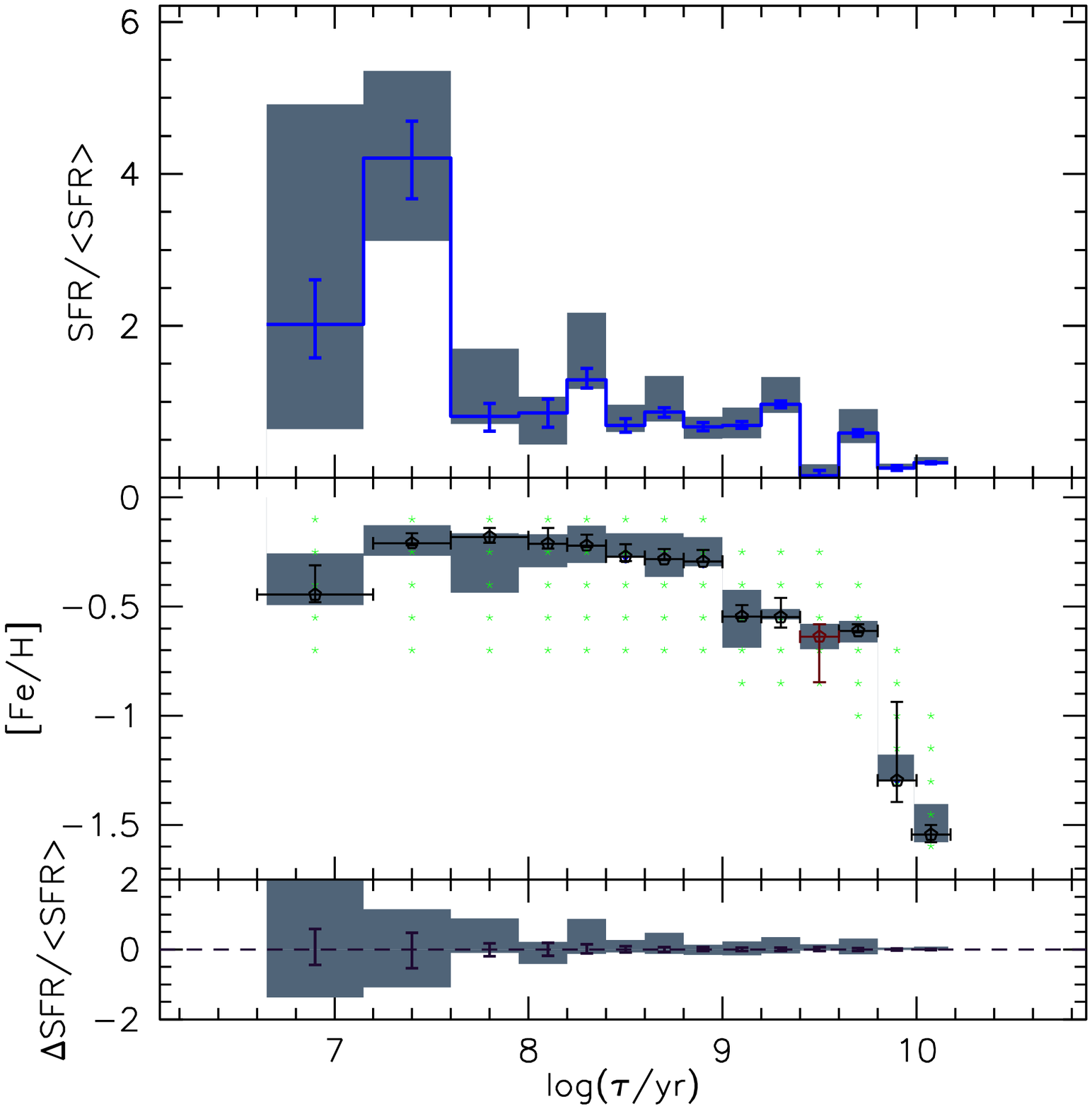}}
}
\end{minipage}
\hfill
\begin{minipage}{\textwidth}
\subfigure[\SFRt\ and AMR of subregions G2, G6 and G10]{
%\subfigure[SFR of sub region G7, G8 and G9]{
\resizebox{0.30\hsize}{!}{\includegraphics{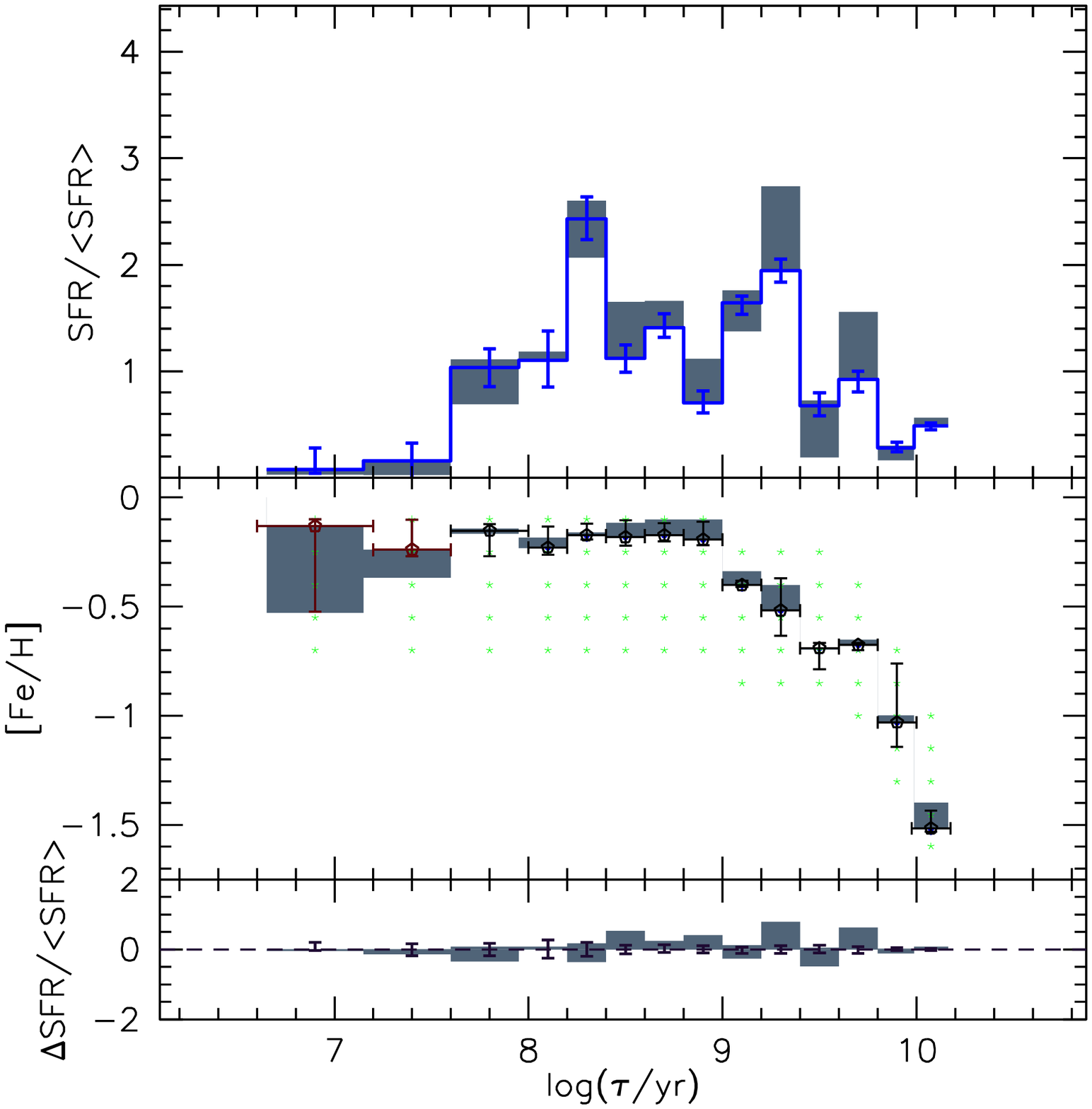}}
\resizebox{0.30\hsize}{!}{\includegraphics{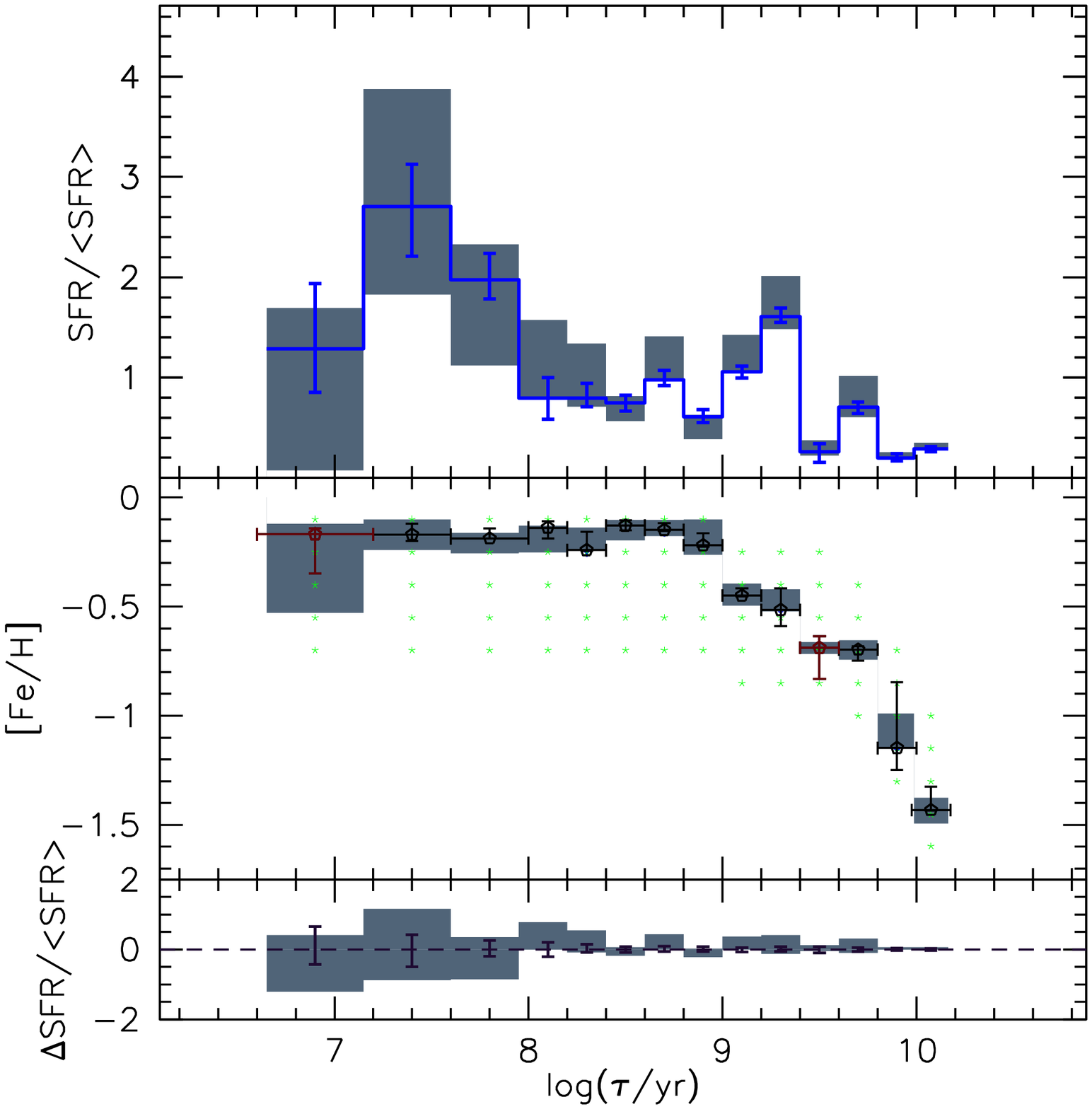}}
\resizebox{0.30\hsize}{!}{\includegraphics{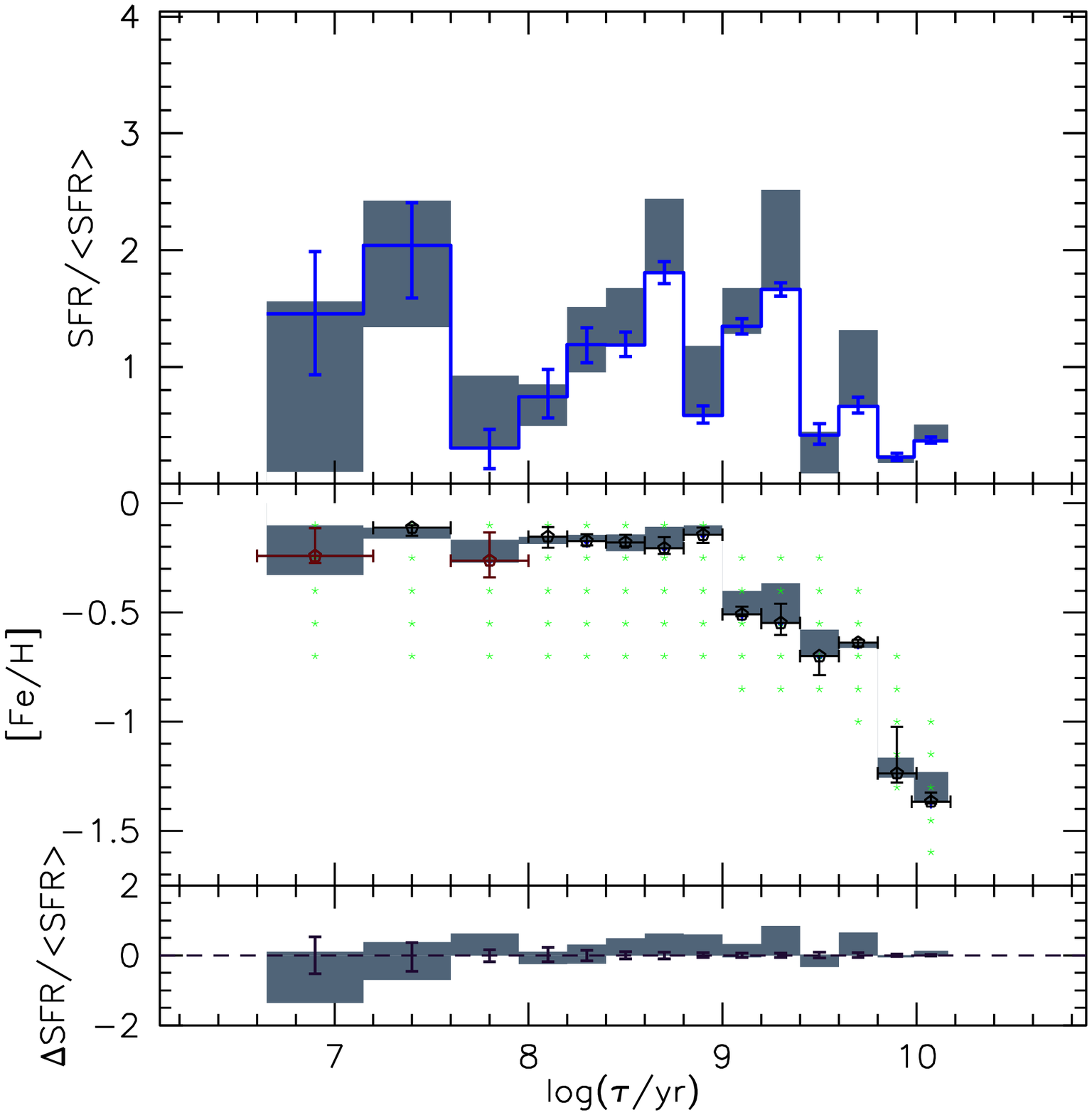}}
}
\end{minipage}
\hfill
\begin{minipage}{\textwidth}
\subfigure[\SFRt\ and AMR of subregions G1, G5 and G9]{
%\subfigure[SFR of sub region G10, G11 and G12]{
\resizebox{0.30\hsize}{!}{\includegraphics{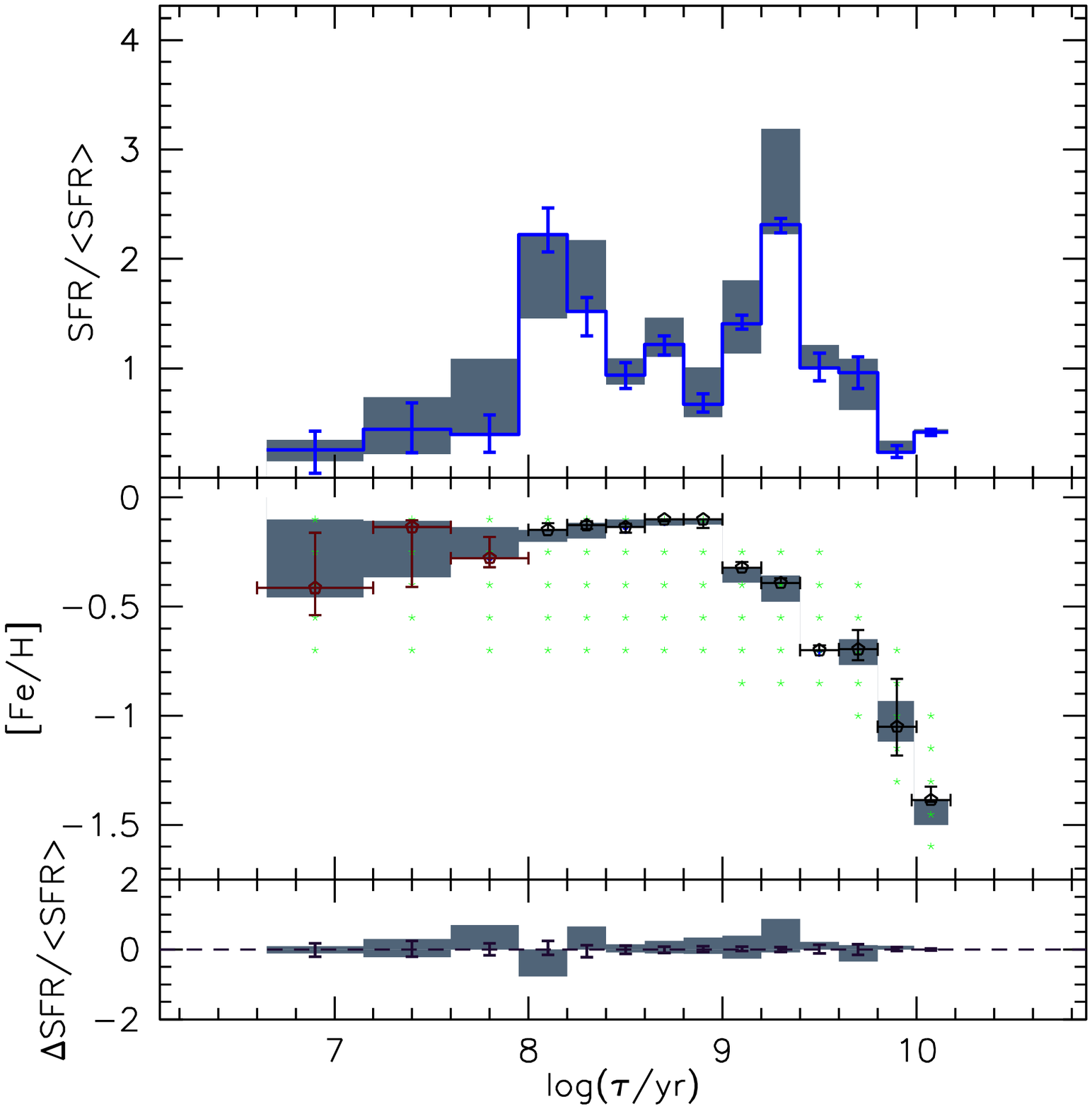}}
\resizebox{0.30\hsize}{!}{\includegraphics{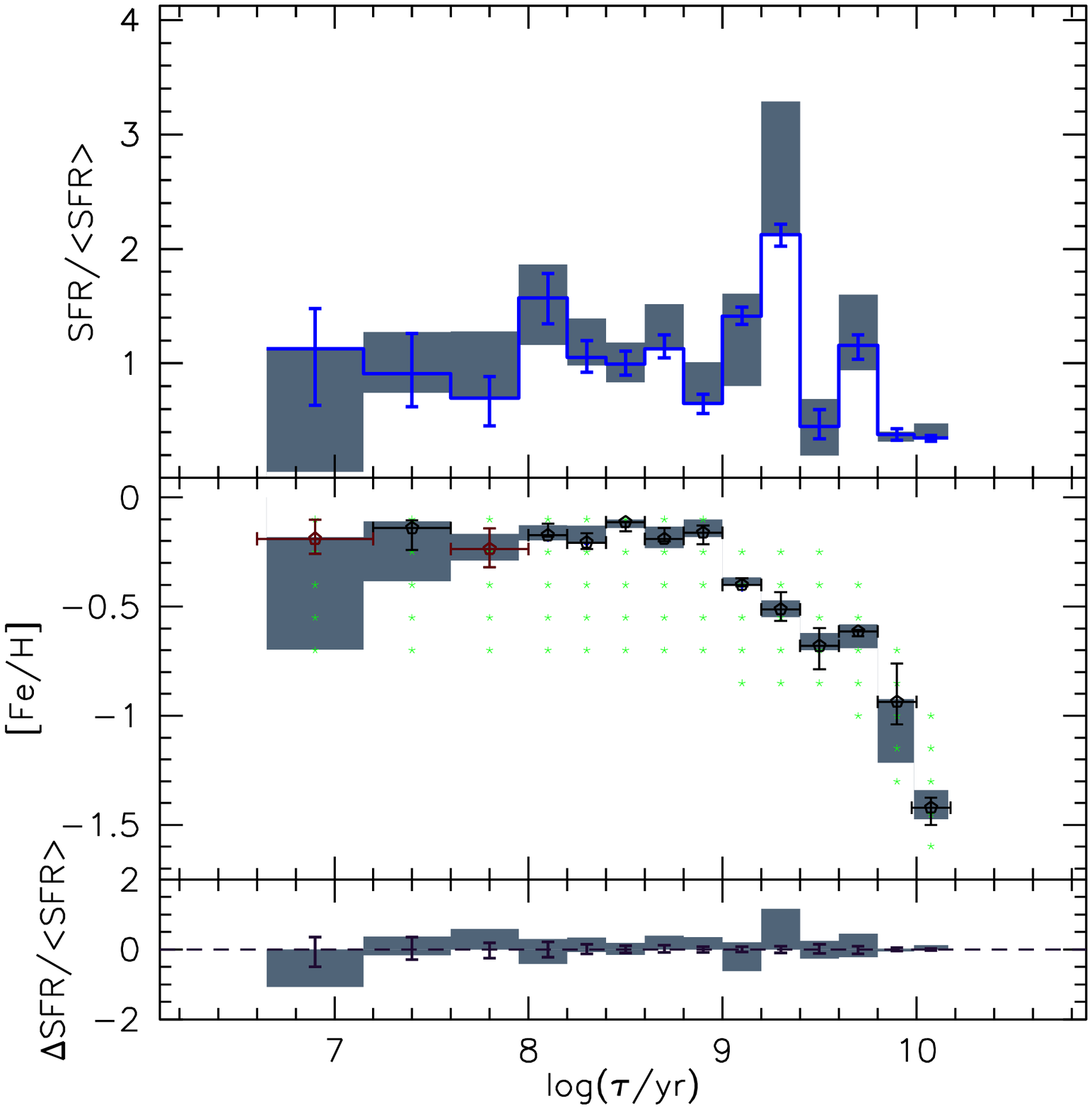}}
\resizebox{0.30\hsize}{!}{\includegraphics{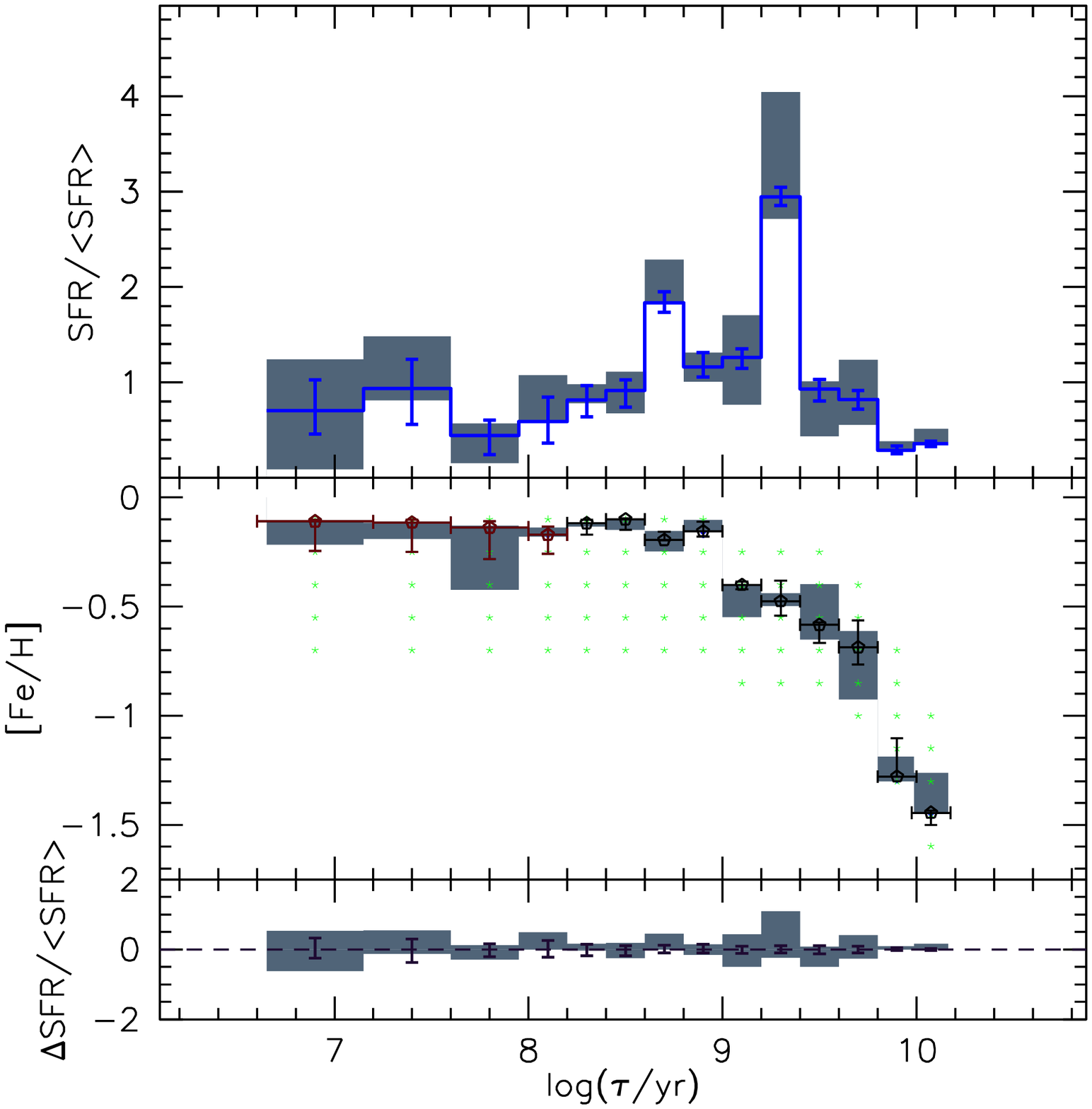}}
}
\end{minipage}
\caption{Same as in Fig.~\ref{sfr88}, but for the SFH for the 8\_3
  tile. The bottom left panels represent subregions closer to the LMC
  centre.}
\label{sfr83}
\end{figure*}

\begin{figure*}[p]
\begin{minipage}{\textwidth}
\subfigure[\SFRt\ and AMR of subregions G4, G8 and G12]{
%\subfigure[SFR of sub region G1, G2, and G3]{
\resizebox{0.30\hsize}{!}{\includegraphics{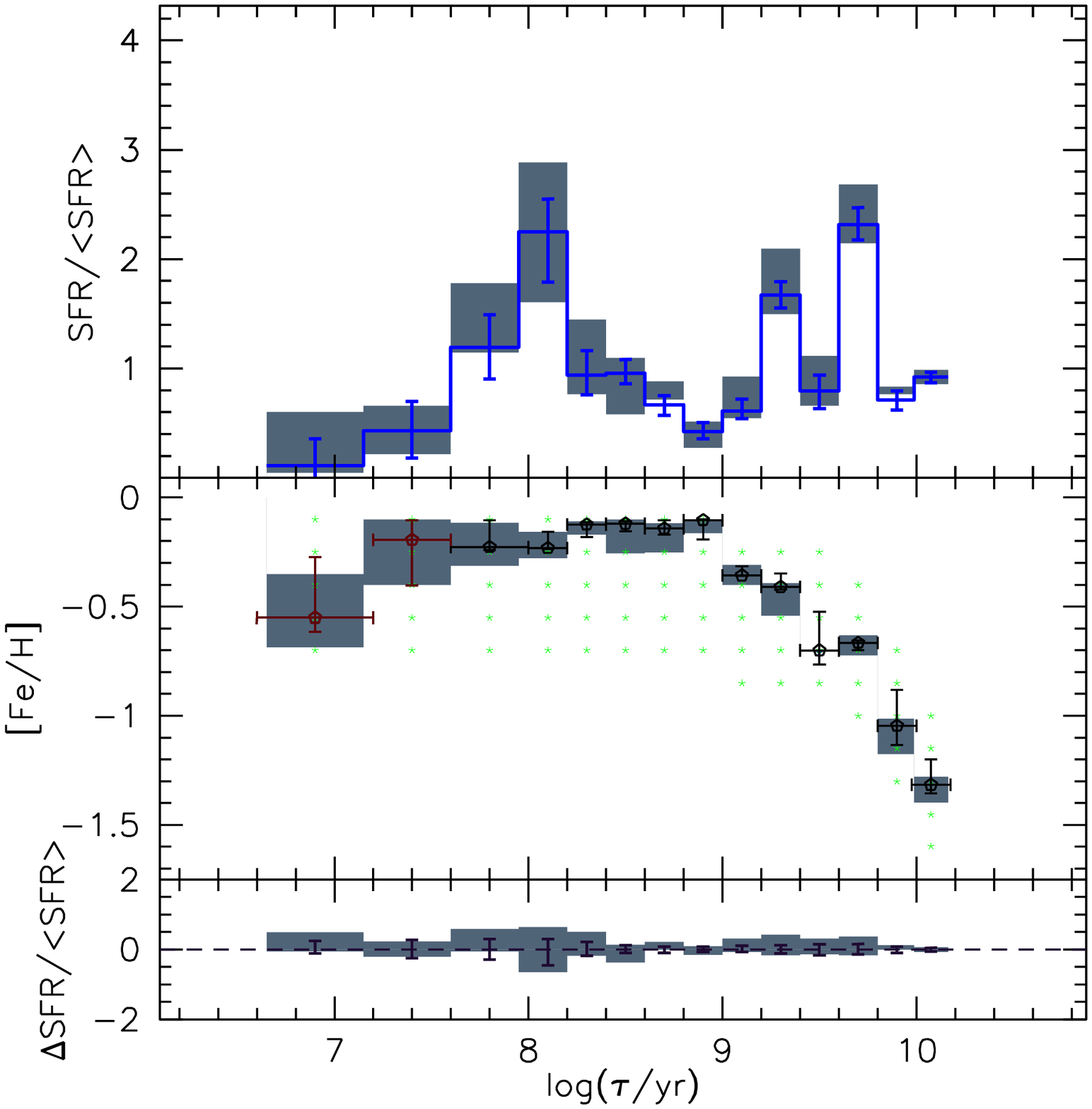}}
\resizebox{0.30\hsize}{!}{\includegraphics{sfr43_G8_18.525_0.42.eps}}
\resizebox{0.30\hsize}{!}{\includegraphics{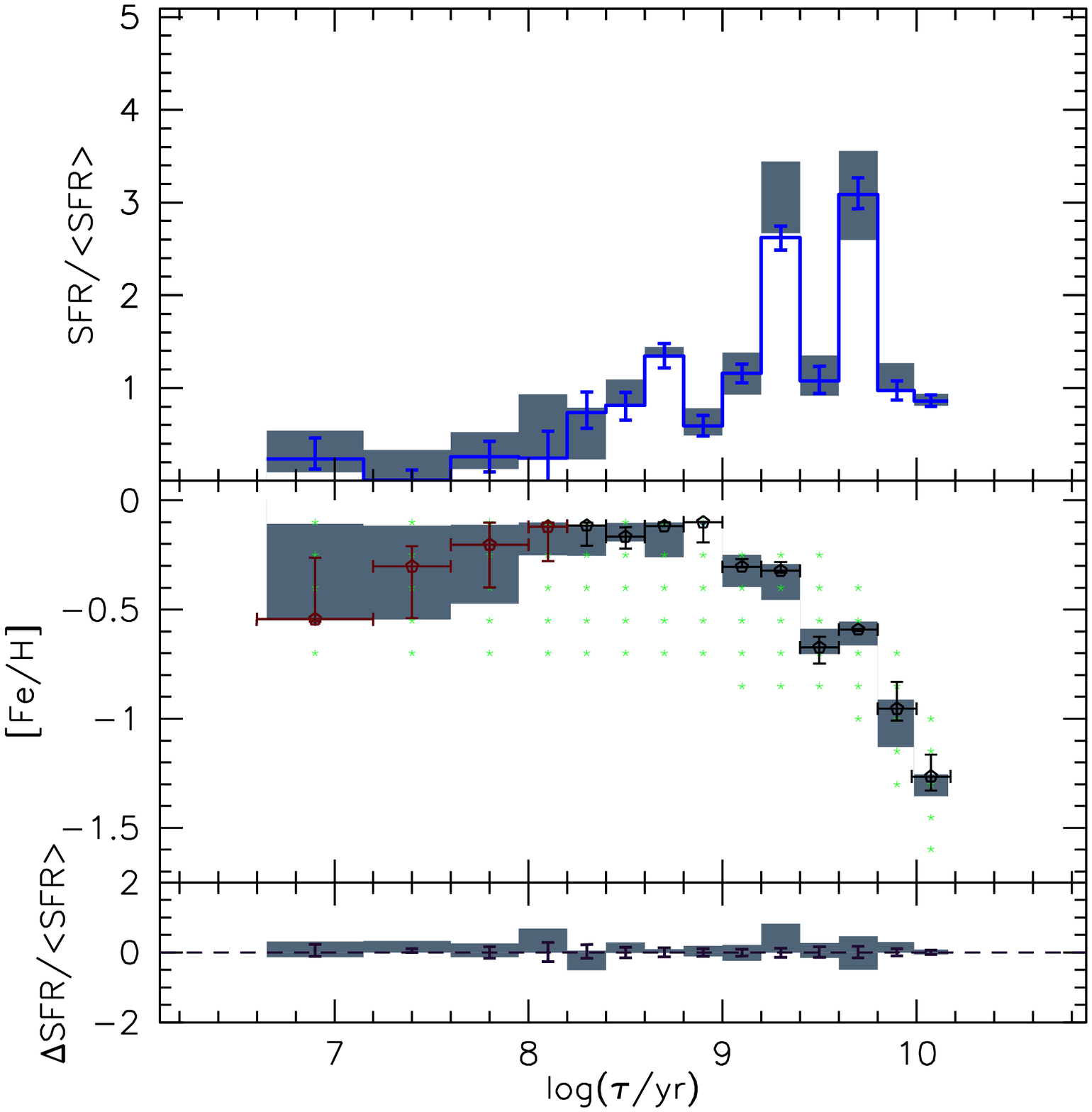}}
}
\end{minipage}
\hfill
\begin{minipage}{\textwidth}
\subfigure[\SFRt\ and AMR of subregions G3, G7 and G11]{
%\subfigure[SFR of sub region G4, G5 and G6]{
\resizebox{0.30\hsize}{!}{\includegraphics{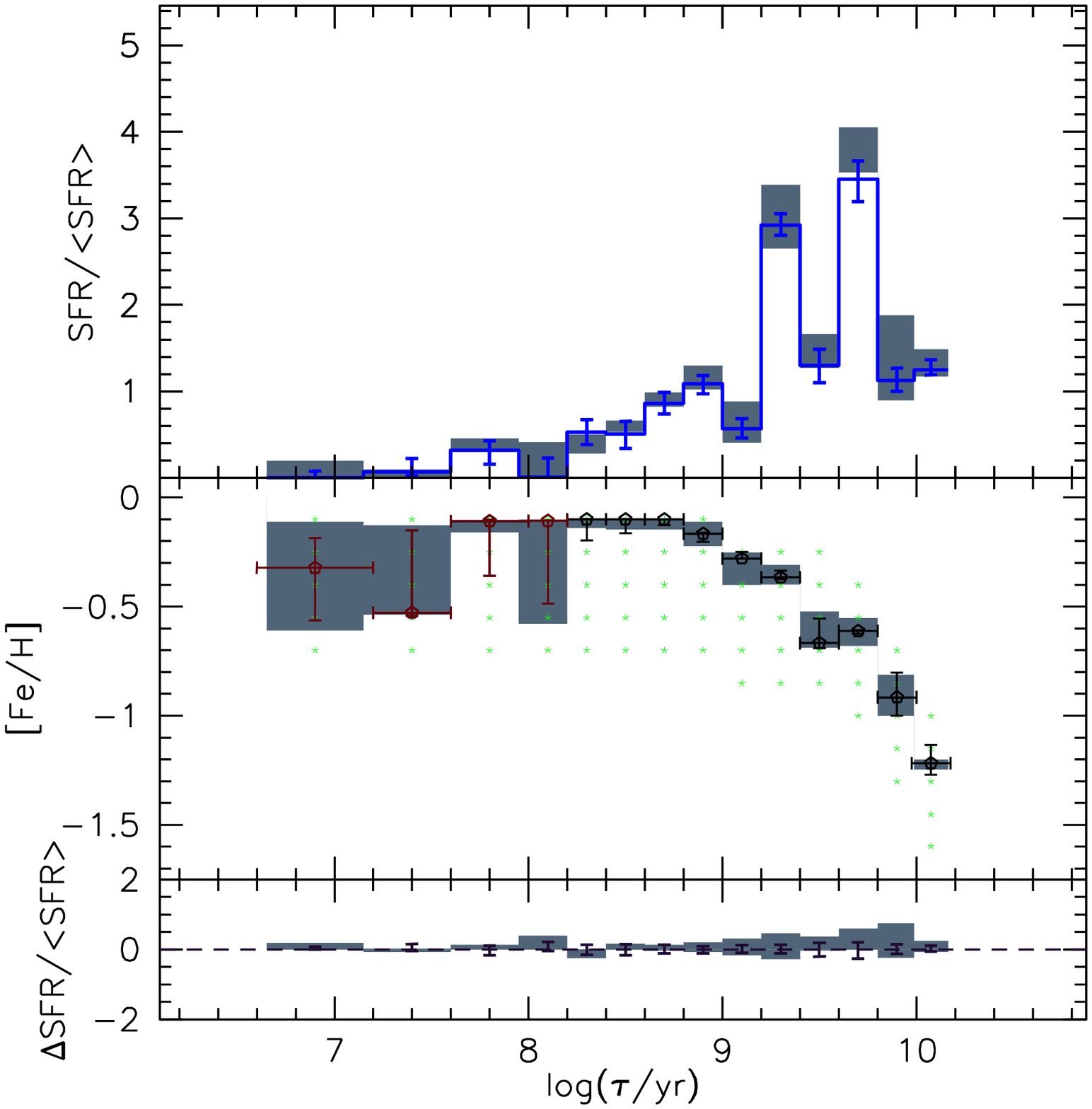}}
\resizebox{0.30\hsize}{!}{\includegraphics{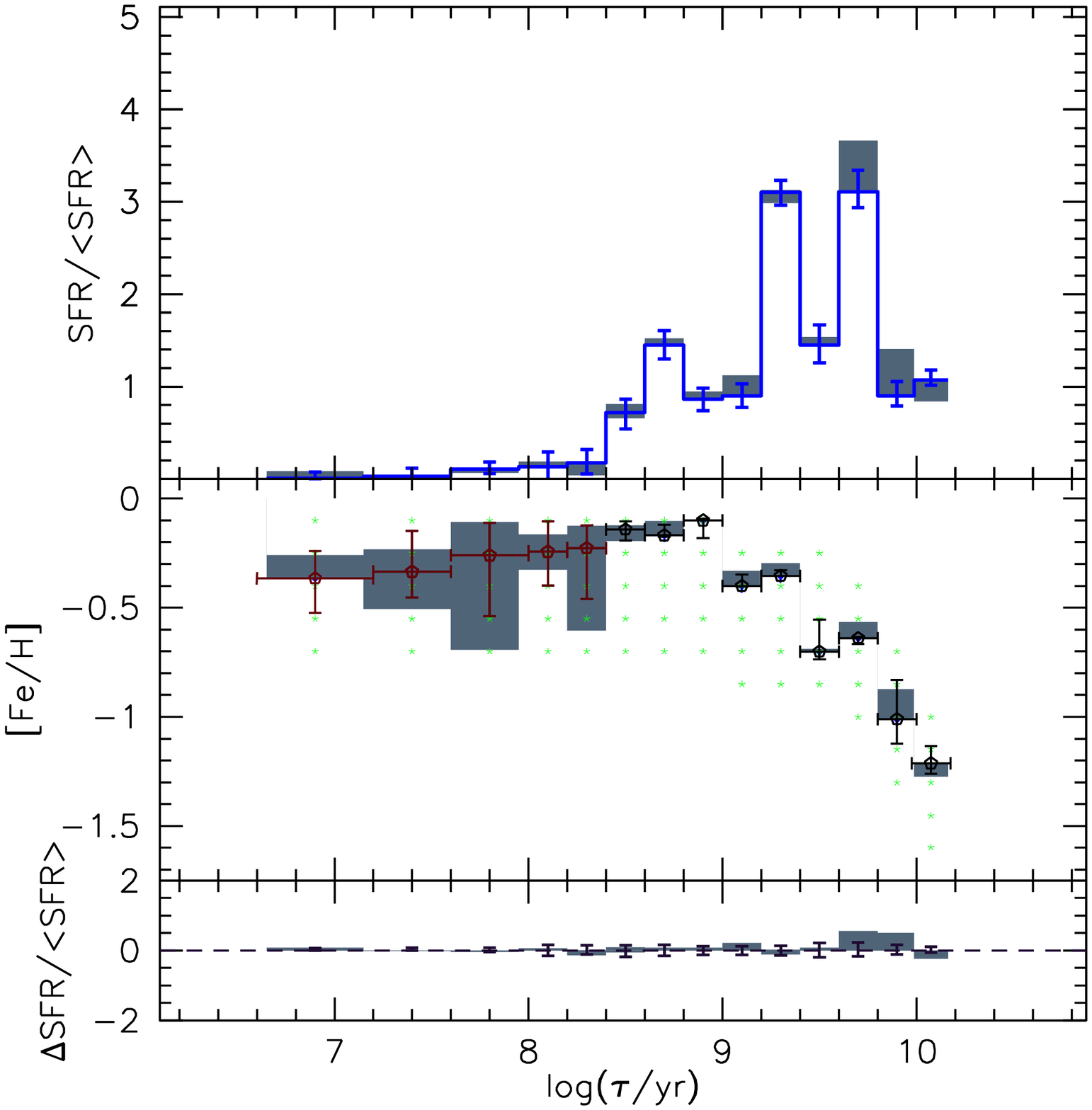}}
\resizebox{0.30\hsize}{!}{\includegraphics{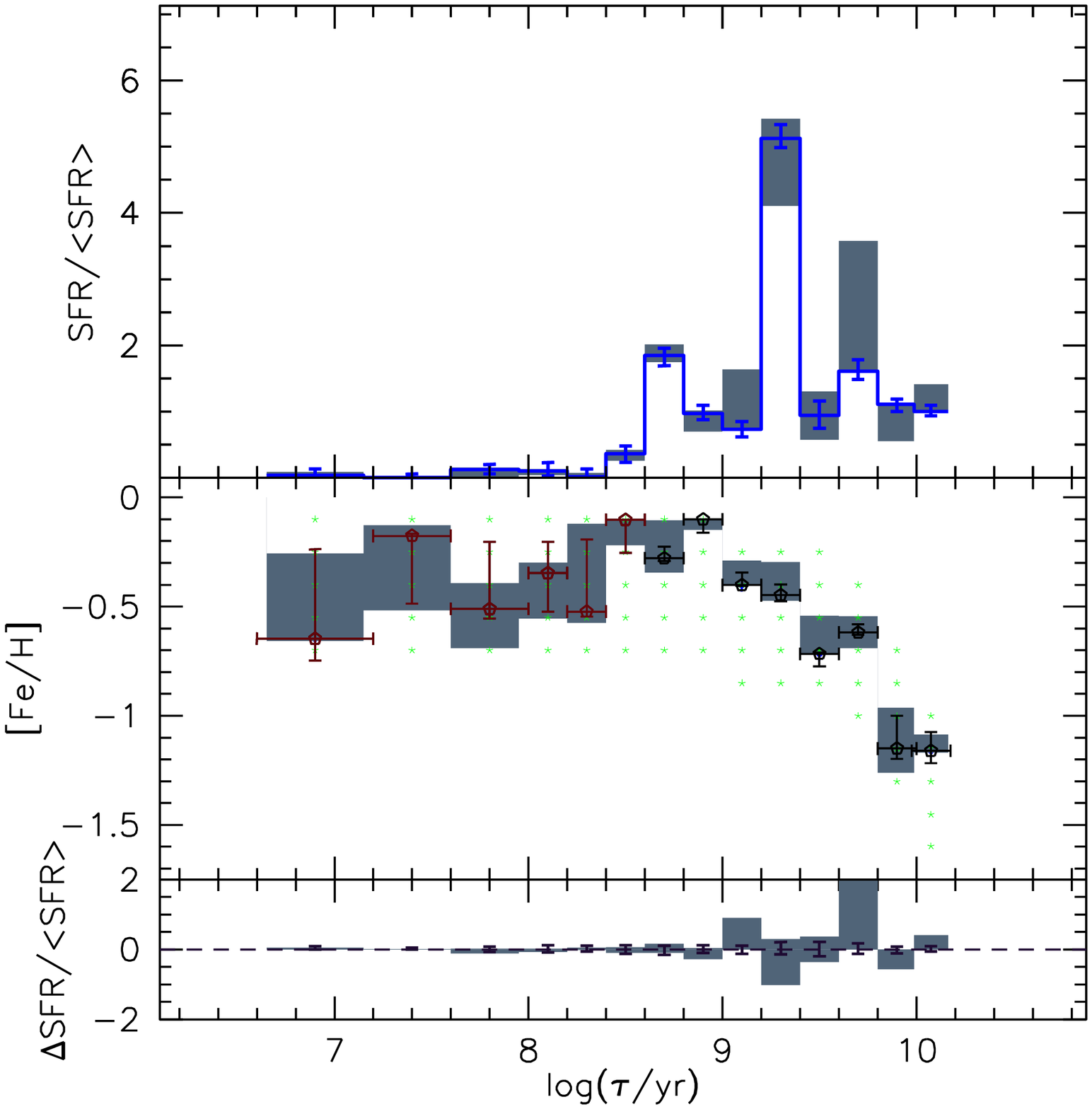}}
}
\end{minipage}
\hfill
\begin{minipage}{\textwidth}
\subfigure[\SFRt\ and AMR of subregions G2, G6 and G10]{
%\subfigure[SFR of sub region G7, G8 and G9]{
\resizebox{0.30\hsize}{!}{\includegraphics{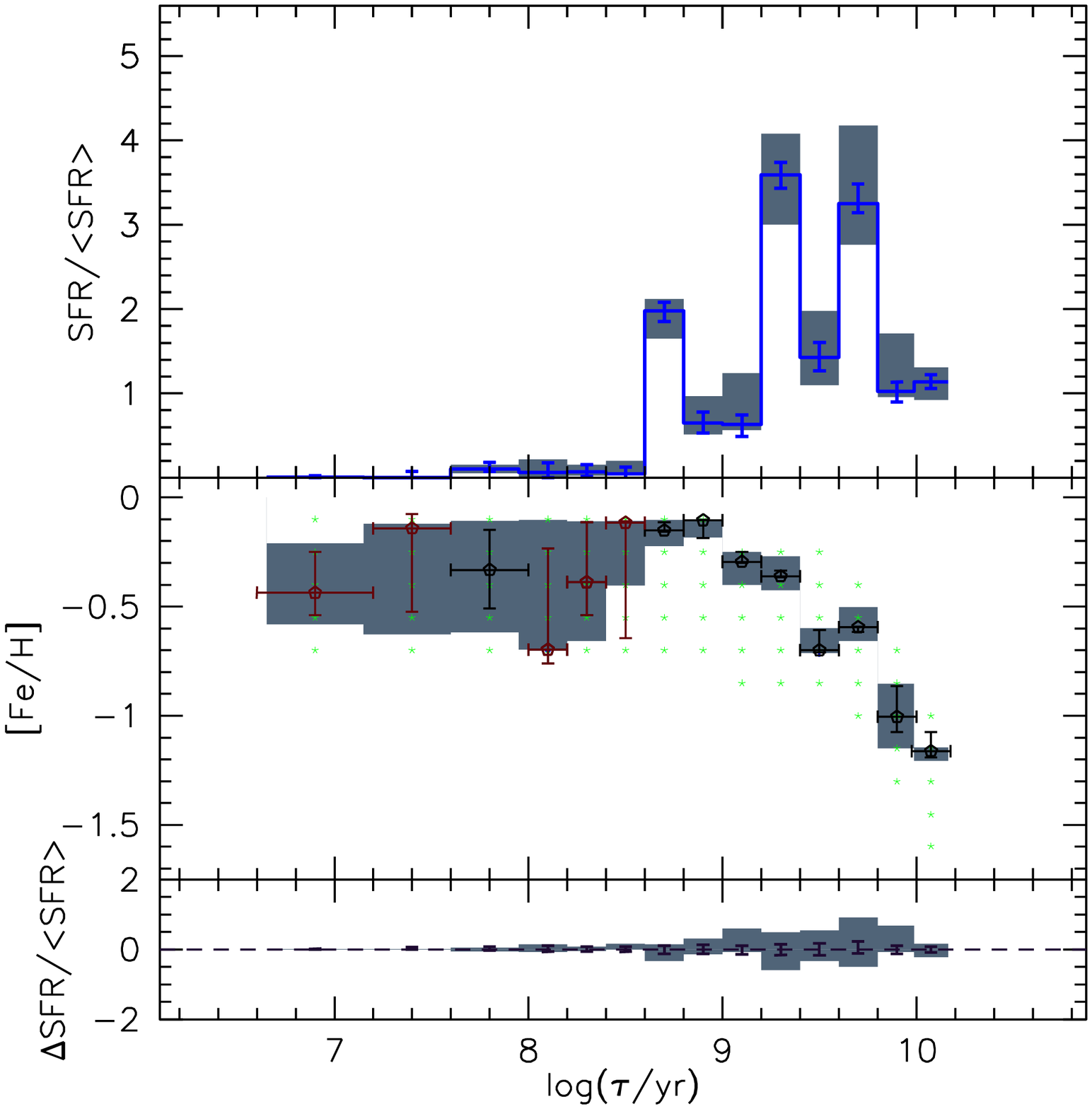}}
\resizebox{0.30\hsize}{!}{\includegraphics{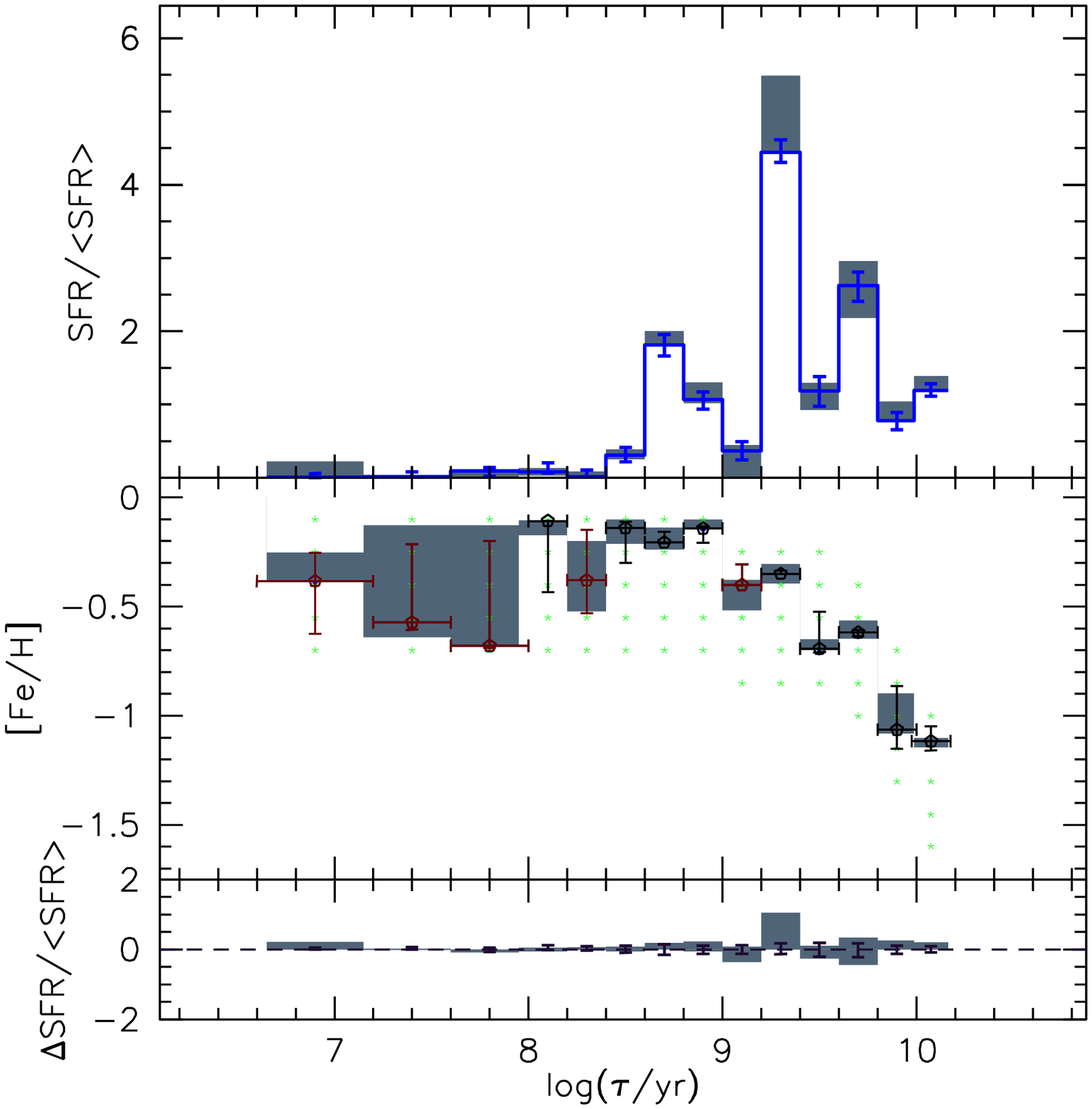}}
\resizebox{0.30\hsize}{!}{\includegraphics{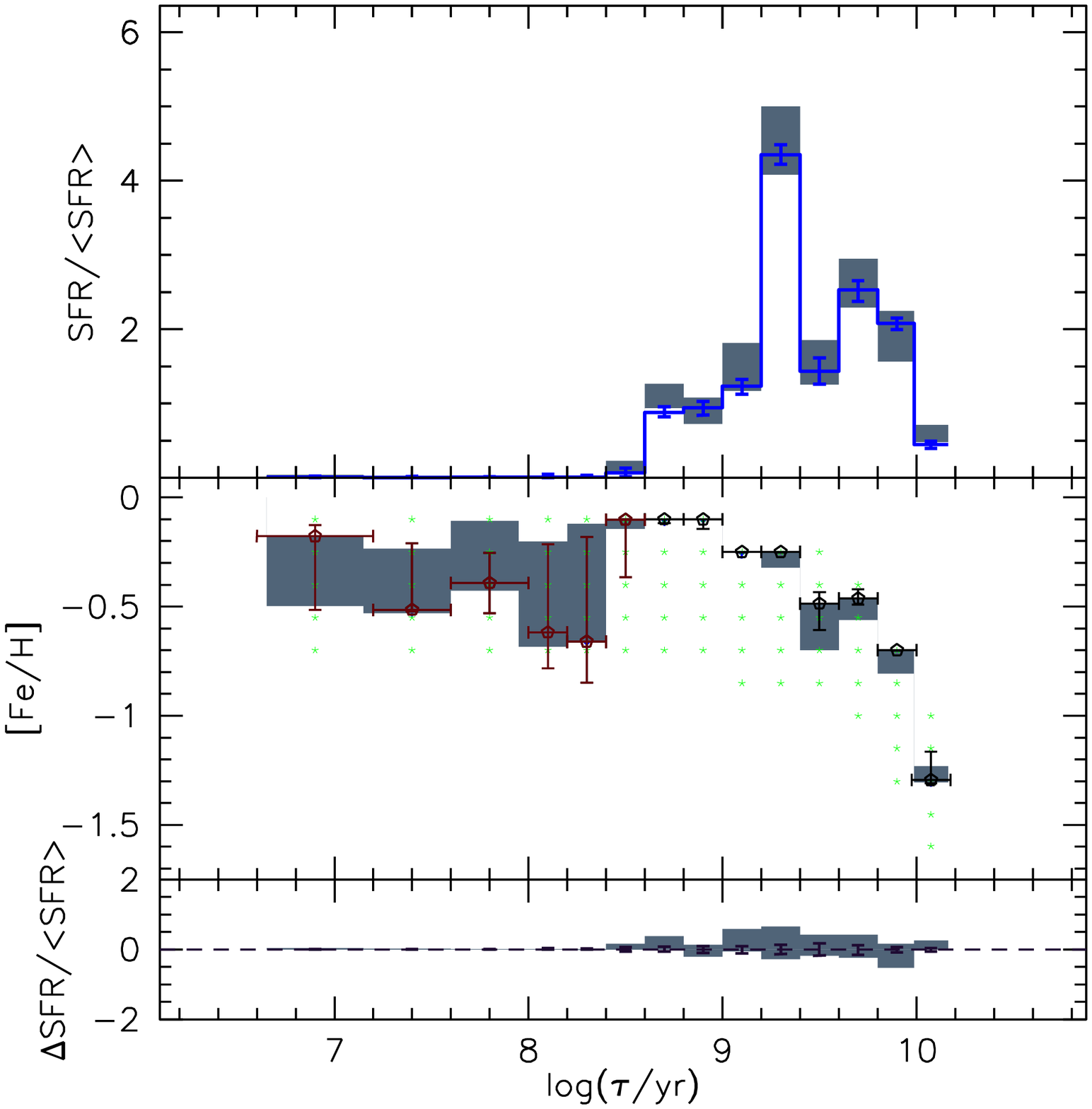}}
}
\end{minipage}
\hfill
\begin{minipage}{\textwidth}
\subfigure[\SFRt\ and AMR of subregions G1 and G5]{
%\subfigure[SFR of sub region G10, G11 and G12]{
\resizebox{0.30\hsize}{!}{\includegraphics{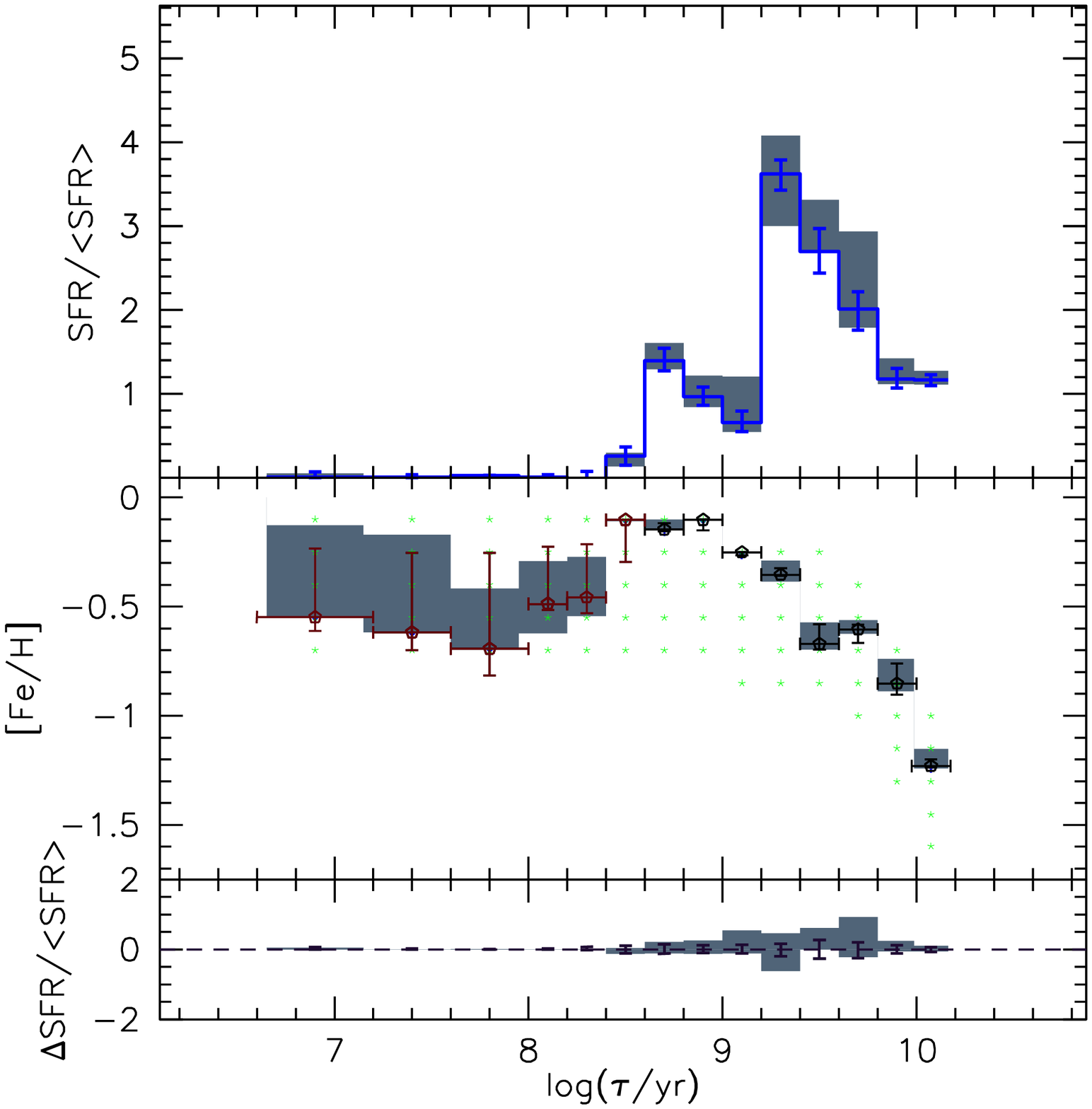}}
\resizebox{0.30\hsize}{!}{\includegraphics{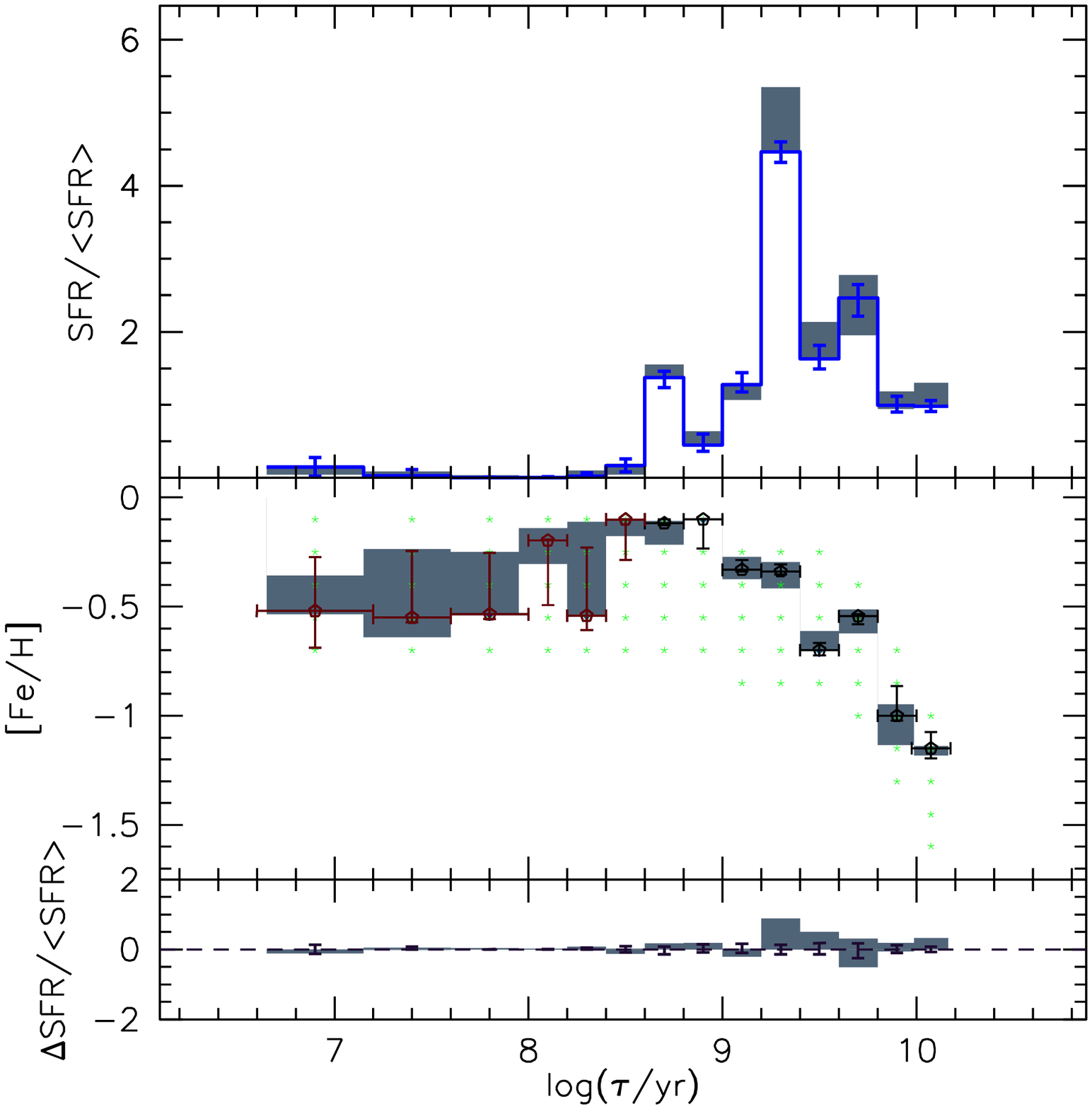}}
}
\end{minipage}
\caption{Same as in Fig.~\ref{sfr88}, but for the SFH for the 4\_3
  tile. The top left panels represent subregions closer to the LMC
  centre.}
\label{sfr43}
\end{figure*}

%\end{appendix}

\end{document}